\documentclass[11pt,a4paper]{article}
\usepackage{jcappub}
\usepackage{mathrsfs}     
\usepackage{bm}     
\usepackage{hyperref} 
\hypersetup{
    colorlinks=true,       
    linkcolor=red,          
    citecolor=blue,        
    filecolor=magenta,      
    urlcolor=blue           
}
\usepackage[all]{hypcap} 
\usepackage{physics} 
\usepackage{tensor} 
\usepackage{bbm} 
\usepackage[retainorgcmds]{IEEEtrantools} 
\usepackage{xcolor}
\usepackage{multirow}
\usepackage[multiple]{footmisc}

\newcommand{\ba}[1]{\begin{align} #1 \end{align}}
\newcommand{\bes}[1]{\begin{equation}\begin{split} #1 \end{split}\end{equation}}
\newcommand{\bsa}[2]{\begin{subequations}\label{#1}\begin{align} #2 \end{align}\end{subequations}}
\newcommand{\kvec}{{\bm k}}
\newcommand{\xvec}{{\bm x}}

\newcommand{\Lcal}{\mathcal{L}}

\newcommand{\Xbb}{\mathbb{X}}
\newcommand{\Mpl}{M_P}
\newcommand{\MP}{M_P}
\newcommand{\GeV}{{\; \mathrm{GeV}}}
\newcommand{\bea}{\begin{eqnarray}}  
\newcommand{\eea}{\end{eqnarray}}  
\newcommand{\LT}{L_T}
\newcommand{\LV}{L_V}
\newcommand{\LS}{L_S}

\newcommand{\LVk}{L_{V,\kvec}}
\newcommand{\LSk}{L_{S,\kvec}}

\newcommand{\fref}[1]{figure~\ref{#1}}
\newcommand{\Fref}[1]{Figure~\ref{#1}}
\newcommand{\tref}[1]{table~\ref{#1}}

\newcommand{\sref}[1]{section~\ref{#1}}

\newcommand{\aref}[1]{appendix~\ref{#1}}

\newcommand{\eref}[1]{eq.~(\ref{#1})}
\newcommand{\erefs}[2]{eqs.~(\ref{#1})~and~(\ref{#2})}

\newcommand{\rref}[1]{ref.~\cite{#1}}
\newcommand{\rrefs}[2]{refs.~\cite{#1}~and~\cite{#2}}

\title{Cosmological gravitational particle production of massive spin-2 particles}
\author[a]{Edward~W.~Kolb,}
\emailAdd{Rocky.Kolb@uchicago.edu}
\author[b]{\ Siyang~Ling,} 
\emailAdd{siyang.ling@rice.edu}
\author[b]{\\ Andrew~J.~Long,}
\emailAdd{andrewjlong@rice.edu}
\author[c]{\ and \ Rachel~A.~Rosen}
\emailAdd{rarosen@cmu.edu}

\affiliation[a]{Kavli Institute for Cosmological Physics \\ and Enrico Fermi Institute, The University of Chicago, \\ 5640 South Ellis Avenue, Chicago, IL 60637, U.S.A.}
\affiliation[b]{Department of Physics and Astronomy, Rice University, \\ 6100 Main Street, Houston, TX 77005, U.S.A.}
\affiliation[c]{Department of Physics, Carnegie Mellon University, \\ 5000 Forbes Avenue, Pittsburgh, PA 15213, U.S.A.}

\abstract{
The phenomenon of cosmological gravitational particle production (CGPP) is expected to occur during the period of inflation and the transition into a hot big bang cosmology.  Particles may be produced even if they only couple directly to gravity, and so CGPP provides a natural explanation for the origin of dark matter.  In this work we study the gravitational production of massive spin-2 particles assuming two different couplings to matter.  We evaluate the full system of mode equations, including the helicity-0 modes, and by solving them numerically we calculate the spectrum and abundance of massive spin-2 particles that results from inflation on a hilltop potential.  We conclude that CGPP might provide a viable mechanism for the generation of massive spin-2 particle dark matter during inflation, and we identify the favorable region of parameter space in terms of the spin-2 particle's mass and the reheating temperature.  As a secondary product of our work, we identify the conditions under which such theories admit ghost or gradient instabilities, and we thereby derive a generalization of the Higuchi bound to Friedmann-Robertson-Walker (FRW) spacetimes.  
}

\keywords{
dark matter, inflation, massive spin-2 particles, bigravity
}

\begin{document}
 \maketitle
\flushbottom

\section{Introduction}
\label{sec:intro}

The study of inflationary perturbations of the massless gravitational field are a standard part of any graduate course on cosmology~\cite{Baumann:2022mni}.  (See also \rref{Schiappacasse:2016nei}.)  Additionally, the study of perturbations in massive lower-spin fields, either during inflation or after inflation during reheating, i.e., cosmological gravitational particle production (CGPP), has its own long history~\cite{Parker:1969au,Parker:1971pt,Ford:2021syk} with applications to cosmological relics such as dark matter~\cite{Chung:1998zb,Chung:1998ua}.  However, there have not been any {\it{comprehensive}} studies of CGPP with massive spin-2 particles. (Although there have been steps in that direction, see \rref{Alexander:2020gmv}).  In this work we investigate the evolution of cosmological perturbations in a massive spin-2 field during the epoch of inflation and the period of reheating after inflation, and we assess the implications for spin-2 particle dark matter.

Such a systematic investigation of CGPP is well motivated: if these massive spin-2 particles have a lifetime greater than the age of the universe they provide a candidate for the dark matter. If their lifetime is less than the age of the universe, their late decays might have interesting cosmological implications.  Moreover, in the context of bigravity which contains one massless spin-2 field (i.e., the graviton) and one massive spin-2 field, one cannot `turn off' the gravitational interactions that lead to CGPP.  The phenomenology of spin-2 dark matter has been studied in refs.~\cite{Aoki:2014cla,Aoki:2016zgp,Babichev:2016bxi,Marzola:2017lbt,GonzalezAlbornoz:2017gbh,Armaleo:2019gil,Armaleo:2020yml,Alexander:2020gmv,Manita:2022tkl}, and our work offers a production mechanism for these particles.

In this article, to perform our analysis, we consider two different constructions of a free massive spin-2 particle on a Friedmann-Robertson-Walker (FRW) background. In the first, more straightforward construction, we generalize the Fierz-Pauli theory \cite{Fierz:1939ix} of a free massive spin-2 particle to an FRW background in a way that maintains the absence of the Boulware-Deser ghost \cite{Boulware:1972yco}.  Though not necessary for the construction, we show how this theory can be obtained from ghost-free bigravity \cite{Hassan:2011zd}.  In the second construction, we consider an exotic nonminimal coupling \cite{deRham:2014naa} of the massive spin-2 field to matter which allows for an alternative free Lagrangian on an FRW background.  We also show how this theory can be obtained from ghost-free bigravity.

An extensive body of literature has explored cosmological solutions and perturbations in several theories of bigravity; see refs.~\cite{DeFelice:2013bxa,deRham:2014zqa} for a review.  
Most of this work has focused on a class of theories where matter only couples to one of the two dynamical metrics~\cite{Hassan:2011zd}.  These studies find that there exist homogeneous and isotropic solutions at the background level~\cite{Volkov:2011an,vonStrauss:2011mq,Comelli:2011zm,Volkov:2012cf,Akrami:2012vf,Koennig:2013fdo}, which is a required feature for a viable cosmological model.  However, the perturbations around these solutions (on at least some branches) are unstable~\cite{Comelli:2012db} (see also refs.~\cite{Koennig:2014ods,Lagos:2014lca,Akrami:2015qga}).  
The instabilities can be evaded if both metrics couple to matter.  One such theory employing two distinct matter sectors was proposed in \rref{Comelli:2012db} and studied in \rref{Comelli:2014bqa}, and another such theory with a single matter sector coupled to a composite effective metric was proposed in \rref{deRham:2014naa} and its cosmology was studied in refs.~\cite{Enander:2014xga,EmirGumrukcuoglu:2014uog,Gumrukcuoglu:2015nua}.  These latter two theories are relevant for our interests here.  In regard to the study of cosmological perturbations in bigravity, our work is distinct from previous studies insofar as we focus on matter couplings that admit equal FRW background solutions for the two metrics, in order to derive the simplest free Lagrangians for the massive spin-2 field.  

Earlier work on inflationary perturbations, e.g., for cosmic microwave background observables, has typically assumed that the additional spectator field is light (mass much smaller than the expansion rate during inflation) and the perturbation amplitude is set when modes leave the horizon during inflation.  However, studies on CGPP have also extended these calculations to models in which the spectator is heavy (mass larger or comparable to the expansion rate during inflation) and  particle production happens near the end of inflation or after inflation during the epoch of reheating.  For massive gravity on a de Sitter background, the Higuchi bound constrains the spin-2 particle's mass $m^2 > 2 H^2$, as otherwise the theory would propagate a ghost~\cite{Higuchi:1986py}.  A similar bound is expected to be realized in theories of massive gravity and bigravity on an FRW background~\cite{Fasiello:2012rw,Fasiello:2013woa}.  As part of our analysis we find a generalization of the Higuchi bound to FRW spacetimes: $m^2 > 2 H^2 (1 - \epsilon)$ where $\epsilon$ is the first slow roll parameter. Since we find the spin-2 particle's mass must exceed the Hubble scale during inflation, particle production happens primarily at the end of inflation and during reheating, which motivates the numerical analysis that we pursue here.  

The remainder of this article is organized as follows.  In section \sref{sec:theories}, we use ghost-free bigravity to derive two distinct theories of a free massive spin-2 particle on an curved background that we will consider.  
Focusing on FRW cosmologies, we perform a scalar-vector-tensor decomposition and present the resultant mode equations in \sref{sec:SVT}.  Some of the mode equations exhibit instabilities, which we discuss in \sref{sec:instabilities}, where we also present an FRW generalization of the Higuchi bound.  Our numerical results appear in \sref{sec:GPP}, including the spectra and relic abundance of gravitationally-produced particles.  In \sref{sec:summary} we summarize and conclude.  The article is supplemented by several appendices: \aref{app:discussion_of_eoms} offers an analytical understanding of the long-wavelength spectrum; \aref{app:stueck} presents an alternative derivation of our FRW-generalized Higuichi bound; and finally, \aref{app:decay} contains a discussion of the stability and decay of massive spin-2 particles in our two theories. 

\section{Massive Spin-2 Fields in an FRW Background}
\label{sec:theories}

 To study the gravitational production of a spin-2 particle species of mass $m$ (which is not the massless graviton) during inflation, we desire an effective field theory that describes a free, massive, spin-2 field and a scalar inflaton field on a fixed curved spacetime background. This free theory with the usual minimally coupled matter sector can be obtained trivially from General Relativity (GR): one can simply expand the Einstein-Hilbert action plus matter sector to quadratic order in perturbations around any background that satisfies the GR equations of motion and then add the Fierz-Pauli mass term \cite{Fierz:1939ix}. The resulting theory of a free massive spin-2 field on a fixed background will be ghost-free, while non-linearities will generically introduce a ghost.

This same free Lagrangian can also be derived from non-linear ghost-free massive gravity \cite{deRham:2010kj}, which describes a self-interacting massive spin-2 particle, and also from ghost-free bigravity \cite{Hassan:2011zd} which describes an interacting massive spin-2 particle and massless spin-2 particle. In addition to giving the free theory, these formulations allow one to consider higher order perturbations while maintaining the contraint that removes the ghost, as well as specific ghost-free nonminimal couplings to matter, and, in the bigravity case, interactions between the massive spin-2 particle and the graviton (i.e., the massless spin-2 particle).  The virtue of the particular interacting theories given in \rrefs{Hassan:2011zd}{deRham:2010kj}, in contrast to, say, Kaluza-Klein theories containing massive spin-2 particles, is that the mass of the spin-2 particle is parametrically lower than the cutoff of the effective field theory so that no new states need to be introduced into the low energy theory beyond a single massive spin-2.

In this section we show how the free Lagrangian of a massive spin-2 field on a fixed background can be derived from bigravity.  In addition to considering the usual minimal matter coupling, we consider the free theory that arises from an exotic nonminimal coupling to matter that has been shown to be ghost-free below the strong coupling scale of the non-linear effective theory $\Lambda_3=(m^2\Mpl)^{1/3}$, where $\Mpl$ is the Planck mass \cite{deRham:2014naa}.  We will consider both the minimally-coupled theory and the nonminimally-coupled theory when studying the gravitational production of massive spin-2 particles.  

\subsection{Ghost-Free Bigravity}
\label{sub:bigravity}

We will construct two different free Lagrangians for a massive spin-2 particle on an FRW background starting from ghost-free bigravity \cite{Hassan:2011zd}.  Bigravity is an interacting theory describing one massive and one massless spin-2 particle, with possible couplings to additional matter fields.\footnote{Massive gravity can be considered as a limit of the bigravity theory when there is a large hierarchy between the Planck masses of the two particles: the massless spin-2 eigenstate effectively freezes out and one is left with only the massive degree of freedom.} In the first construction that we consider, the bigravity theory is minimally coupled to matter and one gets the expected result for the free massive spin-2 Lagrangian: it's simply the free Lagrangian of a massless spin-2 particle (i.e., General Relativity at quadratic order) plus the Fierz-Pauli mass term.  The bigravity formulation allows one to also consider exotic ghost-free nonminimal couplings to the matter sector \cite{deRham:2014naa}. In the second construction, we will derive the free action the nonminimally-coupled theory.

We start from the most general non-linear ghost-free bimetric action.  The action contains an Einstein-Hilbert term for each of the two metrics $g_{\mu\nu}$ and $f_{\mu\nu}$, a non-derivative potential term that mixes them, and matter terms:\footnote{We use the $(+,+,+)$ Misner-Thorne-Wheeler sign convention~\cite{MTW:1973}, with mostly plus signs in the Minkowski metric.}
\bes{
\label{bigL}
	S
	& = \int \! \mathrm{d}^4 x \biggl[ \frac{M_g^2}{2} \sqrt{-g}\,R[g]+ \frac{M_f^2}{2} \sqrt{-f}\,R[f] -m^2 M_*^2 \sqrt{-g} \,V(\Xbb; \beta_n) \\
	& \qquad \qquad 
	+ \sqrt{-g} \,\Lcal_{g}(g,\phi_g)+\sqrt{-f} \,\Lcal_{f}(f,\phi_f)+\sqrt{-g_\star} \,\Lcal_{\star}(g_\star,\phi_\star) \biggr] 
	\;.
}
The parameters $M_g$ and $M_f$ determine the effective mass $M_\ast^2 = \bigl( M_g^{-2} + M_f^{-2} \bigr)^{-1}$ and the reduced Planck mass $\MP^2 = M_g^2+M_f^2$.  The mass parameter $m$ sets the mass of the spin-2 particle.  The metric interaction potential can be written as 
\bea
\label{pot}
	V(\Xbb; \beta_n) \equiv \sum_{n=0}^{4} \beta_n S_n(\Xbb) \, , \quad 
	S_n(\Xbb) \equiv \Xbb^{\mu_1}_{[ \mu_1} \hdots \Xbb^{\mu_n}_{\mu_n ]} \, , \quad 
	\Xbb\indices{^\mu_\sigma}\Xbb\indices{^\sigma_\nu} \equiv g\indices{^\mu^\lambda} f\indices{_\lambda_\nu} 
	\;,
\eea
which depends on the five parameters $\beta_0$ through $\beta_4$.  
A potential of this form guarantees that the classical theory propagates only the correct five degrees of freedom of the massive spin-2 \cite{Hassan:2011hr} and no additional Boulware-Deser ghost \cite{Boulware:1972yco}. 
The matter sectors are discussed further below, including $g_\star$, which is a composite metric defined in \eref{geff}.  

The $\beta$ parameters determine the mass of the spin-2 particle and the cosmological constants $\Lambda_g$ and $\Lambda_f$, and parametrize higher-order interactions between the two metrics.  We take 
\bea
	\label{mass}
	\beta_1 + 2 \beta_2 + \beta_3 = 1
	\;,
\eea
which normalizes the Fierz-Pauli mass to be $m$.  The cosmological constants are
\bea
\label{cc}
	\Lambda_g = m^2 \bigl( \beta_0 + 3 \beta_1 + 3 \beta_2 + \beta_3 \bigr) 
	\qquad \text{and} \qquad 
	\Lambda_f = m^2 \bigl( \beta_1 + 3 \beta_2 + 3 \beta_3 + \beta_4 \bigr) 
	\;.
\eea
Only these linear combinations of the five $\beta_n$ parameters appear in the free action (quadratic in perturbations); the remaining combinations only enter through higher-order interactions.  

There are three matter sectors: one that couples minimally to the metric $g$, one that couples minimally to the metric $f$, and one that couples to a composite metric $g_\star$ given by
\bea
\label{geff}
(g_\star)_{\mu\nu} = \frac{a^2}{(a+b)^2}\,g_{\mu\nu}+\frac{ab}{(a+b)^2}\left( g_{\mu\lambda} \bigl( \sqrt{g^{-1}f} \bigr)^\lambda_{~\nu} + \bigl( \sqrt{g^{-1}f} \bigr)_\mu^{~\lambda} \, g_{\lambda \nu} \right)+\frac{b^2}{(a+b)^2}\,f_{\mu\nu}\, ,
\eea
with free parameters $a$ and $b$.  This form guarantees a ghost-free matter coupling below the strong coupling scale of the nonlinear effective theory $\Lambda_3=(m^2\Mpl)^{1/3}$ \cite{deRham:2014naa}. 
We take the matter sectors to be that of three independent scalar fields $\phi_g$, $\phi_f$ and $\phi_\star$, that are coupled to gravity with Lagrangians of the form:
\bsa{eq:Lagrangians}{
    {\cal L }_g(g,\phi_g) & = -\tfrac{1}{2}g^{\mu\nu} \nabla_\mu \phi_g \nabla_\nu \phi_g - V_g(\phi_g) \, ,\\
    {\cal L }_f(f,\phi_f) & = -\tfrac{1}{2}f^{\mu\nu} \nabla_\mu \phi_f \nabla_\nu \phi_f - V_f(\phi_f) \, ,\\
    {\cal L }_\star(g_\star,\phi_\star) & = -\tfrac{1}{2}g_\star^{\mu\nu} \nabla_\mu \phi_\star \nabla_\nu \phi_\star - V_\star(\phi_\star) \, .
}
These fields will source the background FRW metrics seen by the massive spin-2 fields. In the minimally-coupled theory, a combination of $\phi_g$ and $\phi_f$ will play the role of the inflaton, whereas in the nonminimally-coupled theory the inflaton is identified with $\phi_\star$ alone.  

\subsection{Minimal Matter Coupling}
\label{sub:minimal_theory}

In the absence of the composite metric, i.e. $\Lcal_\star = 0$, the $g$ and $f$ metrics are minimally coupled to their respective matter fields.  Taking this bigravity theory and expanding both metrics and their corresponding matter fields around the same background solutions, we recover the free Lagrangian for the massive degrees of freedom that is equivalent to linearized General Relativity plus a Fierz-Pauli mass term.  

In particular, to derive the free Lagrangian, we expand the metrics and scalar fields around backgrounds (denoted by a bar): 
\begin{subequations}\label{eq:model1_bkg}
\ba{
    g_{\mu\nu} & = \bar{g}_{\mu\nu} + \frac{2}{M_g} h_{\mu\nu} 
    \, , \quad 
    f_{\mu\nu} = \bar{f}_{\mu\nu} + \frac{2}{M_f} k_{\mu\nu} 
    \, , \quad 
    \phi_g = \bar{\phi}_g + \varphi_g 
    \, , \quad \text{and} \quad 
    \phi_f = \bar{\phi}_f + \varphi_f 
    \, .
}
We seek solutions of the background equations of motion with 
\ba{\label{eq:mirrored}
    \bar{g}_{\mu\nu} = \bar{f}_{\mu\nu} 
    \qquad \text{and} \qquad 
    \frac{1}{M_g} \, \bar{\phi}_g = \frac{1}{M_f} \, \bar{\phi}_f \equiv \frac{1}{\MP} \, \bar{\phi}
    \, .
}
\end{subequations}
The existence of such solutions imposes a stringent constraint on the model, which requires the two matter sectors to be mirrored in the sense that 
\bea\label{eq:mirroring_conditions}
    \frac{1}{M_g^2} \, V_g\left(\tfrac{M_g}{\MP} \phi\right) 
    = \frac{1}{M_f^2} \, V_f\left(\tfrac{M_f}{\MP} \phi\right) 
    \equiv \frac{1}{\MP^2} \, V(\phi) 
    \qquad \text{and} \qquad 
    \frac{\Lambda_g}{M_g^2} = \frac{\Lambda_f}{M_f^2} \equiv \frac{\Lambda}{\MP^2}
    \;.
\eea
In other words, if the $g$-sector contains a term $V_g(\phi_g) \supset c_g \phi_g^n$, then the $f$-sector must contain a term $V_f(\phi_f) \supset c_f \phi_f^n$ with $c_f = c_g (M_g / M_f)^{n-2}$.  This relation links the masses and couplings of the two scalar fields.  We keep track of the cosmological constant $\Lambda$ for the analytic expressions, but we set $\Lambda = 0$ for our numerical analysis; the inflationary phase is driven by $V(\bar{\phi}) > 0$.

There are several virtues to expanding around the same background for both metrics.\footnote{In the literature ``proportional" solutions with $\bar{g}_{\mu\nu} = c^2 \bar{f}_{\mu\nu}$ are also often considered.  However, in our setup the constant parameter $c$ would be a rescaling, which can be absorbed into redefinitions of the fields and parameters and does not constitute an independent free parameter.}\footnote{We note that, from the point of view of the bigravity theory, insisting on equal background solutions and the corresponding mirroring of the matter sectors amounts to a tuning. Our incentive here is simply to show how the simplest free Lagrangian, i.e., the generalization of Fierz-Pauli to FRW, can arise from the ghost-free bigravity theory, which necessitates equal backgrounds.  We also note that mirroring can appear in dimensional deconstruction models with two-site discretization.  For a review see \rref{deRham:2014zqa}.}  For one, it makes the expansion of the square-root matrix $\mathbb{X}$ that appears in \eref{pot} simple.  Furthermore, the background equations of motion are simply 
\bsa{eq:model1_bkg_EOM}{
    \bar{R}_{\mu\nu} - \frac{1}{2}\bar{g}_{\mu\nu}\bar{R} + \Lambda \bar{g}_{\mu\nu} & = \frac{1}{\MP^2} \, \bar{T}_{\mu\nu} \label{Einstein_eqn} \, , \\ 
    \Box \bar{\phi} - V'(\bar{\phi}) = 0 \label{phi_eqn} \, ,
}
where the mass term, i.e., the potential term multiplied by $m^2$ in \eref{bigL}, has dropped out.  Here we have used the background stress-energy tensor: 
\ba{
    \bar{T}_{\mu\nu} & = \nabla_\mu \bar{\phi} \nabla_\nu \bar{\phi} + \bar{g}_{\mu\nu} \bar{\Lcal}(\bar{g},\bar{\phi}) \, , \\ 
    \bar{\Lcal}(\bar{g}, \bar{\phi}) & = -\tfrac{1}{2}\bar{g}^{\mu\nu} \nabla_\mu \bar{\phi} \nabla_\nu \bar{\phi} - V(\bar{\phi}) \label{eq:Lbar}
    \, . 
}
(In these expressions all derivatives are taken with respect to the background metric $\bar{g}_{\mu\nu}$.)  As a result, the background solutions for both metrics are what we expect from General Relativity.  For example, if the scalar background is homogeneous, $\bar{\phi}(t,\xvec) = \bar{\phi}(t)$, it induces a homogeneous and isotropic expansion, described by the FRW metric $\bar{g}_{\mu\nu} = \mathrm{diag}(-1, a(t)^2, a(t)^2, a(t)^2)$.  
The temporal and spatial components of the stress-energy tensor, 
\bea
    \bar{T}_{00} 
    = \tfrac{1}{2} \dot{\bar{\phi}}^2 + V(\bar{\phi}) = \bar{\rho} 
    \qquad \text{and} \qquad 
    \bar{T}_{ij} = \bigl( \tfrac{1}{2} \dot{\bar{\phi}}^2 - V(\bar{\phi}) \bigr) \, a^2(t) \, \delta_{ij} = \bar{p} \, a^2 \, \delta_{ij} \, ,
\eea
can be identified with the homogeneous energy density $\bar{\rho}(t)$ and pressure $\bar{p}(t)$.  

The free Lagrangian is obtained by expanding the action \eqref{bigL} around the background \eqref{eq:model1_bkg} keeping terms that are second order in the field perturbations.  We can write 
\ba{
	S = \int \! \mathrm{d}^4x \, \Bigl[ \sqrt{-\bar{g}} \, \bar{\mathcal{L}}(\bar{g},\bar{\phi}) + \sqrt{-\bar{g}} \, \mathcal{L}_\mathrm{massless}^{(2)} + \sqrt{-\bar{g}} \, \mathcal{L}_\mathrm{massive}^{(2)} + \text{interactions} \Bigr] 
	\;,
}
where the massive spin-2 and massless spin-2 sectors of the free Lagrangian decouple.  This decoupling is manifest with the appropriate choice of basis for the metric perturbations~\cite{Hassan:2011zd}
\bea\label{eq:uv_to_hk}
	\frac{u_{\mu\nu}}{M_*} = \frac{h_{\mu\nu}}{M_f}+ \frac{k_{\mu\nu}}{M_g} \, , ~~~~\frac{v_{\mu\nu}}{M_*} = \frac{h_{\mu\nu}}{M_g}- \frac{k_{\mu\nu}}{M_f} \, ,
\eea
and the scalar perturbations 
\bea\label{eq:phigftophiuv}
	\frac{\varphi_u}{M_*} = \frac{\varphi_g}{M_f} + \frac{ \varphi_f }{M_g}\, , ~~~~  \frac{\varphi_v}{M_*} = \frac{\varphi_g}{M_g}  - \frac{\varphi_f}{M_f}  
	\;.
\eea
We identify $u_{\mu\nu}$ as the massless metric perturbation and $v_{\mu\nu}$ as the massive perturbation.  We find, as expected, for the massless sector: 
\begin{subequations}\label{eq:model1_L2_massless}
\begin{align}
	\Lcal^{(2)}_{\rm massless} 
	= \Lcal^{(2)}_{uu} + \Lcal^{(2)}_{u\,\varphi_u} + \Lcal^{(2)}_{\varphi_u\varphi_u}
	\;,
\end{align}
where
\ba{
	\Lcal^{(2)}_{uu} 
	& = 
	- \tfrac{1}{2} \nabla_\lambda u_{\mu\nu} \nabla^\lambda u^{\mu\nu} 
	+ \nabla_\mu u^{\nu\lambda} \nabla_\nu u\indices{^\mu_\lambda} 
	- \nabla_\mu u^{\mu\nu} \nabla_\nu u 
	+ \tfrac{1}{2} \nabla_\mu u  \nabla^\mu u 
	\\ & \qquad 
	+ \Bigl( \bar{R}_{\mu\nu} - \MP^{-2}\, \nabla_\mu \bar{\phi} \nabla_\nu \bar{\phi} \Bigr) 
	\Bigl( u^{\mu\lambda} u_\lambda^{~\nu} - \tfrac{1}{2} u^{\mu\nu} u \Bigr) 
	\;, \nonumber \\
	\Lcal^{(2)}_{u\,\varphi_u} 
	& = 
	\MP^{-1} \Bigl[ 
	\bigl( \nabla_\mu \bar{\phi} \nabla_\nu \varphi_u + \nabla_\nu \bar{\phi} \nabla_\mu \varphi_u \bigr) 
	\bigl( u^{\mu\nu} - \tfrac{1}{2} \bar{g}^{\mu\nu}u \bigr) - V'(\bar{\phi} \bigr) \varphi_u u \Bigr] 
	\;, \\
	\Lcal^{(2)}_{\varphi_u \varphi_u} 
	& = 
	- \tfrac{1}{2} \nabla_\mu \varphi_u \nabla^\mu \varphi_u 
	- \tfrac{1}{2} V''(\bar{\phi}) \varphi_u^2 
	\label{eq:phi_uphi_u}
	\;.
}
\end{subequations}
Indices are raised and lowered with the background metric $\bar{g}$ and $u = \bar{g}^{\mu\nu} u_{\mu\nu}$.  This is equivalent to the Einstein-Hilbert Lagrangian plus a minimally coupled scalar field expanded to quadratic order in perturbations $u_{\mu\nu}$ and $\varphi_u$.  The massive sector has the identical form, plus the Fierz-Pauli mass term:
\begin{subequations}\label{eq:model1_L2}
\ba{
	\Lcal^{(2)}_{\rm massive} = \Lcal^{(2)}_{vv}+\Lcal^{(2)}_{v\,\varphi_v}+\Lcal^{(2)}_{\varphi_v\varphi_v}
    \;,
}
where
\ba{
	\Lcal^{(2)}_{vv} 
	& = 
	- \tfrac{1}{2} \nabla_\lambda v_{\mu\nu} \nabla^\lambda v^{\mu\nu} 
	+ \nabla_\mu v^{\nu\lambda} \nabla_\nu v\indices{^\mu_\lambda} 
	- \nabla_\mu v^{\mu\nu} \nabla_\nu v 
	+ \tfrac{1}{2} \nabla_\mu v \nabla^\mu v 
	\\ & \qquad 
	+ \Bigl( \bar{R}_{\mu\nu} - \MP^{-2}\, \nabla_\mu \bar{\phi} \nabla_\nu \bar{\phi} \Bigr) 
	\Bigl( v^{\mu\lambda} v_\lambda^{~\nu} - \tfrac{1}{2} v^{\mu\nu} v \Bigr) 
	\nonumber \\ & \qquad 
	- \tfrac{1}{2} m^2 \bigl( v^{\mu\nu} v_{\mu\nu} - v^2 \bigr) 
	\;, \nonumber \\
	\Lcal^{(2)}_{v\,\varphi_v} 
	& = \MP^{-1} \Bigl[ \bigl( \nabla_\mu \bar{\phi} \nabla_\nu \varphi_v + \nabla_\nu \bar{\phi} \nabla_\mu \varphi_v \bigr) \bigl( v^{\mu\nu} - \tfrac{1}{2} \bar{g}^{\mu\nu} v \bigr) - V'(\bar{\phi}) \varphi_v v \Bigr]  
	\;, \\
	\Lcal^{(2)}_{\varphi_v \varphi_v} 
	& = 
	- \tfrac{1}{2} \nabla_\mu \varphi_v \nabla^\mu \varphi_v 
	- \tfrac{1}{2} V''(\bar{\phi}) \varphi_v^2
	\;.
}
\end{subequations}
As mentioned above, $\Lcal^{(2)}_{\rm massive}$ is what you would get by starting from the General Relativistic expression, finding the free Lagrangian and adding a Fierz-Pauli mass term.  Alternatively, our bigravity approach allows one to also consider higher order terms or couplings between the massive and massless spin-2 particles, and the expressions will be ghost-free by construction.

\subsection{Nonminimal Matter Coupling}
\label{sub:nonminimal_theory}

The avoidance of ghosts typically forbids both metrics $g$ and $f$ from interacting with the same matter sector.  However, there is an exotic exception \cite{deRham:2014naa} that we will refer to as the nonminimal matter coupling.  Setting  $\Lcal_g = \Lcal_f = 0$ in the action \eqref{bigL}, let us consider a coupling of both metrics to a single matter sector, containing a scalar field $\phi_\star$, via the composite metric $g_\star$ in \eref{geff}.  For this nonminimal matter coupling, the free Lagrangian differs from the minimal coupling.  

To determine the free theory, the metrics and scalar field are expanded around their backgrounds as follows, 
\begin{subequations}\label{eq:model2_bkg}
\ba{
    g_{\mu\nu} = \bar{g}_{\mu\nu} + \frac{2}{M_g} h_{\mu\nu} 
    \, , \quad 
    f_{\mu\nu} = \bar{f}_{\mu\nu} +  \frac{2}{M_f}  k_{\mu\nu}
    \, , \quad \text{and} \quad 
    \phi_\star = \bar{\phi}_\star + \varphi_\star 
    \, , 
}
and we seek solutions with equal backgrounds for the metrics, 
\ba{
    \bar{g}_{\mu\nu} = \bar{f}_{\mu\nu} 
    \qquad \text{and} \qquad 
    \bar{\phi}_\star \equiv \bar{\phi} 
    \, .
}
\end{subequations}
The existence of such backgrounds imposes a constraint on the parameters $a$ and $b$ of the composite metric \eqref{geff}, as well as a mirroring condition on the cosmological constants: 
\bea
	\label{eq:a_b_condition}
	\frac{a}{M_g^2} = \frac{b}{M_f^2} 
	\qquad \text{and} \qquad 
	\frac{\Lambda_g}{M_g^2} = \frac{\Lambda_f}{M_f^2} \equiv \frac{\Lambda}{\MP^2}
	\;.
\eea
The composite metric \eqref{geff} is expanded, up to second order in the metric perturbations, as 
\bes{\label{geffex}
    (g_\star)_{\mu\nu} 
    & = \bar{g}_{\mu\nu}+\frac{2}{\MP} u_{\mu\nu}+\frac{\frac{M_f}{M_g}a-\frac{M_g}{M_f}b}{a+b} \frac{2}{\MP} v_{\mu\nu}-\frac{ab}{(a+b)^2}\frac{1}{M_*^2}v_{\mu\lambda}v^\lambda_{~\nu}  , \\ 
    & = \bar{g}_{\mu\nu}+\frac{2}{\MP} u_{\mu\nu}-\frac{1}{\MP^2}v_{\mu\lambda}v^\lambda_{~\nu}  \, .
}
where we have used \eref{eq:uv_to_hk} to express the result in terms of the massless and massive metric perturbations, $u_{\mu\nu}$ and $v_{\mu\nu}$.  For general $a$ and $b$ a quadratic term $u\indices{_\mu_\lambda}u\indices{^\lambda_\nu}$ is absent.  Choosing $a$ and $b$ to respect \eref{eq:a_b_condition}, the massive mode $v_{\mu\nu}$ is removed from the effective metric at linear order. Moreover, using the mass eigenstates, all the dependence on $M_g$ and $M_f$ separately drops out and one is left with only one coupling scale, $\MP$.

We expand the action \eqref{bigL} in powers of the perturbations \eqref{eq:model2_bkg} to obtain 
\ba{
	S = \int \! \mathrm{d}^4x \, \Bigl[ \sqrt{-\bar{g}} \, \bar{\mathcal{L}}(\bar{g},\bar{\phi}) + \sqrt{-\bar{g}} \, \mathcal{L}^{(2)} + \text{interactions} \Bigr] 
	\;,
}
Assuming equal backgrounds for the two metrics, and using the condition in \eref{eq:a_b_condition}, the background equations of motion for this nonminimally-coupled theory are equivalent to the equations of motion for the minimally-coupled theory, which appear in \eref{eq:model1_bkg_EOM}.  

The free Lagrangian $\Lcal^{(2)}$ is obtained by expanding the metrics and scalar field to second order in their perturbations.  Doing so gives 
\begin{subequations}\label{eq:model2_L2}
\bea
	\Lcal^{(2)} = \Lcal^{(2)}_{uu} + \Lcal^{(2)}_{u\varphi_\star} + \Lcal^{(2)}_{\varphi_\star \varphi_\star} + \Lcal^{(2)}_{vv} 
	\;,
\eea
where 
\ba{
	\Lcal^{(2)}_{uu} 
	& = 
	-\tfrac{1}{2} \nabla_\lambda u_{\mu\nu} \nabla^\lambda u^{\mu\nu} +  \nabla_\mu u^{\nu\lambda} \nabla_\nu u^\mu_{~ \lambda} - \nabla_\mu u^{\mu\nu} \nabla_\nu u+\tfrac{1}{2} \nabla_\mu u  \nabla^\mu u 
	\\ & \qquad 
    + \Bigl( \bar{R}_{\mu\nu} - \MP^{-2}\, \nabla_\mu \bar{\phi} \nabla_\nu \bar{\phi} \Bigr)
	\Bigl( u^{\mu\lambda} u_\lambda^{~\nu} - \tfrac{1}{2} u^{\mu\nu} u \Bigr)
	\;, \nonumber \\
	\Lcal^{(2)}_{u\varphi_\star}
	& = 
	\MP^{-1} \Bigl[ \bigl( \nabla_\mu \bar{\phi} \nabla_\nu \varphi_\star+\nabla_\nu \bar{\phi} \nabla_\mu \varphi_\star \bigr) \bigl( u^{\mu\nu} - \tfrac{1}{2} \bar{g}^{\mu\nu} u \bigr) - V'(\bar{\phi}) \varphi_\star u 
	\Bigr] 
	\;, \\ 
	\Lcal^{(2)}_{\varphi_\star\varphi_\star} 
	& = 
	- \tfrac{1}{2} \partial_\mu \varphi_\star \partial^\mu \varphi_\star 
	- \tfrac{1}{2} V''(\bar{\phi}) \varphi_\star^2 
	\;, \\
	\label{nonminimal}
	\Lcal^{(2)}_{vv} 
	& = 
	- \tfrac{1}{2} \nabla_\lambda v_{\mu\nu} \nabla^\lambda v^{\mu\nu} 
	+ \nabla_\mu v^{\nu\lambda} \nabla_\nu v^\mu_{~ \lambda} 
	- \nabla_\mu v^{\mu\nu} \nabla_\nu v 
	+ \tfrac{1}{2} \nabla_\mu v \nabla^\mu v 
	\\ & \qquad 
    + \Bigl( \bar{R}_{\mu\nu} + \tfrac{1}{2} \MP^{-2} \bigl( \nabla_\mu \bar{\phi} \nabla_\nu \bar{\phi} + \bar{g}_{\mu\nu} \bar{\Lcal}(\bar{g},\bar{\phi}) \bigr) \Bigr) 
	v^{\mu\lambda} v_\lambda^{~\nu}
    \nonumber \\ & \qquad 
    - \tfrac{1}{2} \Bigl( \bar{R}_{\mu\nu} + \MP^{-2} \bigl( \nabla_\mu \bar{\phi} \nabla_\nu \bar{\phi} + \bar{g}_{\mu\nu} \bar{\Lcal}(\bar{g},\bar{\phi}) \bigr) \Bigr) 
	v^{\mu\nu} v
	\nonumber \\ & \qquad 
	- \tfrac{1}{2}m^2 \bigl( v_{\mu\nu} v^{\mu\nu} - v^2 \bigr) 
	\;. \nonumber
}
\end{subequations}
Moreover, as observed previously in \rref{Schmidt-May:2014xla}, at quadratic order the massive mode $v_{\mu\nu}$ and the scalar perturbation $\varphi_\star$ decouple entirely.  Despite the similar notation, note that $\Lcal_{vv}^{(2)}$ here is different from the expression appearing in \eref{eq:model1_L2} for the minimally-coupled model, whereas $\Lcal_{uu}^{(2)}$ is identical to \eref{eq:model1_L2_massless}.   

We note that the nonminimal coupling to matter defines a theory that does not yield expected results in several regards.  For example, taking the de Sitter limit of the FRW background does not give the usual free action of a massive spin-2 particle on de Sitter.  The reason for this is straightforward to see.  Normally when one considers bigravity in de Sitter, one adds a cosmological constant for each metric 
\bea
\label{bidS}
	S = \int \! \mathrm{d}^4x \biggl[ 
	\frac{M_g^2}{2} \sqrt{-g}\, \bigl( R[g]- 2 \Lambda \bigr) 
	+ \frac{M_f^2}{2} \sqrt{-f}\, \bigl( R[f]-2 \Lambda \bigr) 
	+ \ldots 
	\biggr] 
	\;.
\eea
Alternatively, one could introduce a cosmological constant via a constant scalar potential.  In this case, using the nonminimal matter coupling in terms of $g_\star$ to couple the two metrics to the scalar field gives:
\bea
	S = \int \! \mathrm{d}^4x \biggl[ 
	\frac{M_g^2}{2} \sqrt{-g} \, R[g] + \frac{M_f^2}{2} \sqrt{-f} \, R[f] 
	+ \sqrt{-g_\star} \, \bigl( -\tfrac{1}{2} g_\star^{\mu\nu} \partial_\mu \phi_\star \partial_\nu \phi_\star + V(\phi_\star) \bigr) 
	+ \ldots 
	\biggr] 
	\;.
\eea
But for $\partial_\mu \phi_\star = 0$ and $V(\phi_\star) = const $ this Lagrangian does not give rise to the same quadratic Lagrangian as \eqref{bidS}.  That is to say
\bea
    - M_g^2 \sqrt{-g} \, \Lambda - M_f^2 \sqrt{-f} \, \Lambda \neq \sqrt{-g_\star} \, V \, .
\eea
This is thus a truly exotic coupling which will give results that do not reproduce those of a usual massive spin-2 particle in the appropriate limits. 

\section{Cosmological perturbations}
\label{sec:SVT}

To study gravitational production of massive spin-2 particles in an inflationary cosmology we require the background fields to describe a homogeneous and isotropic FRW spacetime.  We write the background metric $\bar{g}_{\mu\nu}$ and background scalar field $\bar{\phi}$ as 
\ba{\label{eq:FRW_background}
    \bar{g}\indices{_\mu_\nu}(\eta,\xvec) 
    = \bar{g}\indices{_\mu_\nu}(\eta) 
    = a^2(\eta) \, \mathrm{diag}(-1, 1, 1, 1) 
    \qquad \text{and} \qquad 
    \bar{\phi}(\eta,\xvec) = \bar{\phi}(\eta) 
    \;,
}
where $\eta$ is the conformal time coordinate, $\xvec$ is the comoving spatial coordinate, and $a(\eta)$ is the scale factor.  
The background equations of motion \eqref{eq:model1_bkg_EOM} become 
\ba{
    \label{eq:friedmann_eqn}
    \Mpl^2 (3 H^2 - \Lambda) = V(\bar{\phi}) + (\bar{\phi}')^2 / (2 a^2)
    \quad \text{and} \quad 
    \bar{\phi}'' + 2 a H \bar{\phi}' + a^2 V'(\bar{\phi}) = 0
    \;,
}
where $H = a'/a^2$ is the Hubble parameter, $V^\prime(\bar{\phi}) = dV/d\bar{\phi}$ is the potential gradient, and other primes denote derivatives with respect to conformal time. 

The polarization modes of the spin-2 fields decouple at quadratic order in the homogeneous and isotropic FRW spacetime, and the equations of motion are easily studied using a scalar-vector-tensor (SVT) decomposition.  The SVT decomposition allows a 4-tensor to be represented by variables that transform as 3-scalars/vectors/tensors under spatial rotations. For the massive spin-2 field $v\indices{_\mu_\nu}(\eta,\xvec)$ the SVT decomposition is written as~\cite{Baumann:2022mni} 
\begin{subequations}\label{eq:SVT_decom_def}
\ba{
    v\indices{_0_0} = a^2 E,\quad
    v\indices{_0_i} = a^2 (\partial_i F + G_i),\quad
    v\indices{_i_j} = a^2 (\delta\indices{_i_j} A + \partial_i \partial_j B + \partial_i C_j + \partial_j C_i + D\indices{_i_j})
    \;,
}
where $i,j=1,2,3$ are spatial indices. 
We call $D\indices{_i_j}(\eta,\xvec)$ the \emph{tensor} component of $v\indices{_\mu_\nu}$; we call $C_i(\eta,\xvec)$ and $G_i(\eta,\xvec)$ the \emph{vector} components; and we call $A(\eta,\xvec)$, $B(\eta,\xvec)$, $E(\eta,\xvec)$, and $F(\eta,\xvec)$ the \emph{scalar} components, since they transform accordingly under spatial rotations.  The vector components $C_i$ and $G_i$ are required to be transverse, while the tensor component $D\indices{_i_j}$ is required to be transverse and traceless; these constraints are summarized as 
\ba{
    \partial_i C_i = 0,\quad
    \partial_i G_i = 0,\quad
    \partial_i D\indices{_i_j} = 0,\quad \text{and} \quad 
    D\indices{_i_i} = 0 \;,
}
\end{subequations}
where repeated indices are summed.  Since $v\indices{_\mu_\nu}$ is symmetric, the tensor component is also symmetric $D\indices{_i_j} = D\indices{_j_i}$. 

Upon implementing the SVT decomposition \eqref{eq:SVT_decom_def}, the action \eqref{bigL} breaks up into separate scalar, vector, and tensor sectors that are unmixed at quadratic order in perturbations: 
\ba{\label{eq:S_to_LS_LV_LT}
    S = \int \! \dd{\eta} \dd[3]{\xvec} \, \bigl( \LS + \LV + \LT \bigr) + {\cal O}^3
    \;,
}
where $\LS$, $\LV$ and $\LT$ are the quadratic-order scalar/vector/tensor sector Lagrangians, respectively. 
Note that $L = \sqrt{-g} \, \Lcal = a^4 \, \Lcal$.  In the following subsections, we present each of these terms and provide the corresponding equations of motion for the field variables.  

\subsection{Minimal matter coupling}
\label{sub:minimal_matter_SVT}

For the theory with a minimal coupling to matter, we perform the SVT decomposition on the massive spin-2 field $v\indices{_\mu_\nu}$ and isolate the corresponding quadratic-order Lagrangians $\LS$, $\LV$, and $\LT$.  The covariant action for the spectator fields, $v\indices{_\mu_\nu}$ and $\varphi_v$, was given by \eref{eq:model1_L2} at quadratic order. Using the SVT decomposition \eqref{eq:SVT_decom_def} on an FRW background causes the scalar, vector, and tensor sectors to decouple at quadratic order.  This is expected, since one can check that all bi-linear cross terms from two different sectors can be eliminated by a combination of integration by parts and SVT constraints. There are 2 degrees of freedom in the tensor sector, corresponding to the $\pm 2$ polarization modes of $v\indices{_\mu_\nu}$; there are 2 degrees of freedom in the vector sector, corresponding to the $\pm 1$ polarization modes of $v\indices{_\mu_\nu}$; and there are $1+1$ degrees of freedom in the scalar sector, corresponding to a mixture of the $0$-polarization mode of $v\indices{_\mu_\nu}$ as well as the additional $\varphi_v$. We shall present the Lagrangians for each sector, and derive the corresponding mode equations in appropriate variables.

\subsubsection{Tensor sector}
\label{sub:model1_tensor_sector}

The tensor sector Lagrangian is given by:
\ba{\label{eq:model1_LT}
    \LT = \frac{1}{2} a^2\Big[D_{ij}^\prime D_{ij}^\prime 
    - \partial\indices{_k} D\indices{_i_j} \partial\indices{_k} D\indices{_i_j}
    - a^2 m^2 D\indices{_i_j}D\indices{_i_j} \Big] 
    \;.
}
The kinetic term is rendered canonically normalized by the change of variable $\chi\indices{_i_j} = a D\indices{_i_j}$, which leads to 
\ba{\label{eq:model1_LT_chi}
    \LT = \frac{1}{2} \Big[\chi_{ij}^\prime \chi_{ij}^\prime 
    - \partial\indices{_k} \chi\indices{_i_j} \partial\indices{_k} \chi\indices{_i_j}
    - a^2 \bigl( m^2 - 2 H^2 - a^{-1} H' \bigr) \chi\indices{_i_j}\chi\indices{_i_j}
    \Big]
}
where we have dropped total derivatives. We use $\tilde\chi\indices{_i_j}(\eta,\kvec)$ to denote the Fourier modes of $\chi\indices{_i_j}(\eta,\xvec)$.  Since the Lagrangian is isotropic, we can take $\kvec = (0,0,k)$ without loss of generality, and the transverse/traceless conditions~\eqref{eq:SVT_decom_def} let us write 
\ba{\label{eq:chi_plus_cross}
    [\tilde{\chi}\indices{_i_j}] 
    = \mqty[\tilde{\chi}_+ & \tilde{\chi}_\times & 0 \\ \tilde{\chi}_\times & - \tilde{\chi}_+ & 0 \\ 0 & 0 & 0] \;,
}
which isolates the plus and cross mode functions, $\tilde{\chi}_+(\eta,\kvec)$ and $\tilde{\chi}_\times(\eta,\kvec)$.  
The corresponding mode equations are written as 
\bes{\label{eq:model1_tensor_mode_eqn}
    & \tilde{\chi}_s^{\prime\prime}(\eta,\kvec) + \omega_k^2(\eta) \, \tilde{\chi}_s(\eta,\kvec) = 0 \qquad \text{for $s=+,\times$} \\
    \qq{where} & \omega_k^2(\eta) = k^2 + a^2 m^2 - 2 a^2 H^2 - a H^\prime 
    \;.
}
This expression reveals that the tensor sector consists of 2 propagating degrees of freedom, which can be identified with the $\pm 2$ polarization modes of the spin-2 field $v\indices{_\mu_\nu}$.  The mode equation that results upon setting $m=0$ is equivalent to the mode equation for a gravitational wave propagating on an FRW background, which is familiar from studies of tensor perturbations in an inflationary cosmology~\cite{Mukhanov:2005,Baumann:2009ds}.  Note that the effective squared mass can be written as $m^2 - 2 H^2 - a^{-1} H^\prime = m^2 - \bar{R}/6$ where $\bar{R} = 6 a^{\prime\prime}/a^3$ is the Ricci scalar in the FRW spacetime.  The squared angular frequency $\omega_k^2(\eta)$ can be either positive or negative, depending on whether $m^2$ or $\bar{R}/6$ is larger.  It is useful to observe that a free scalar field (minimally coupled to gravity) has the same mode equation as the one in \eref{eq:model1_tensor_mode_eqn}, and we leverage this similarity to develop intuition about gravitational particle production. 

\subsubsection{Vector sector}
\label{sub:model1_vector_sector}

The vector sector Lagrangian is given by:
\ba{\label{eq:model1_LV}
    \LV = a^2 \Big[ \partial\indices{_j}\bigl( G\indices{_i} - C_i' \bigr) \partial\indices{_j} \bigl( G\indices{_i} - C_i' \bigr)
    + a^2 m^2 \bigl( G\indices{_i}G\indices{_i} - \partial\indices{_j}C\indices{_i}\partial\indices{_j}C\indices{_i} \bigr) \Big] 
    \;.
}
Moving to Fourier space, we let $\tilde{C}_i(\eta, \kvec)$ and $\tilde{G}_i(\eta, \kvec)$ denote the Fourier modes of $C_i(\eta, \xvec)$ and $G_i(\eta, \xvec)$, respectively. The action $\int \! \dd\eta \dd[3]{\xvec} \LV = \int \! \dd\eta \dd[3]{\kvec} \LVk / (2\pi)^3$ defines the Lagrangian in Fourier space: 
\ba{
    \LVk = a^2 k^2 | \tilde{G}_i - \tilde{C}_i' |^2 + a^4 m^2 | \tilde{G}_i |^2 - a^4 k^2 m^2 | \tilde{C}_i |^2 
    \;.
}
Using the constraint $(k^2 + a^2 m^2) \tilde{G}_i = k^2 \tilde{C}_i'$ to integrate out $\tilde{G}_i$ leads to 
\ba{\label{eq:model1_LVk}
    \LVk = \frac{a^4 k^2 m^2}{k^2 + a^2 m^2} | {\tilde C}_i' |^2 - a^4 k^2 m^2 | {\tilde C}_i |^2
    \;.
}
Note that for $m = 0$ the Lagrangian would vanish trivially, indicating that the massless theory does not propagate any vector modes.  For theories with $m > 0$ and modes of finite wavelength, $k > 0$, the kinetic term may be rendered canonically normalized by a change of variables:
\ba{\label{eq:model1_C_to_chi}
    \tilde{\chi}_i = \sqrt{2\frac{a^4 k^2 m^2}{k^2 + a^2 m^2}} \, \tilde{C}_i 
    \;.
}
Without loss of generality we take $\kvec = (0,0,k)$ and the transverse constraint $\partial_i C_i = 0$ implies $\tilde{\chi}_3 = 0$.  From the two remaining mode functions we define $\tilde{\chi}_{\pm}(\eta, \kvec) = (\tilde{\chi}_1 \mp i \tilde{\chi}_2) / \sqrt{2}$.  Their mode equations are found to be
\bes{\label{eq:model1_vector_mode_eqn}
    & \tilde{\chi}_s^{\prime\prime}(\eta,\kvec) + \omega_k^2(\eta) \, \tilde{\chi}_s(\eta,\kvec) = 0 \qquad \text{for $s = +,-$} \\
    \qq{where} & \omega_k^2(\eta) = k^2 + a^2 m^2 - f^{\prime\prime}/f,\quad
    f = a^2 / \sqrt{k^2 + a^2 m^2}
    \;.
}
For $m \neq 0$ the vector sector consists of 2 degrees of freedom, which can be identified with the $\pm 1$ polarization modes of the spin-2 field $v\indices{_\mu_\nu}$.  For modes that are non-relativistic at conformal time $\eta$ we have $k \ll a(\eta) m$ and $\omega_k^2 \approx a^2 m^2 - 2 a^2 H^2 - a H^\prime$, which is the same effective mass appearing in the tensor-sector mode \eref{eq:model1_tensor_mode_eqn}.  The sign of $\omega_k^2$ may be either positive or negative depending on how $m^2$ compares with $k$, $H^2$ and $H^\prime/a$.

\subsubsection{Scalar sector}
\label{sub:model1_scalar_sector}

The analysis of the scalar sector is substantially more challenging than either the tensor or vector sectors.  There are several sources of difficulty.  First, there are more field variables in the scalar sector.  In addition to the scalar field perturbation $\varphi_u$, the massive spin-2 field contains four scalar perturbations ($A$, $B$, $E$, and $F$) for a total of five field variables.  Second, all but two of these fields are restricted by a combination of gauge symmetry and constraints.  It is necessary to eliminate the constrained fields in order to isolate the two propagating fields.  Third, the two propagating fields experience a time-dependent mixing in an FRW spacetime.  
Care must be taken to identify appropriate initial conditions and extract physical observables.  Fourth and finally, many more terms in the quadratic action \eqref{eq:model1_L2} contribute to the scalar sector than either the tensor or vector sector.  In this subsection, we only discuss the key steps in the calculation and present our final results.  The algebra was checked using the Mathematica package \texttt{xTensor}.  

We implement the SVT decomposition \eqref{eq:SVT_decom_def} in the quadratic action \eqref{eq:model1_L2} and take an FRW background \eqref{eq:FRW_background}.  Setting to zero the tensor and vector sector fields leaves the scalar sector Lagrangian: 
\ba{\label{eq:model1_LS}
    \LS(A, B, E, F, \varphi_v; \ A^\prime, B^\prime, E^\prime, F^\prime, \varphi_v^\prime;  \partial_i A, \partial_i B, \partial_i E, \partial_i F, \partial_i \varphi_v; \ \eta) 
    \;.
}
Each term in $\LS$ is bi-linear in the five fields. Upon integration by parts, one can show that the Lagrangian does not contain second-order time or spatial derivatives in any of the fields.  It is useful to define 
\ba{\label{eq:model1_phiv_to_phivhat}
    \hat{\varphi}_v = \varphi_v - \frac{a^{-1} \bar{\phi}^\prime}{\Mpl H} A
}
and to eliminate $\varphi_v$ for $\hat{\varphi}_v$ through that relation. The hatted field is invariant under gauge transformations, making it more closely connected with the `physical' propagating degrees of freedom.  Since $\LS$ is bilinear in each of the five fields, it is convenient to move to Fourier space where the Lagrangian density is written as $\LSk(\tilde{A}, \tilde{B}, \tilde{E}, \tilde{F}, \tilde{\hat{\varphi}}_v; \ \tilde{A}^\prime, \tilde{B}^\prime, \tilde{E}^\prime, \tilde{F}^\prime, \tilde{\hat{\varphi}}_v^\prime; \ \eta) $. An explicit computation reveals that $\LSk$ does not contain kinetic terms for either $\tilde{E}$ nor $\tilde{F}$, which may be identified as non-dynamical variables. The corresponding Euler-Lagrange equations are constraints that can be solved to express $\tilde{E}$ and $\tilde{F}$ in terms of the other variables.  Upon doing so, the Lagrangian can be written as $\LSk(\tilde{A}, \tilde{B}, \tilde{\hat{\varphi}}_v; \ \tilde{A}^\prime, \tilde{B}^\prime, \tilde{\hat{\varphi}}_v^\prime; \ \eta)$.  With these transformations, the kinetic term for $\tilde{A}$ has dropped out of the Lagrangian, and its Euler-Lagrange equation is a constraint that can be solved to eliminate $\tilde{A}$ in terms of the other variables.  Upon doing so, we arrive at a concrete expression for the scalar-sector Lagrangian, which takes the form 
\ba{\label{eq:model1_LSk}
    \LSk = 
    K_{\varphi} \, | \tilde{\hat{\varphi}}_v' |^2 
    - M_{\varphi} \, | \tilde{\hat{\varphi}}_v |^2 
    + K_{B} \, | \tilde{B}' |^2 
    - M_{B} \, | \tilde{B} |^2 
    + L_2 \, \tilde{\hat{\varphi}}_v^{\ast\prime} \tilde{B}' 
    + L_1 \, \tilde{\hat{\varphi}}_v^\ast \tilde{B}' 
    - L_0 \, \tilde{\hat{\varphi}}_v^\ast \tilde{B}
    \;.
}
Note that the mixed terms may be complex, but they lead to a real action because of the reality condition $L_{S,-\kvec} = \LSk^\ast$.  
The seven coefficients, which are time-dependent and real, can be expressed as 
\begin{subequations}\label{eq:scalar_sector_coeffs}
\ba{
    K_\varphi 
    & = \frac{a^2}{2} 
    \frac{
    H^2 k^4 
    + 3 a^2 \bigl( m^2 - m_H^2 \bigr) H^2 k^2 
    + \tfrac{9}{4} a^4 m^2 \bigl( m^2 - m_H^2 \bigr) H^2 
    }{
    H^2 k^4 
    + 3 a^2 \bigl( m^2 - m_H^2 \bigr) H^2 k^2 
    + \tfrac{3}{8} a^4 m^2 \bigl( 6 m^2 H^2 - 4 H^2 m_H^2 - m_H^4 \bigr)
    } 
    \\
    M_\varphi 
    & = \frac{a^2}{2} 
    \frac{
    c_{10} k^{10} 
    + c_8 k^8 
    + c_6 k^6 
    + c_4 k^4 
    + c_2 k^2 
    + c_0
    }{\bigl[ 
    H^2 k^4 
    + 3 a^2 \bigl( m^2 - m_H^2 \bigr) H^2 k^2 
    + \tfrac{3}{8} a^4 m^2 \bigl( 6 m^2 H^2 - 4 H^2 m_H^2 - m_H^4 \bigr) 
    \bigr]^2} \\ 
    & \qquad c_{10} = H^4 \nonumber \\ 
    & \qquad c_8 = \tfrac{1}{2} a^2 H^2 \bigl[ \bigl( 12 m^2 H^2 + 8 H^4 - 14 H^2 m_H^2 - m_H^4 \bigr) + 4 \tfrac{H V^\prime(\bar{\phi}) \bar{\phi}^\prime}{a \MP^2} + 2 H^2 V^{\prime\prime}(\bar{\phi}) \bigr] \nonumber \\ 
    & \qquad c_6 = \tfrac{3}{8} a^4 H^2 \bigl[ \bigl( 36 m^4 H^2 + 72 m^2 H^4 - 82 m^2 H^2 m_H^2 - 64 H^4 m_H^2 
    \nonumber \\ & \qquad \qquad \qquad  
    - 7 m^2 m_H^4 + 40 H^2 m_H^4 + 8 m_H^6 \bigr) 
    \nonumber \\ & \qquad \qquad \qquad  
    + 8 \bigl( 3 m^2 - 4 m_H^2 \bigr) \tfrac{H V^\prime(\bar{\phi}) \bar{\phi}^\prime}{a \MP^2} 
    \nonumber \\ & \qquad \qquad \qquad  
    + 16 \bigl( m^2 - m_H^2 \bigr) H^2 V^{\prime\prime}(\bar{\phi}) \bigr] \nonumber \\ 
    & \qquad c_4 = \tfrac{3}{8} a^6 \bigl[ 4 H^2 \bigl( 9 m^6 H^2 + 36 m^4 H^4 + 16 m^2 H^6 - 30 m^4 H^2 m_H^2 - 76 m^2 H^4 m_H^2 
    \nonumber \\ & \qquad \qquad \qquad  
    - 3 m^4 m_H^4 + 31 m^2 H^2 m_H^4 + 24 H^4 m_H^4 + 6 m^2 m_H^6 - 6 H^2 m_H^6 - 3 m_H^8 \bigr) 
    \nonumber \\ & \qquad \qquad \qquad  
    - 4 m^2 H^2 \bigl( H^2 - m_H^2 \bigr) \tfrac{V^\prime(\bar{\phi})^2}{\MP^2} 
    \nonumber \\ & \qquad \qquad \qquad  
    + \bigl( 36 m^4 H^2 + 8 m^2 H^4 - 94 m^2 H^2 m_H^2 + m^2 m_H^4 + 48 H^2 m_H^4 \bigr) \tfrac{H V^\prime(\bar{\phi}) \bar{\phi}^\prime}{a \MP^2} 
    \nonumber \\ & \qquad \qquad \qquad  
    + \bigl( 36 m^4 H^2 - 58 m^2 H^2 m_H^2 - m^2 m_H^4 + 24 H^2 m_H^4 \bigr) H^2 V^{\prime\prime}(\bar{\phi}) \bigr] \nonumber \\ 
    & \qquad c_2 = \tfrac{9}{32} a^8 m^2 \bigl[ H^2 \bigl( 18 m^6 H^2 + 120 m^4 H^4 + 128 m^2 H^6 - 78 m^4 H^2 m_H^2 - 384 m^2 H^4 m_H^2 
    \nonumber \\ & \qquad \qquad \qquad  
    - 9 m^4 m_H^4 + 132 m^2 H^2 m_H^4 + 128 H^4 m_H^4 + 23 m^2 m_H^6 - 32 H^2 m_H^6 - 16 m_H^8 \bigr) 
    \nonumber \\ & \qquad \qquad \qquad  
    - 8 H^2 \bigl( 2 m^2 H^2 - 2 m^2 m_H^2 + m_H^4 \bigr) \tfrac{V^\prime(\bar{\phi})^2}{\MP^2} 
    \nonumber \\ & \qquad \qquad \qquad  
    + 4 \bigl( 6 m^4 H^2 - 22 m^2 H^2 m_H^2 + m^2 m_H^4 + 14 H^2 m_H^4 \bigr) \tfrac{H V^\prime(\bar{\phi}) \bar{\phi}^\prime}{a \MP^2} 
    \nonumber \\ & \qquad \qquad \qquad  
    + 4 \bigl( m^2 - m_H^2 \bigr) \bigl( 12 m^2 H^2 - 10 H^2 m_H^2 - m_H^4 \bigr) H^2 V^{\prime\prime}(\bar{\phi}) 
    \bigr] \nonumber \\ 
    & \qquad c_0 = \tfrac{27}{32} a^{10} m^4 \bigl[ - 2 H^2 \bigl( 2 m^2 H^2 - 2 m^2 m_H^2 + m_H^4 \bigr) \tfrac{V^\prime(\bar{\phi})^2}{\MP^2} 
    \nonumber \\ & \qquad \qquad \qquad  
    - m^2 \bigl( 2 H^2 - m_H^2 \bigr) \bigl( 4 H^2 + m_H^2 \bigr) \tfrac{H V^\prime(\bar{\phi}) \bar{\phi}^\prime}{a \MP^2} 
    \nonumber \\ & \qquad \qquad \qquad  
    + \bigl( m^2 - m_H^2 \bigr) \bigl( 6 m^2 H^2 - 4 H^2 m_H^2 - m_H^4 \bigr) H^2 V^{\prime\prime}(\bar{\phi}) 
    \bigr] \nonumber 
}
\ba{
    K_B 
    & = \frac{a^6 m^2}{8} \, 
    \frac{
    \bigl( 8 m^2 H^2 - 6 H^2 m_H^2 - m^2 m_H^2 \bigr) k^4
    }{
    H^2 k^4 
    + 3 a^2 \bigl( m^2 - m_H^2 \bigr) H^2 k^2 
    + \tfrac{3}{8} a^4 m^2 \bigl( 6 m^2 H^2 - 4 H^2 m_H^2 - m_H^4 \bigr)
    } \\
    M_B 
    & = \frac{a^6 m^2}{8}
    \frac{
    c_{10} k^{10} 
    + c_8 k^8 
    + c_6 k^6 
    + c_4 k^4
    }{\bigl[ 
    H^2 k^4 
    + 3 a^2 \bigl( m^2 - m_H^2 \bigr) H^2 k^2 
    + \tfrac{3}{8} a^4 m^2 \bigl( 6 m^2 H^2 - 4 H^2 m_H^2 - m_H^4 \bigr) 
    \bigr]^2} \\ 
    & \qquad c_{10} = H^2 \bigl( 8 m^2 H^2 - 8 H^4 - 2 H^2 m_H^2 - m^2 m_H^2 \bigr) \nonumber \\ 
    & \qquad c_8 = a^2 H^2 \bigl[ \bigl( 30 m^4 H^2 + 32 m^2 H^4 - 96 H^6 - 3 m^4 m_H^2 - 56 m^2 H^2 m_H^2
    \nonumber \\ & \qquad \qquad \qquad  
    + 48 H^4 m_H^2 + 5 m^2 m_H^4 + 6 H^2 m_H^4 \bigr) 
    \nonumber \\ & \qquad \qquad \qquad  
    + \bigl( 4 m^2 - 24 H^2 \bigr) \tfrac{H V^\prime(\bar{\phi}) \bar{\phi}^\prime}{a \Mpl^2} \bigr] \nonumber \\ 
    & \qquad c_6 = \tfrac{3}{8} a^4 m^2 \bigl[ \bigl( 96 m^4 H^4 + 144 m^2 H^6 - 6 m^4 H^2 m_H^2 - 252 m^2 H^4 m_H^2 - 192 H^6 m_H^2 
    \nonumber \\ & \qquad \qquad \qquad 
    + 8 m^2 H^2 m_H^4 + 200 H^4 m_H^4 - 10 H^2 m_H^6 - m^2 m_H^6 \bigr) 
    \nonumber \\ & \qquad \qquad \qquad 
    + \bigl( 8 m^2 m_H^2 - 16 H^2 m_H^2 \bigr) \tfrac{H V^\prime(\bar{\phi}) \bar{\phi}^\prime}{a \Mpl^2} \bigr] \nonumber \\ 
    & \qquad c_4 = \tfrac{3}{8} a^6 m^4 \bigl[ \bigl( 36 m^4 H^4 - 48 m^2 H^6 + 64 H^8 - 12 m^2 H^4 m_H^2 - 32 H^6 m_H^2 
    \nonumber \\ & \qquad \qquad \qquad 
    - 12 m^2 H^2 m_H^4 + 4 H^4 m_H^4 + 12 H^2 m_H^6 - 3 m^2 m_H^6 + 2 m_H^8 \bigr) 
    \nonumber \\ & \qquad \qquad \qquad 
    - \bigl( 24 m^2 H^2 - 16 H^4 - 12 m^2 m_H^2 - 8 H^2 m_H^2 + 8 m_H^4 \bigr) \tfrac{H V^\prime(\bar{\phi}) \bar{\phi}^\prime}{a \Mpl^2} \bigr] \nonumber 
}
\ba{
    L_2 
    & = \frac{a^3 m^2 \bar{\phi}^\prime}{2 \Mpl H} 
    \frac{
    H^2 k^4 
    + \tfrac{3}{2} a^2 \bigl( m^2 - m_H^2 \bigr) H^2 k^2
    }{
    H^2 k^4 
    + 3 a^2 \bigl( m^2 - m_H^2 \bigr) H^2 k^2 
    + \tfrac{3}{8} a^4 m^2 \bigl( 6 m^2 H^2 - 4 H^2 m_H^2 - m_H^4 \bigr)
    } \\
    L_1 
    & = - \frac{a^4 m^2 \bar{\phi}^\prime}{\Mpl} 
    \frac{
    \bigl( H^2 - \tfrac{1}{4} m_H^2 - \tfrac{1}{2} \tfrac{a H V^\prime(\bar{\phi})}{\bar{\phi}^\prime} \bigr) k^4 
    - \tfrac{3}{2} a^2 \bigl( m^2 - m_H^2 \bigr) \bigl( H^2 + \tfrac{1}{4} m_H^2 
    + \tfrac{1}{2} \tfrac{a H V^\prime(\bar{\phi})}{\bar{\phi}^\prime} \bigr) k^2
    }{
    H^2 k^4 
    + 3 a^2 \bigl( m^2 - m_H^2 \bigr) H^2 k^2 
    + \tfrac{3}{8} a^4 m^2 \bigl( 6 m^2 H^2 - 4 H^2 m_H^2 - m_H^4 \bigr)
    } \\
    L_0 
    & = \frac{a^3 m^2 \bar{\phi}^\prime}{2 \Mpl H} 
    \frac{
    c_{10} k^{10} 
    + c_8 k^8 
    + c_6 k^6 
    + c_4 k^4 
    + c_2 k^2
    }{\bigl[ 
    H^2 k^4 
    + 3 a^2 \bigl( m^2 - m_H^2 \bigr) H^2 k^2 
    + \tfrac{3}{8} a^4 m^2 \bigl( 6 m^2 H^2 - 4 H^2 m_H^2 - m_H^4 \bigr) 
    \bigr]^2} \\ 
    & \qquad c_{10} = H^4 \\ 
    & \qquad c_8 = \tfrac{1}{2} a^2 H^4 \bigl[ \bigl( 9 m^2 + 12 H^2 - 13 m_H^2 \bigr) - 4 \tfrac{a H V^\prime(\bar{\phi})}{\bar{\phi}^\prime} \bigr] \nonumber \\ 
    & \qquad c_6 = \tfrac{3}{8} a^4 H^2 \bigl[ \bigl( 18 m^4 H^2 + 32 m^2 H^4 + 64 H^6 - 48 m^2 H^2 m_H^2 - 64 H^4 m_H^2 
    \nonumber \\ & \qquad \qquad \qquad  
    + m^2 m_H^4 + 28 H^2 m_H^4 \bigr) 
    \nonumber \\ & \qquad \qquad \qquad  
    + 8 \bigl( - 4 m^2 H^2 + 4 H^4 + m^2 m_H^2 \bigr) \tfrac{a H V^\prime(\bar{\phi})}{\bar{\phi}^\prime} \bigr] \nonumber \\ 
    & \qquad c_4 = \tfrac{3}{16} a^6 m^2 H^2 \bigl[ \bigl( 18 m^4 H^2 - 24 m^2 H^4 + 256 H^6 - 54 m^2 H^2 m_H^2 - 160 H^4 m_H^2 
    \nonumber \\ & \qquad \qquad \qquad  
    + 9 m^2 m_H^4 + 60 H^2 m_H^4 - 7 m_H^6 \bigr) 
    \nonumber \\ & \qquad \qquad \qquad  
    + 4 \bigl( - 30 m^2 H^2 + 32 H^4 + 12 m^2 m_H^2 + 4 H^2 m_H^2 - 7 m_H^4 \bigr) \tfrac{a H V^\prime(\bar{\phi})}{\bar{\phi}^\prime} \bigr] \nonumber \\ 
    & \qquad c_2 = \tfrac{9}{16} a^8 m^4 H^2 \bigl( 2 H^2 - m_H^2 \bigr) \bigl[ - \bigl( 4 H^2 + m_H^2 \bigr) \bigl( 3 m^2 - 4 H^2 - m_H^2 \bigr)
    \nonumber \\ & \qquad \qquad \qquad  
    + 4 \bigl( -3 m^2 + 2 H^2 + 2 m_H^2 \bigr) \tfrac{a H V^\prime(\bar{\phi})}{\bar{\phi}^\prime} \bigr] \nonumber 
}
\end{subequations}
where we've defined a time-dependent squared mass parameter 
\ba{\label{eq:mHsq_def}
    m_H^2(\eta) = 2 H^2 - (\bar{\phi}^\prime)^2 / (a \Mpl)^2
    \;.
}
The field variables $\tilde{\hat{\varphi}}_v$ and $\tilde{B}$ have both kinetic mixing and a mass mixing.  The kinetic mixing can be eliminated by a change of variables: 
\ba{
    \tilde{\hat{\varphi}}_v = \tilde{\Pi} + \kappa(\eta) \, \tilde{\mathcal{B}}
    \qquad \text{and} \qquad 
    \tilde{B} = k^{-2} \tilde{\mathcal{B}}
}
where the time-dependent coefficient is 
\ba{\label{eq:kappa}
    \kappa(\eta) 
    & = - \frac{L_2}{2 k^2 K_\varphi} 
    = - \frac{a m^2 \bar{\phi}^\prime}{2 \Mpl H} \, \frac{k^2 + \tfrac{3}{2} a^2 \bigl( m^2 - m_H^2 \bigr)}{k^4 + 3 a^2 \bigl( m^2 - m_H^2 \bigr) k^2 + \tfrac{9}{4} a^4 m^2 \bigl( m^2 - m_H^2 \bigr)} 
    \;.
}
In terms of the new field variables, the scalar sector Lagrangian is finally written as 
\ba{\label{eq:model1_LSk_diagonal}
    \LSk = 
    K_{\Pi} \, | \tilde{\Pi}' |^2 
    - M_{\Pi} \, | \tilde{\Pi} |^2 
    + K_{\mathcal{B}} \, | \tilde{\mathcal{B}}' |^2 
    - M_{\mathcal{B}} \, | \tilde{\mathcal{B}} |^2 
    + \lambda_1 \, \tilde{\Pi}^\ast \tilde{\mathcal{B}}' 
    - \lambda_0 \, \tilde{\Pi}^\ast \tilde{\mathcal{B}}
    \;,
}
where we have also used integration by parts and dropped total derivative terms.  In this new field basis, there is no kinetic mixing.  
The kinetic term coefficients are given by: 
\ba{
	K_\Pi
	& = K_\varphi
	= \frac{a^2}{2} \frac{H^2 k^4 + 3 a^2 \bigl( m^2 - m_H^2 \bigr) H^2 k^2 + \tfrac{9}{4} a^4 m^2 \bigl( m^2 - m_H^2 \bigr) H^2}{H^2 k^4 + 3 a^2 \bigl( m^2 - m_H^2 \bigr) H^2 k^2 + \tfrac{3}{8} a^4 m^2 \bigl( 6 m^2 H^2 - 4 H^2 m_H^2 - m_H^4 \bigr)} \\ 
	K_{\mathcal{B}} 
	& = \frac{4 K_\varphi K_B - L_2^2}{4 k^4 K_\varphi} 
	= \frac{3 a^6 m^2 (m^2 - m_H^2)}{4 k^4 + 12 a^2 (m^2 - m_H^2) k^2 + 9 a^4 m^2 (m^2 - m_H^2)}
	\;,
}
and the other coefficients are easily derived, but too unwieldy to reproduce here.  

Note that the various kinetic and mass coefficients may be either positive or negative, allowing for either ghost-like or tachyon-like instabilities.  For instance $K_B < 0$ for $m^2 < m_H^2$.  We analyze these instabilites further in \sref{sub:ghost_instability}.  

The scalar-sector quadratic action \eqref{eq:model1_LSk_diagonal} contains a wealth of information about this system.  
It reveals that the scalar sector contains two propagating (`physical') degrees of freedom, which are identified with $\tilde{\Pi}$ and $\tilde{\mathcal{B}}$.  However, the presence of time-dependent mixed terms (with coefficients $\lambda_1$ and $\lambda_0$) prevents one from immediately associating $\tilde{\Pi}$ with inflaton particles and $\tilde{\mathcal{B}}$ with helicity-0 polarization, massive spin-2 particles.  In light of this mixing and its impact on our gravitational particle production calculation, we take care to identify the appropriate initial conditions and to extract physical observables.

\begin{figure}[t]
	\centering
	\includegraphics[width=0.8\textwidth]{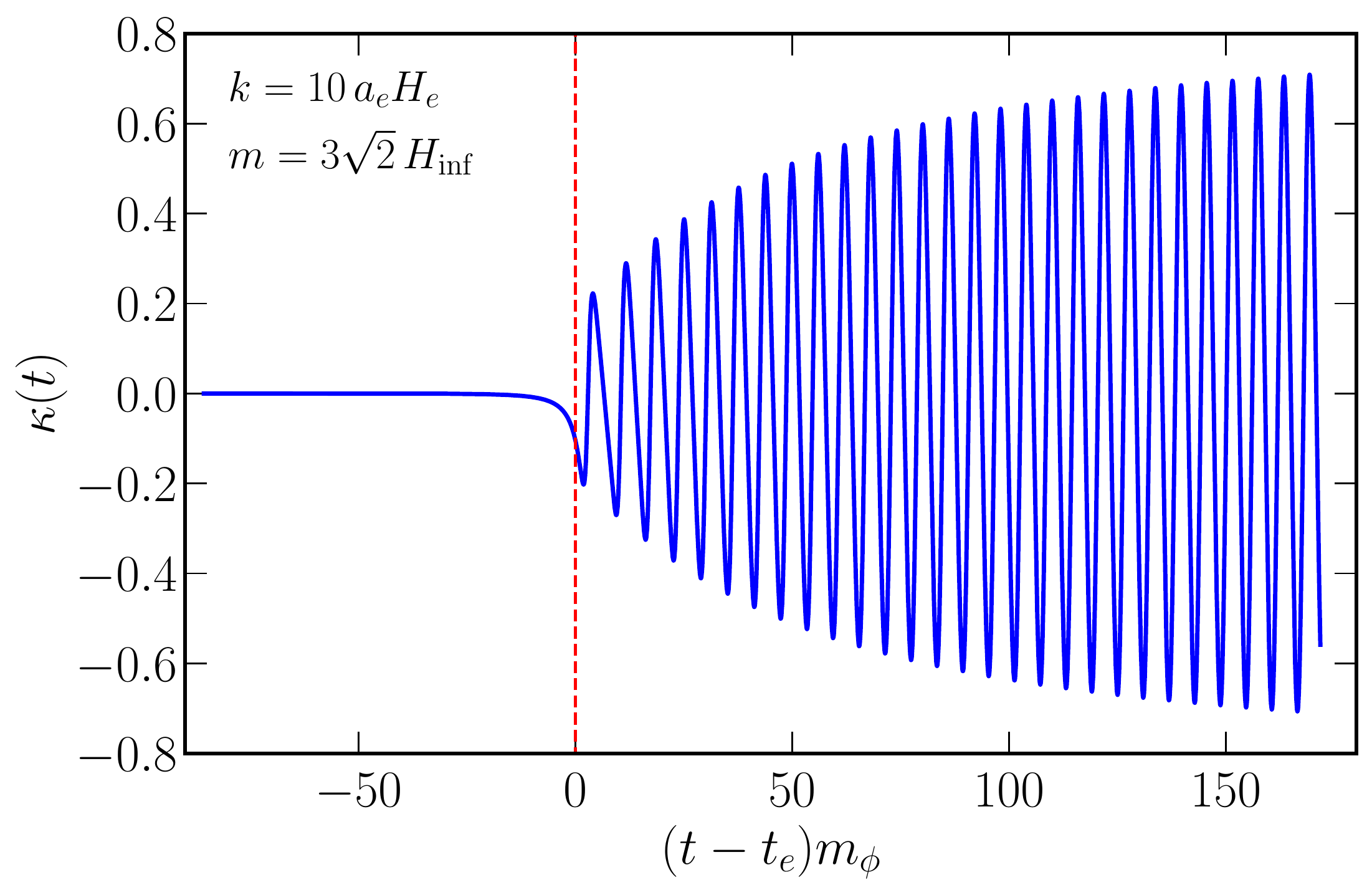} 
	\caption{\label{fig:kappa} The evolution of the mixing parameter $\kappa(\eta)$ near the end of inflation, plotted in coordinate time.   The end of inflation is indicated by the vertical dashed line.
	}
\end{figure}

We study the evolution of the mixing by investigating the time dependence of $\kappa(\eta)$ (\fref{fig:kappa} shows the evolution of $\kappa$ near to the end of inflation).  At early times when the background is inflating,  for relativistic modes inside the horizon we have the relations 
\ba{
	\abs{ \frac{\bar{\phi}^\prime}{a H \Mpl} } \ll 1
	\ , \quad
	\frac{k}{a m} \gg 1
	\ , \quad
	\frac{k}{a H} \gg 1 
	\ , \quad \text{and} \quad 
	\abs{ \kappa(\eta) } \approx \frac{m^2 a^2}{2 k^2 } \abs{ \frac{\bar{\phi}^\prime}{a H \Mpl} } \ll 1
	\;.
}
As such, initially the kinetic terms are diagonalized in either basis, since $\tilde{\Pi} \approx \tilde{\hat{\varphi}}_v$ and $\tilde{\mathcal{B}} = k^2 \tilde{B}$, and the kinetic mixing is negligible.  At late times, when $H \ll m$ and for non-relativistic modes, the FRW equations imply:  
\ba{
	H \ll m
	\ , \quad 
	\frac{k}{a m} \ll 1
	\ , \quad \text{and} \quad 
	\kappa(\eta) \approx - \frac{\bar{\phi}^\prime}{3 a H \Mpl}
	\;.
}
At late times after inflation $\bar{\phi}'/a$ oscillates about zero with magnitude $\sqrt{2/3}\sim 0.8$. As such, there is an $\order{1}$ kinetic mixing in the $(\tilde{\hat{\varphi}}_v, \tilde{B})$ basis, which motivates our move to the $(\tilde{\Pi}, \tilde{\mathcal{B}})$ basis where there is no kinetic mixing. In the new basis, it is illuminating to evaluate the late-time behavior of the scalar sector Lagrangian.  This is accomplished by expressing $\LSk$ as a series in powers of $H/m$, which takes values $H/m \ll 1$ at late times.  Doing so yields\footnote{The FRW equations and the inflaton EOM imply: $\bar{\phi}^\prime / a m \Mpl = \order{H/m}$, $\bar{\phi}^\prime / k \Mpl = \order{H/m}$, and $V'(\bar{\phi}) \approx m_\phi^2 (\bar{\phi} - v) \approx \pm m_\phi (6 \Mpl^2 H^2 - (\bar{\phi}^\prime)^2 / a^2)^{1/2} = \order{H/m}$.}
\bes{
\label{eq:model1_LSk_diagonal_late}
    \LSk & =
    \frac{a^2}{2} \Big[ |\tilde{\Pi}'|^2 - \bigl( k^2 + a^2 V''(\bar{\phi}) \bigr) |\tilde{\Pi}|^2 \Big]
    \\ & \quad 
    + \frac{3 a^6 m^4}{(2 k^2 + 3 a^2 m^2)^2} \Big[ |\tilde{\mathcal{B}}'|^2 - \bigl( k^2 + a^2 m^2 \bigr) |\tilde{\mathcal{B}}|^2 \Big]
    + \order{H/m}
    \;.
}
In particular, note that the mixings in this basis are $\lambda_1, \lambda_0 = \order{H/m}$ at late times.  The absence of mixings and the presence of familiar mass terms, allow us to interpret $\tilde{\Pi}$ as the inflaton perturbation and $\tilde{\mathcal{B}}$ as the helicity-0 mode of the massive spin-2 field.  We use this basis to calculate observables in our study of gravitational particle production. 

Provided that $K_\pi, K_\mathcal{B} > 0$, the kinetic terms in the Lagrangian \eqref{eq:model1_LSk_diagonal} can be canonically normalized.  
The change of variables
\ba{\label{eq:chi_Pi_and_B}
    \tilde{\Pi} = \frac{1}{\sqrt{2 K_\Pi}} \tilde{\chi}_\Pi,\quad
    \tilde{\mathcal{B}} = \frac{1}{\sqrt{2 K_\mathcal{B}}} \tilde{\chi}_\mathcal{B}
    \;,
}
allows the Lagrangian to be written as 
\ba{\label{eq:L_chi_Pi_and_B}
    \LSk = 
    \frac12 \, | \tilde{\chi}_\Pi' |^2 
    - \frac12 \omega_\Pi^2 \, | {\tilde{\chi}_\Pi} |^2 
    + \frac12 \, | \tilde{\chi}_\mathcal{B}' |^2 
    - \frac12 \omega_\mathcal{B}^2 \, | {\tilde{\chi}_\mathcal{B}} |^2 
    + \sigma_1 \, \tilde{\chi}_\Pi^\ast \tilde{\chi}_\mathcal{B}' 
    - \sigma_0 \, \tilde{\chi}_\Pi^\ast \tilde{\chi}_\mathcal{B}
    \;,
}
where total derivatives have been dropped, and where 
\bes{
    \omega_\Pi^2 & = \frac{4 K_\Pi M_\Pi + (K_\Pi')^2 - 2 K_\Pi K_\Pi''}{4 K_\Pi^2} 
    \;, \qquad 
    \omega_\mathcal{B}^2 = \frac{4 K_\mathcal{B} M_\mathcal{B} + (K_\mathcal{B}')^2 - 2 K_\mathcal{B} K_\mathcal{B}''}{4 K_\mathcal{B}^2} 
    \;, \\
    \sigma_1 & = \frac{\lambda_1}{2 \sqrt{K_\Pi} \sqrt{K_\mathcal{B}} } 
    \; \qquad \text{and} \qquad 
    \sigma_0 = \frac{2 K_\mathcal{B} \lambda_0 + \lambda_1 K_\mathcal{B}' }{4 \sqrt{K_\Pi} (K_\mathcal{B})^{3/2} } 
    \;.
}
The mode equations for $\tilde{\chi}_\Pi$ and $\tilde{\chi}_\mathcal{B}$ are given by:
\bes{\label{eq:EOM_chi_Pi_and_B}
    & \tilde{\chi}_\Pi'' + \omega_\Pi^2 \tilde{\chi}_\Pi - \sigma_1 \tilde{\chi}_\mathcal{B}' + \sigma_0 \tilde{\chi}_\mathcal{B} = 0 \\
    & \tilde{\chi}_\mathcal{B}'' + \omega_\mathcal{B}^2 \tilde{\chi}_\mathcal{B} + \sigma_1 \tilde{\chi}_\Pi' + \sigma_0 \tilde{\chi}_\Pi = 0
    \;.
}
At late times, the modes $\tilde{\chi}_\Pi$ and $\tilde{\chi}_\mathcal{B}$ decouple as in \eref{eq:model1_LSk_diagonal_late}:
\ba{
    \LSk & =
    \frac{1}{2} \Big[ |\tilde{\chi}_\Pi'|^2 - \bigl( k^2 + a^2 V''(\bar{\phi}) \bigr) |\tilde{\chi}_\Pi|^2 \Big]
    + \frac12 \Big[ |\tilde{\chi}_\mathcal{B}'|^2 - \bigl( k^2 + a^2 m^2 \bigr) |\tilde{\chi}_\mathcal{B}|^2 \Big]
    + \order{H/m}
    \;.
}

\subsection{Nonminimal matter coupling}
\label{sub:nonminimal_matter_SVT}

For the theory with a nonminimal coupling to matter, the covariant Lagrangian appears in \eref{eq:model2_L2}, and we implement the SVT decomposition using \eref{eq:SVT_decom_def}.  The resultant scalar, vector, and tensor sector Lagrangians appearing in \eref{eq:S_to_LS_LV_LT} are presented here, along with the corresponding mode equations.  

\subsubsection{Tensor sector}
\label{sub:model2_tensor_sector}

The tensor sector Lagrangian is given by: 
\ba{\label{eq:model2_LT}
    \LT = \frac{1}{2} a^2\Big[{D'}\indices{_i_j} {D'}\indices{_i_j}
    - \partial\indices{_k} D\indices{_i_j} \partial\indices{_k} D\indices{_i_j}
    - a^2 \bigl( m^2 - \Lambda + 3 H^2 + 2 a^{-1} H' \bigr) D\indices{_i_j}D\indices{_i_j} \Big] 
    \;.
}
Performing the change of variable $\chi\indices{_i_j} = a D\indices{_i_j}$ yields a Lagrangian with canonically normalized kinetic terms: 
\ba{\label{eq:model2_LT_can}
    \LT = \frac{1}{2} \Big[{\chi'}\indices{_i_j} {\chi'}\indices{_i_j}
    - \partial\indices{_k} \chi\indices{_i_j} \partial\indices{_k} \chi\indices{_i_j}
    - a^2 \bigl( m^2 - \Lambda + H^2 + a^{-1} H' \bigr) \chi\indices{_i_j}\chi\indices{_i_j}
    \Big]
    \;.
}
Moving to the Fourier domain and identifying the plus and cross modes \eqref{eq:chi_plus_cross} yields the mode equations
\bes{\label{eq:model2_tensor_mode_eqn}
    & \tilde{\chi}_s^{\prime\prime}(\eta, \kvec) + \omega_k^2(\eta) \, \tilde{\chi}_s(\eta, \kvec) = 0 \qquad \text{for $s=+,\times$} \\
    \qq{where} & \omega_k^2(\eta) = k^2 + a^2 m^2 - a^2 \Lambda + a^2 H^2 + a H^\prime 
    \;.
}
Comparing this mode equation with the minimally-coupled theory \eqref{eq:model1_tensor_mode_eqn}, the two expressions differ only in the cosmological constant and the Hubble-dependent terms appearing in the effective mass.  

\subsubsection{Vector sector}
\label{sub:model2_vector_sector}

The vector sector Lagrangian is given by:
\ba{\label{eq:model2_LV}
    \LV &= a^2 \Big[ 
    \partial\indices{_j}(G\indices{_i} - C_i') \partial\indices{_j} \bigl( G\indices{_i} - C_i' \bigr) 
    + a^2 \mu_1^2 G\indices{_i} G\indices{_i} 
    - a^2 \mu_2^2 \partial\indices{_j}C\indices{_i}\partial\indices{_j}C\indices{_i} 
    \Big] 
}
where we've defined the time-dependent squared mass parameters 
\bsa{}{
    \mu_1^2(\eta) & = m^2 - \Lambda + 3 H^2 - a^{-1}H' \\ 
    \mu_2^2(\eta) & = m^2 - \Lambda + 3 H^2 + 2 a^{-1} H' 
    \;.
}
This Lagrangian admits a Fourier representation, which is 
\bes{
    \LVk &= 
    a^2 k^2 | \tilde{G}_i - \tilde{C}_i' |^2 
    + a^4 \mu_1^2 | \tilde{G}_i |^2 
    - a^4 k^2 \mu_2^2 | \tilde{C}_i |^2 
    \;.
}
By using the constraint equation, $(k^2 + a^2 \mu_1^2) \tilde{G}_i = k^2 \tilde{C}_i^\prime$, the variable $\tilde{G}_i$ is eliminated giving 
\ba{\label{eq:model2_LVk}
    \LVk &= K_C \, |\tilde{C}_i'|^2 - M_C \,|\tilde{C}_i|^2 
    \;,
}
where the time-dependent kinetic and mass term coefficients are 
\ba{\label{eq:KC_MC_def}
    K_C(\eta) = \frac{a^4 k^2 \mu_1^2}{k^2 + a^2 \mu_1^2} 
    \qquad \text{and} \qquad 
    M_C(\eta) = a^4 k^2 \mu_2^2 
    \;.
}
In contrast with the minimally-coupled theory from \eref{eq:model1_LVk}, this Lagrangian does not vanish for $m=0$, and the theory still propagates vector modes even if the spin-2 field is massless.\footnote{This can potentially be understood from the fact that, because of the nonminimal matter coupling, there is no enhanced diffeomorphism invariance in the $m\rightarrow 0$ limit which would remove additional degrees of freedom, in contrast to the case of the minimal matter coupling where there are two independent diffeomorphism invariances of the bigravity theory when $m\rightarrow 0$.}  Note that the time-dependent coefficient of the kinetic term remains non-negative for cosmologies with $\Lambda = 0$, $H^\prime \leq 0$ and models with non-tachyonic mass $m^2 \geq 0$.  Therefore the kinetic term can be canonically normalized by the transformation $\tilde{\chi}_i(\eta,\kvec) = \sqrt{2} K_C(\eta)^{1/2} \tilde{C}_i(\eta,\kvec)$, and the Lagrangian becomes 
\ba{\label{eq:fouriervector}
    \LVk & = 
    \frac{1}{2} |\tilde{\chi}_i'|^2 
    - \frac{1}{2} \omega_k^2 \,|\tilde{\chi}_i|^2 
    \;,
}
up to a total derivative term that is dropped.  The squared comoving angular frequency is given by 
\bes{\label{model1_vector_omegaksq}
    \omega_k^2 
    & = \frac{4 K_C M_C + (K_C^\prime)^2 - 2 K_C K_C^{\prime\prime}}{4 K_C^2} 
    = c_s^2 \, k^2 + a^2 m_k^2
    \;,
}
where 
\ba{\label{eq:cs_mk_def}
	c_s^2(\eta) 
	& = \frac{\mu_2^2}{\mu_1^2} 
	= \frac{m^2 - \Lambda + 3 H^2 + 2 a^{-1} H'}{m^2 - \Lambda + 3 H^2 - a^{-1} H'} 
	\;.
}
The squared sound speed $c_s^2$ controls the high-$k$ behavior of $\omega_k^2$ while the squared effective mass $m_k^2$ goes as $k^0$ as $k \to \infty$.  The transverse condition \eqref{eq:SVT_decom_def} implies $k_i \tilde{C}_i = 0$, which eliminates one degree of freedom, such that the vector sector has only 2 propagating degrees of freedom, which can be identified with the $\pm 1$ polarization modes of the spin-2 field $v\indices{_\mu_\nu}$.  The equations or motion are 
\bes{\label{eq:model2_vector_mode_eqn}
    & \tilde{\chi}_s^{\prime\prime}(\eta,\kvec) + \omega_k^2(\eta) \, \tilde{\chi}_s(\eta,\kvec) = 0 \qquad \text{for $s = +,-$} 
    \;,
}
where $\omega_k^2(\eta)$ is given by \eref{model1_vector_omegaksq}.  

In the Minkowski spacetime we have $c_s^2 \to 1$ and and $m_k^2 \to m^2$.  (A time-dependent squared sound speed $c_s^2(\eta)$ also arises in the mode equation for a spin-3/2 field on an FRW background~\cite{Hasegawa:2017hgd,Kolb:2021xfn}.)   For models with $m^2 > \Lambda$ and cosmologies with $H^\prime < 0$, the mass parameter $\mu_1^2$ is positive at all times, and the sign of $c_s^2$ is controlled by the sign of $\mu_2^2$.  If $H^\prime$ becomes sufficiently large and negative, which may happen at the end of inflation, then $\mu_2^2$ and $c_s^2$ may be temporarily negative, and the mode equation admits an exponentially growing solution.  We explore this gradient instability in \sref{sub:gradient_instability}. 

\subsubsection{Scalar sector}
\label{sub:model2_scalar_sector}

The analysis of the scalar sector in this model of bigravity with a nonminimal coupling to matter is simpler than the minimally-coupled model.  This is mainly because the scalar field perturbation $\varphi_\star$ does not couple to the massive metric perturbation $v_{\mu\nu}$ at quadratic order, which can be seen from the free Lagrangian in \eref{eq:model2_L2}, and there is only a single propagating degree of freedom in the scalar sector.  Otherwise, the analysis here runs parallel to the discussion in \sref{sub:model1_scalar_sector} for the scalar sector of the minimally-coupled model.  In Fourier space on an FRW background, the scalar sector Lagrangian is written as 
\ba{
    \LSk(\tilde{A}, \tilde{B}, \tilde{E}, \tilde{F}; \ \tilde{A}^\prime, \tilde{B}^\prime, \tilde{E}^\prime, \tilde{F}^\prime; \ \eta) 
    \;. 
}
Neither $\tilde{E}(\eta,\kvec)$ nor $\tilde{F}(\eta,\kvec)$ have kinetic terms, and their Euler-Lagrange equations are constraints that can be solved to eliminate these variables.  Upon doing so, the kinetic term for $\tilde{A}(\eta,\kvec)$ also drops out, and this variable too can be eliminated by solving its constraint equation.  We are left with only a single variable $\tilde{B}(\eta,\kvec)$, and the scalar sector Lagrangian takes the form 
\ba{\label{eq:model2_LSk}
    \LSk = K_B | \tilde{B}' |^2 - M_B | \tilde{B} |^2
}
up to total derivatives that are dropped.  The time-dependent, real coefficients are\footnote{The coefficients above are presented in terms of $H$ and its derivatives instead of $m_H$ as in \eref{eq:model1_LSk}, since $m_H$ is irrelevant to the nonminimally coupled model. We could also present the coefficients in terms of $H$, $\bar{\phi}$ and $\bar{\phi}^\prime$ by systematically substituting out the derivatives of $H$ via rules such as $H' \to - \bar{\phi}^{\prime 2} / (2 a \Mpl^2)$ and $H'' \to \bar{\phi}^\prime (2 a V'(\bar{\phi}) + 5 H \bar{\phi}^\prime) / (2 \Mpl^2)$. These rules can be derived from the field equation \eref{eq:friedmann_eqn} for $\bar{\phi}$ and the Friedmann equations; they reflect the fact that the background equation is a 2nd-order ODE whose solution is completely determined by $\bar{\phi}$ and $\bar{\phi}^\prime$ at a given time.}
\bsa{eq:model2_KB_MB_def}{
    K_B & = \tfrac{a^4}{P} \big[ c_6 k^6 + c_4 k^4 \big] \\
    & \qquad c_6 = -4 \dot{H}^2 \nonumber \\
    & \qquad c_4 = 3 a^2 \bigl( m^2 + H^2 \bigr) \bigl( m^2 + 3 H^2 - \dot{H} \bigr) \bigl( m^2 + 3 H^2 + 2 \dot{H} \bigr) \nonumber \\
    M_B & = \tfrac{a^6}{P^2} \big[ c_{10} k^{10} + c_8 k^8 + c_6 k^6 + c_4 k^4 \big] \\
    & \qquad c_{10} = 
    12 \bigl( m^2 + H^2 \bigr) \bigl( m^2 + 3 H^2 \bigr)^3 
    + 16 \bigl( m^2 + 3 H^2 \bigr)^2 \bigl( 6 m^2 + 7 H^2 \bigr) \dot{H} 
    \nonumber \\ & \qquad \qquad 
    + 4 \bigl( m^2 + 3 H^2 \bigr) \bigl( 63 m^2 + 71 H^2 \bigr) \dot{H}^2 
    + 8 \bigl( 25 m^2 + 27 H^2 \bigr) \dot{H}^3 
    - 48 \dot{H}^4 
    \nonumber \\ & \qquad \qquad 
    - 32 H \bigl( m^2 + 3 H^2 \bigr) \dot{H} \ddot{H} 
    - 48 H \dot{H}^2 \ddot{H}
    \nonumber \\
    & \qquad c_{8} = 12 a^2 \bigl( m^2 + 3 H^2 + 2 \dot{H} \bigr) 
    \times \bigl[ 2 \bigl( m^2 + H^2 \bigr) \bigl( m^2 + 3 H^2 \bigr)^2 \bigl( 2 m^2 + 5 H^2 \bigr) 
    \nonumber \\ & \qquad \qquad 
    + \bigl( m^2 + 3 H^2 \bigr) \bigl( 19 m^4 + 64 m^2 H^2 + 49 H^4 \bigr) \dot{H} 
    + 2 \bigl( 7 m^4 + 20 m^2 H^2 + 17 H^4 \bigr) \dot{H}^2 
    \nonumber \\ & \qquad \qquad 
    - \bigl( 23 m^2 + 25 H^2 \bigr) \dot{H}^3
    + 2 \dot{H}^4
    - 2 H \bigl( m^2 + H^2 \bigr) \bigl( m^2 + 3 H^2 \bigr) \ddot{H}
    \nonumber \\ & \qquad \qquad 
    - 4 H \bigl( m^2 + H^2 \bigr) \dot{H} \ddot{H} 
    \bigr] \nonumber \\
    & \qquad c_{6} = 9 a^4 \bigl( m^2 + H^2 \bigr) \bigl( m^2 + 3 H^2 - \dot{H} \bigr) \bigl( m^2 + 3 H^2 + 2 \dot{H} \bigr)^2 
    \nonumber \\ & \qquad \qquad 
    \times \bigl[ 7 \bigl( m^2 + H^2 \bigr) \bigl( m^2 + 3 H^2 \bigr) + 17 \bigl( m^2 + H^2 \bigr) \dot{H} - 8 \dot{H}^2 \bigr] \nonumber \\
    & \qquad c_{4} = 27 a^6 \bigl( m^2 + H^2 \bigr)^2 \bigl(  m^2 + 3 H^2 - \dot{H} \bigr)^2 \bigl( m^2 + 3 H^2 + 2 \dot{H} \bigr)^3 \nonumber 
}
where $\dot{H} \equiv a^{-1} H^\prime$, $\ddot{H} \equiv a^{-2} H^{\prime\prime} - a^{-1} H H^\prime$, and 
\bes{
    P & = 
    4 \bigl[ m^2 + 3 H^2 + 3 \dot{H} \bigr] \, k^4 
    \\ & \quad 
    + 12 a^2 \bigl[ \bigl( m^2 + H^2 \bigr) \bigl( m^2 + 3 H^2 \bigr) + 2 \bigl( m^2 + H^2 \bigr) \dot{H} - \dot{H}^2 \bigr] \, k^2
    \\ & \quad 
    +9 a^4 \bigl( m^2 + H^2 \bigr) \bigl( m^2 + 3 H^2 - \dot{H} \bigr) \bigl( m^2 + 3 H^2 + 2 \dot{H} \bigr)
    \;.
}
Note that these coefficients may be either positive or negative depending on the FRW background and the mass $m$ and comoving wavenumber $k$ of the massive spin-2 mode.  Negative values for $K_B$ and positive values for $M_B$ would indicate the presence of an instability in the system, which we explore further in \sref{sec:instabilities}.  Assuming that $K_B(\eta) > 0$, the kinetic term can be canonically normalized by the change of variables $\tilde{\chi} = \sqrt{2} K_B^{1/2} B$, which allows the Lagrangian to be written as 
\ba{\label{eq:nonminvectorfourier}
    \LSk = \frac12 | \tilde{\chi}' |^2 - \frac12 \omega_k^2 | \tilde{\chi} |^2
}
up to total derivatives, which are dropped, and where 
\bes{\label{eq:model2_scalar_omegak}
    \omega_k^2 
    & = \frac{4 K_B M_B + (K_B^\prime)^2 - 2 K_B K_B^{\prime\prime}}{4 K_B^2}
    \;.
}
The corresponding mode equation is written as 
\ba{\label{eq:nonminvecmode}
    \tilde{\chi}^{\prime\prime}(\eta,\kvec) + \omega_k^2(\eta) \, \tilde{\chi}(\eta,\kvec) = 0 
    \;.
}
Note that the squared angular frequency $\omega_k^2(\eta)$ may be either positive or negative.  At early times $\omega_k^2(\eta) \to k^2$ for relativistic modes inside the horizon.  At high-$k$ there is a singularity in $\omega_k^2(\eta)$, associated with a ghost instability ($K_B = 0$), which we discuss further in \sref{sub:model2_ghost_instability}. 

\section{Instabilities}
\label{sec:instabilities}

In this section we discuss instabilities that can arise in these two theories of bigravity on an FRW background.  

\subsection{Ghost instability and FRW-generalized Higuchi bound (minimally-coupled theory)}
\label{sub:ghost_instability}

For massive gravity on a de Sitter background, there is a unitarity bound that constrains the spin-2 particle's mass relative to the constant Hubble parameter: $m^2 > 2 H^2$.  This relation is known as the Higuchi bound~\cite{Higuchi:1986py}.  For masses below this bound, the helicity-0 mode of the massive spin-2 field has a wrong-sign kinetic term, corresponding to a ghost instability.  In this section we derive a generalization of this bound on an FRW background that applies for either massive gravity or bigravity with a minimal coupling to matter.

For massive gravity with a minimal coupling to matter, the scalar sector quadratic action is given by \eref{eq:model1_LSk}.  The absence of a ghost requires the two-by-two matrix of kinetic terms, 
\ba{
    L_{S,\kvec} \supset \mqty( \tilde{\hat{\varphi}}_v^{\prime \ast} & \tilde{B}^{\prime \ast} ) \mqty(K_{\varphi} & L_2 / 2 \\ L_2 / 2 & K_B) \mqty( \tilde{\hat{\varphi}}_v^\prime \\ \tilde{B}^\prime )
    \;,
}
to have two positive eigenvalues.  This ensures the positivity of the kinetic terms in the corresponding Hamiltonian.  The matrix coefficients depend on comoving wavenumber $k$ and on conformal time $\eta$ via the scale factor $a(\eta)$, the Hubble parameter $H(\eta)$, and its derivatives.  We find that both eigenvalues are positive, for arbitrary wavenumber $k$, provided that\footnote{This relation generalizes trivially to higher dimensions as $m^2 > (d-1)H^2 (1 - \epsilon)$.} 
\ba{\label{eq:FRW_Higuchi}
    m^2 > m_H^2(\eta) = 2 H(\eta)^2 \bigl[ 1 - \epsilon(\eta) \bigr] 
}
where $m_H^2(\eta)$ was defined in \eref{eq:mHsq_def}, and where $\epsilon(\eta) = - H' / (a H^2)$ is the first slow-roll parameter.  We can write this relation equivalently in several useful ways:
\ba{
    m_H^2 
    = 2 H^2 - \frac{(\bar{\phi}^\prime)^2}{(a \Mpl)^2} 
    = 2 H^2 + 2 a^{-1} H'
    = \frac23 \Lambda - \frac{3p + \rho}{3\Mpl^2}
    = (1 + w) \Lambda - (1 + 3 w) H^2
    \;,
}
where we assume that the cosmological medium consists of a perfect fluid with energy density $\rho(\eta)$, pressure $p(\eta)$, and equation of state $w(\eta) = p(\eta)/\rho(\eta)$. Equation \eqref{eq:FRW_Higuchi} is our FRW-generalized Higuchi bound for massive gravity or bigravity with a minimal coupling to matter. In the de-Sitter limit, sending $\epsilon \to 0$ yields the familiar Higuchi bound~\cite{Higuchi:1986py}.\footnote{We provide an alternative, more straightforward derivation of this result in \aref{app:stueck} using the Stueckelberg approach.  We note that in previous works \cite{Fasiello:2012rw,Fasiello:2013woa} using different criteria, a generalized Higuchi bound was derived for massive gravity and bigravity in the case of two different FRW metrics for $\bar{g}_{\mu\nu}$ and $\bar{f}_{\mu\nu}$.  In the limit that the two FRW metrics are the same, the authors' result reduces to the usual Higuchi bound $m^2 = 2H^2$ with no $\epsilon$ correction.  }

In de Sitter spacetime, if the Higuchi bound is saturated, $m^2 = 2 H^2$, then the helicity-0 mode of the massive spin-2 field drops out entirely from the Lagrangian.  This is due to an enhanced gauge symmetry known as the ``partially massless" symmetry \cite{Deser:1983mm,Deser:1983tm}.  In an FRW spacetime, the generalized Higuchi bound \eqref{eq:FRW_Higuchi} can only be satisfied momentarily, since $H(\eta)$ and $\epsilon(\eta)$ vary in time.  At the time $t_\ast$ when the bound is saturated $m^2 = 2 H(t_\ast)^2 \bigl( 1 - \epsilon(t_\ast) \bigr)$, we find that the coefficient of its kinetic term passes through zero, but the scalar mode is still present in the Lagrangian through the mass and mixing terms.  Thus, we find that there is no analogous gauge symmetry at this point.

In matter-dominated and radiation-dominated universes, the slow-roll parameter is $\epsilon = 3(1+w)/2 \geq 3/2$.  The right-side of the generalized Higuchi bound \eqref{eq:FRW_Higuchi} becomes negative, implying that there is no lower bound on $m$.  In our numerical analysis of gravitational particle production, we choose $m$ such that \eref{eq:FRW_Higuchi} is satisfied at all times, and the ghost instability is avoided.  Since $H$ is monotonically decreasing for inflationary cosmologies, choosing $m^2 > 2 H_\mathrm{inf}^2$ will guarantee that the FRW-generalized Higuchi bound is satisfied during the entire cosmic history. 

\subsection{Gradient instability (nonminimally-coupled theory)}
\label{sub:gradient_instability}

For bigravity with the nonminimal coupling to matter on an FRW background, the vector sector can exhibit a gradient instability~\cite{Comelli:2015pua,Gumrukcuoglu:2015nua} in which the field amplitude grows exponentially at a rate set by the comoving wavenumber $k = |\kvec|$.  This instability is evident from the mode equation \eqref{eq:model2_vector_mode_eqn}: modes with large comoving wavenumber $k$ satisfy 
\ba{
    \tilde{\chi}_r'' \approx - c_s^2 k^2 \tilde{\chi}_r 
    \quad \Rightarrow \quad 
    \tilde{\chi}_r \propto \mathrm{exp}\biggl[ \pm \int^\eta \! \dd{\eta^\prime} k \sqrt{- c_s^2} \biggr]
    \;,
}
where $c_s^2(\eta)$ is the squared sound speed.  Note that $c_s^2 \propto m^2 - \Lambda + 3 H^2 + 2 a^{-1} H^\prime$ may be either positive or negative, since $H^\prime < 0$ in an inflationary cosmology.  A negative squared sound speed $c_s^2(\eta) < 0$, even temporarily, leads to solutions that grow exponentially in time at a rate controlled by the comoving wavenumber $k$, such that smaller-scale modes (larger $k$) grow more quickly.  

The gradient instability is avoided if $c_s^2(\eta) > 0$ at all times, which implies a constraint on the mass $m^2$ and on the cosmology.  
If the cosmological medium consists of a perfect fluid with energy density $\rho(\eta)$, pressure $p(\eta)$, and equation of state $w(\eta) = p(\eta)/\rho(\eta)$, then the condition $c_s^2(\eta) > 0$ translates to:
\ba{\label{eq:gradient_instability}
    m^2 > w(\eta) \,  
    \frac{\rho(\eta)}{\Mpl^2} = w(\eta) \, \bigl( 3 H^2(\eta) - \Lambda \bigr)
    \;.
}
The equation of state is $w \approx -1$ during the quasi-dS period of inflation and $w \approx 0$ during matter domination; at these times \eref{eq:gradient_instability} is satisfied trivially for any non-tachyonic mass. However, during the radiation-dominated epoch we have $w \approx 1/3$, and the avoidance of the instability requires $m > H(\eta)$, neglecting the cosmological constant term $\Lambda$. 
Since the Hubble parameter decreases monotonically with time, the strongest constraint is obtained at the start of the radiation era, namely the reheating period.  The temperature of the plasma at reheating $T_\text{\sc rh}$ is unknown; it can be as large as approximately $10^{16} \ \mathrm{GeV}$ without coming into conflict with the CMB limit on the energy scale of inflation~\cite{Cook:2015vqa}, or it can be as small as about a few $\mathrm{MeV}$ without disrupting nucleosynthesis and cosmic neutrino production~\cite{deSalas:2015glj}.  At reheating, the Friedmann equation implies $3 \Mpl^2 H_\text{\sc rh}^2 = \pi^2 g_{\ast,\text{\sc rh}} T_\text{\sc rh}^4 / 30$, where $g_{\ast,\text{\sc rh}}$ is the effective number of relativistic species in thermal equilibrium at temperature $T_\text{\sc rh}$.  Thus the condition for avoiding a gradient instability during the radiation era is expressed in terms of $g_{\ast,\text{\sc rh}}$ and $T_\text{\sc rh}$ as
\ba{
    m > H_\text{\sc rh} \simeq \bigl( 140 \ \mathrm{GeV} \bigr) \biggl( \frac{g_{\ast,\text{\sc rh}}}{106.75} \biggr)^{1/2} \biggl( \frac{T_\text{\sc rh}}{10^{10} \ \mathrm{GeV}} \biggr)^2
    \;.
}
Since we focus on models with $m \gtrsim H_\mathrm{inf}$ to avoid a ghost instability in the scalar sector, the gradient instability is also avoided since $H_\mathrm{inf} \geq H_\text{\sc rh}$ in general.  

\subsection{Ghost instability (nonminimally-coupled theory)}
\label{sub:model2_ghost_instability}

The theory of bigravity with a nonminimal coupling to matter also exhibits a ghost instability in the scalar sector \cite{Gumrukcuoglu:2015nua}. However, unlike the minimally-coupled theory in which the instability can be avoided with a judicious choice of parameters \eqref{eq:FRW_Higuchi}, the ghost instability in the nonminimally-coupled theory is inevitable for sufficiently high-momentum modes.\footnote{The nonminimally-coupled theory on Minkowski spacetime is known to have a ghost at the scale $\Lambda_3 = (m^2 \Mpl)^{1/3}$~\cite{deRham:2014naa}.  Here we are talking about a lower-scale ghost that is potentially within the regime of validity of the EFT.  Note that the ghost at $\Lambda_3$ does not arise on the FRW background that we study, which is why it doesn't appear in our SVT decomposition; see for example \rref{Schmidt-May:2014xla}.  }  Consequently, the nonminimally-coupled theory must be understood as an EFT with a UV cutoff $p_\mathrm{max}$, where $p=k/a$ denotes physical momentum.

We are interested in the sign of the time-dependent kinetic term coefficient $K_B(\eta)$ in the scalar sector Lagrangian of the nonminimally-coupled theory \eqref{eq:model2_LSk}.  For an FRW cosmology with $\dot{H} = a^{-1} H^\prime \neq 0$, the factor $K_B(\eta)$ takes positive values for small $p=k/a$, negative values for large $p$, and vanishes for $p = p_\mathrm{max}(\eta)$ where 
\ba{
    p_\mathrm{max}(\eta) 
    & = \sqrt{\frac{3}{4}} H \frac{\sqrt{m^2/H^2 + 1}\, \sqrt{m^2/H^2 + 3  + \epsilon} \, \sqrt{m^2/H^2 + 3  - 2 \epsilon}}{|\epsilon|} 
    \;,
}
with $\epsilon = -\dot{H} / H^2$ the first inflationary slow-roll parameter, and 
assuming all square roots are positive. Modes with $p < p_\mathrm{max}(\eta)$ have healthy evolution, whereas modes with $p > p_\mathrm{max}(\eta)$ are ghostly ($K_B < 0$), and modes that cross $p = p_\mathrm{max}(\eta)$ hit a singularity in their evolution.  The vanishing of the kinetic term coefficient indicates that the theory becomes strongly coupled at momenta approaching $p_\mathrm{max}(\eta)$ from below, and thus $p_\mathrm{max}(\eta)$ can be interpreted as the time-dependent UV cutoff.  

The time evolution of $p_\mathrm{max}(\eta)$ depends on the model of inflation, but its limiting behavior is understood as follows. During inflation $\epsilon  \ll 1$ and $p_\mathrm{max} \approx \max(m^3,H_\mathrm{inf}^3) / \epsilon H_\mathrm{inf}^2 \gg m$ is large and roughly constant.  Long after inflation $|\dot{H}| \approx H^2 \ll m^2$ and $p_\mathrm{max} \approx m^3 / H^2 \gg m$.  Generally, $p_\mathrm{max}(\eta)$ reaches a minimum around the end of inflation when $|\dot{H}| \approx H^2$ and 
\ba{
\label{eq:p_max_end_of_inflation}
    p_\mathrm{max} = \order{\frac{m^3}{H_e^2}}
    \;,
}
assuming $m \gg H_e$, and $H_e$ is the Hubble parameter at the end of inflation.  

Modes that are on the Hubble scale at the end of inflation have a comoving wavenumber of $k = a_e H_e$, which is below the cutoff $k/a_e = H_e \ll p_\mathrm{max}$ for $m = \order{10 H_e}$, and within the regime of validity of the EFT.  
Smaller-scale modes with larger $p = k/a$ are above the cutoff, and cannot be described by the effective theory.  
In our numerical analysis of CGPP, we only present spectra corresponding to a range of momenta that are within the EFT at the end of inflation.  

\section{Cosmological gravitational particle production}
\label{sec:GPP}

We are interested in the gravitational production of massive spin-2 particles in an inflationary cosmology and its phenomenological implications for dark matter and cosmological relics.  This section begins by introducing the hilltop model of inflation that we study and by explaining our numerical methods.  Then our main results are presented for models of bigravity with both minimal and nonminimal coupling to matter.  

\subsection{Hilltop inflation}
\label{sub:hilltop_inflation}

For numerical studies it is necessary to select a model of inflation to determine the evolution of the background FRW metric $\bar{g}_{\mu\nu}(\eta)$ and scalar inflaton field $\bar{\phi}(\eta)$.  We assume a hilltop model for two reasons.  First, its predictions for cosmological observables ($A_s$, $n_s$, and $r$) are compatible with current measurements by the \textit{Planck} satellite~\cite{Planck:2018vyg}.  Second, it requires a hierarchy between the inflaton mass and the inflationary Hubble scale, $m_\phi \approx 29.4 H_\mathrm{inf}$, which allows us to explore the parameter space where $H_\mathrm{inf} \ll m \ll m_\phi$.  

The hilltop model of inflation~\cite{Kumekawa:1994gx,Izawa:1996dv} is specified by the scalar potential 
\ba{
    V(\bar{\phi}) = \frac{m_\phi^2 v^2}{72} \bigg(1 - \frac{\bar{\phi}^6}{v^6} \bigg)^2
}
where $v = \Mpl/2$, and the inflaton mass $m_\phi$ is a free parameter.  For bigravity with a minimal coupling to matter, this function is related to the scalar potentials $V_g$ and $V_f$ through the mirroring condition \eqref{eq:mirroring_conditions}, and for the nonminimally-coupled theory this is $V_\star$.  For both theories, $V$ governs the dynamics of the homogeneous background field $\bar{\phi}(\eta)$, which we call the inflaton field; it appears in the background Lagrangian $\bar{\Lcal}$ via \eref{eq:Lbar} and in the Friedmann equations via \eref{eq:friedmann_eqn}, which determine the cosmic expansion history.  In the inflationary scenario that we consider, $\bar{\phi}$ initially takes values in the range $0 < \bar{\phi}(\eta_i) \ll v$, and then it ``slowly rolls" toward the potential's global minimum at $v$, and oscillates about this minimum after the end of inflation.  During inflation, the first slow-roll parameter is small and growing, $\epsilon \equiv -H' / (a H^2) \ll 1$, and we define the end of inflation as the time when $\epsilon = 1$.  We denote the scale factor and Hubble rate at the end of inflation by $a_e$ and $H_e$, respectively.  After the end of inflation the inflaton has a mass $\sqrt{V''(v)} = m_\phi$.

The single parameter $m_\phi$ is chosen such that our hilltop model predicts an amplitude for the scalar power spectrum that is compatible with measurements of this quantity inferred from CMB observations by the \textit{Planck} satellite.  A standard calculation~\cite{Baumann:2009ds} in inflationary cosmology is employed to derive expressions for the energy scale of inflation $H_\mathrm{cmb}$ and the amplitude of the scalar power spectrum $A_s$ in terms of the inflaton mass $m_\phi$ and the number of $e$-foldings $N_\mathrm{cmb}$ between CMB mode crossing and the end of inflation.\footnote{For inflationary bigravity with a minimal coupling to matter, there are two inflaton fields $\varphi_u$ and $\Pi$, and their fluctuations both contribute to the curvature perturbations.  We have verified that the spectra are approximately equal, see \fref{fig:min_spectra_scalar}, which leads to a doubling of $A_s$ as compared with the single-field model.  However, we neglect this factor of $2$ when selecting $m_\phi$ to yield the observed $A_s$.  }  We take $m_\phi = 4.14 \times 10^{12} \GeV$ and $N_\mathrm{cmb} = 60$ such that our hilltop model predicts an $A_s$ at the central value of the \textit{Planck} measurement $\ln(10^{10} A_s) = 3.044 \pm 0.014$~\cite{Planck:2018vyg}.  This implies $H_\mathrm{inf} = 1.41 \times 10^{11} \GeV$ and $H_e = 1.33 \times 10^{11} \GeV$ such that $m_\phi = 29.4 \, H_\mathrm{inf}$ and $H_\mathrm{inf} = 1.06 \, H_e$.  The requirement that inflation lasts for at least $N \geq N_\mathrm{cmb}$ $e$-foldings imposes a bound on the initial inflaton field excursion $0 < \bar{\phi}(\eta_i) \leq \bar{\phi}_\mathrm{cmb} = 0.048 \, \Mpl$.  This bound is compatible with the $\bar{\phi}(\eta_i) \ll v = \order{\Mpl}$.  

\subsection{Numerical methods}
\label{sub:numerical_methods}

We adapt standard methods to study cosmological gravitational particle production in our two theories of bigravity on an FRW background driven by hilltop inflation.  In particular, our method entails the following steps:  First, we identify the equations of motion for each field's Fourier modes.  For the two theories of bigravity discussed in \sref{sec:SVT}, and for each of the scalar, vector, and tensor sectors, we write the equations of motion in the form\footnote{For the scalar sector of the minimally-coupled theory, presented in \sref{sub:model1_scalar_sector}, the inflaton perturbations and massive spin-2 perturbations are mixed.  We discuss this case separately below.  }
\ba{
	\tilde{\chi}^{\prime\prime}(\eta,\kvec) + \omega_k^2(\eta) \, \tilde{\chi}(\eta,\kvec) = 0 
	\;,
}
where the comoving squared angular frequency $\omega_k^2(\eta)$ is a function of the comoving wavenumber $k$, and its time dependence is controlled by the hilltop inflation background.  Second, we impose the Bunch-Davies initial condition.  For an inflationary cosmology, all Fourier modes are initially inside the horizon ($k > aH$) and relativistic ($k > am$).  This observation motivates the Bunch-Davies initial condition 
\ba{\label{eq:bd_condition}
	\lim_{\eta \to -\infty} \tilde{\chi}(\eta,\kvec) = \frac{1}{\sqrt{2k}} \, e^{-i k \eta} 
	\;,
}
which imposes only the positive-frequency mode to be present at early times.  Third, we solve the mode equations along with the Bunch-Davies initial condition using numerical methods,\footnote{All mode equations were transformed into their coordinate time versions and numerically integrated in coordinate time. For producing the spectrum and relic abundance plots, we numerically integrated until $a(\eta) \approx 676 a_e$, by which time most of the Bogoliubov coefficients have stabilized, except for some parameter points with high-$k$ and low-$m$. Step sizes were chosen adaptively with a relative tolerance of $10^{-9}$ and zero absolute tolerance; see chapter II.4 of \rref{hairer_norsett_wanner_2009} for a discussion on adaptive step size. The numerical methods used include the Adams-Moulton method, the BDF method, and DOPRI5; different methods were chosen to solve different equations in order to minimize time usage. }   scanning over values of the comoving wavenumber $k$.  Modes are expected to evolve nearly adiabatically at early and late times when $\omega_k^2(\eta)$ is not changing quickly, but there may be a departure from adiabaticity at intermediate times, typically when modes leave the horizon during inflation ($k = aH$) or near to the end of inflation.  Fourth, we calculate the Bogoliubov coefficient $\beta_\kvec$ that links the early-time vacuum state with the late-time number operator; it corresponds to the amplitude of the negative-frequency mode at late time.  We calculate the Bogoliubov coefficient for modes with comoving wavevector $k=|\kvec|$ as 
\ba{
\label{eq:beta_k_def}
    |\beta_\kvec|^2 = \lim_{\eta \to \infty} \left( \frac{\omega_k}{2} | \tilde{\chi} |^2 + \frac{1}{2\omega_k} | \partial_\eta \tilde{\chi} |^2 -\frac{1}{2} \right) 
    \;.
}
Note that we normalize the mode functions by imposing $\tilde{\chi} \partial_\eta \tilde{\chi}^\ast - \tilde{\chi}^\ast \partial_\eta \tilde{\chi} = i$ such that $\tilde{\chi}(\eta,\kvec)$ is the mode function associated with creation/annihilation operators having canonical commutation relations.   Fifth, and finally, we calculate the spectrum of gravitationally produced particles.  The (physical) number density of particles with comoving momentum $p=k$ is calculated as 
\ba{
\label{eq:n_k_def}
	n_k(\eta) = a(\eta)^{-3} \frac{k^3}{2\pi^2} |\beta_k|^2
	\;,
}
and the total number density is $n(\eta) = \int_0^\infty n_k(\eta) \, \mathrm{d}k / k$.  

For the scalar sector of the minimally-coupled theory presented in \sref{sub:model1_scalar_sector}, the inflaton perturbations and massive spin-2 perturbations are mixed, and the methods presented above require the following modifications:  The equations of motion for the canonically-normalized field variables are given by \eref{eq:EOM_chi_Pi_and_B}: 
\bes{
    & \tilde{\chi}_\Pi'' + \omega_\Pi^2 \tilde{\chi}_\Pi - \sigma_1 \tilde{\chi}_\mathcal{B}' + \sigma_0 \tilde{\chi}_\mathcal{B} = 0 \\
    & \tilde{\chi}_\mathcal{B}'' + \omega_\mathcal{B}^2 \tilde{\chi}_\mathcal{B} + \sigma_1 \tilde{\chi}_\Pi' + \sigma_0 \tilde{\chi}_\Pi = 0
    \;.
}
The two mode functions are coupled through the time-dependent mixing parameters $\sigma_1(\eta)$ and $\sigma_0(\eta)$.  At early times, the mixing parameters go to zero while $\omega_\Pi^2 \approx \omega_\mathcal{B}^2 \approx k^2$, which motivates taking a Bunch-Davies initial condition \eqref{eq:bd_condition} for both mode functions.  The evolution equations mix the two mode functions while also mixing the positive and negative frequency modes.  At late times, the mixing parameters again asymptote to zero, and we evaluate the Bogoliubov coefficients using \eref{eq:beta_k_def} with $\omega_k^2 = \omega_\Pi^2$ and $\omega_\mathcal{B}^2$ as appropriate, and the number densities follow from \eref{eq:n_k_def}.  

\subsection{Stability and relic abundance}
\label{sub:stability}

The gravitational production of massive spin-2 particles during inflation may have various different phenomenological implications on cosmology and particle physics.  If these particles are unstable, their decay may affect the reheating history of the universe.  Depending on how they decay, they may populate a hidden sector, which could have implications for the origin of dark matter, dark radiation, or the matter-antimatter asymmetry of the universe.  If these particles are stable, they would survive in the universe today as all or part of the dark matter~\cite{Babichev:2016bxi}.  We study the stability of the massive spin-2 field, and report on our findings in \aref{app:decay}.  In brief, for the theory of bigravity with a minimal coupling to matter, the helicity-0 mode of the massive spin-2 field has trilinear interactions with the massless graviton and the inflaton perturbations, which can mediate its decay.  If $m > m_\phi$, decays to inflaton perturbations are kinematically accessible, and despite the Planck-suppressed couplings these decays are rapid, since we require $m > \sqrt{2} H_\mathrm{inf}$ to avoid the ghost instability (Higuchi bound).  However, such decays are kinematically blocked for $m < m_\phi \approx 29.4 H_\mathrm{inf}$, which anyway corresponds to most of the parameters presented in \fref{fig:min_spectra}.  If the inflaton were stable, this would ensure the stability of the massive spin-2 particle, but otherwise the issue of stability and the massive spin-2 particle's lifetime entails additional model building, which is beyond the scope of our work.  In order to connect with a potential phenomenological implication of our work, in what follows we assume that the massive spin-2 particle is cosmologically long lived and we calculate its present-day relic abundance.  It is worth remarking that in the theory of bigravity with \textit{nonminimal} coupling to matter, the massive spin-2 particle is stable at tree level, and it provides a natural dark matter candidate.  

We calculate the relic abundance $\Omega h^2$ of gravitationally-produced massive spin-2 particles.  First we integrate the spectra in figures \ref{fig:min_spectra} and \ref{fig:nonmin_spectra}, as well as the spectra for other masses not shown here, to obtain the comoving number densities $a^3 n$ for each sector.  The relation to the relic abundance $\Omega h^2$ depends on the reheating history.  We assume a late reheating scenario~\cite{Kolb:2020fwh}, meaning that the universe is still in the matter-dominated phase of reheating at the time when $m \approx 3 H(\eta)$ and the massive spin-2 field becomes non-relativistic.  This assumption implies an upper bound on the plasma temperature at the start of radiation domination, $T_\mathrm{RH} < (8 \times 10^{13} \GeV) (m / 10^{10} \GeV)^{1/2}$, which is easily satisfied for the parameters of interest.  The relation between comoving number density $a^3 n$ and relic abundance $\Omega h^2$ is given by~\cite{Kolb:2020fwh,Ling:2021zlj}:
\bes{\label{eq:relic_abundance}
    \Omega h^2 
    & \approx 
   0.12
    \left(\frac{m}{10^{10} \GeV}\right)
    \left(\frac{H_e}{10^{10} \GeV}\right)
    \left(\frac{T_{\mathrm{RH}}}{10^{8} \GeV}\right)
    \left(\frac{a^3 n}{a_e^3 H_e^3}\right) 
    \;.
}
Note that the relic abundance is proportional to the reheating temperature $T_\mathrm{RH}$.

Here we have assumed that there is no thermal production contributing to the relic abundance.  This is justified since the mass is large (larger than $H_\mathrm{inf}$ to avoid ghost instabilities) and $T_\mathrm{RH}$ is much below the mass.

\subsection{Minimal matter coupling}
\label{sub:minimal_matter_GPP}

\begin{figure}[t]
	\centering
    \includegraphics[width=\textwidth]{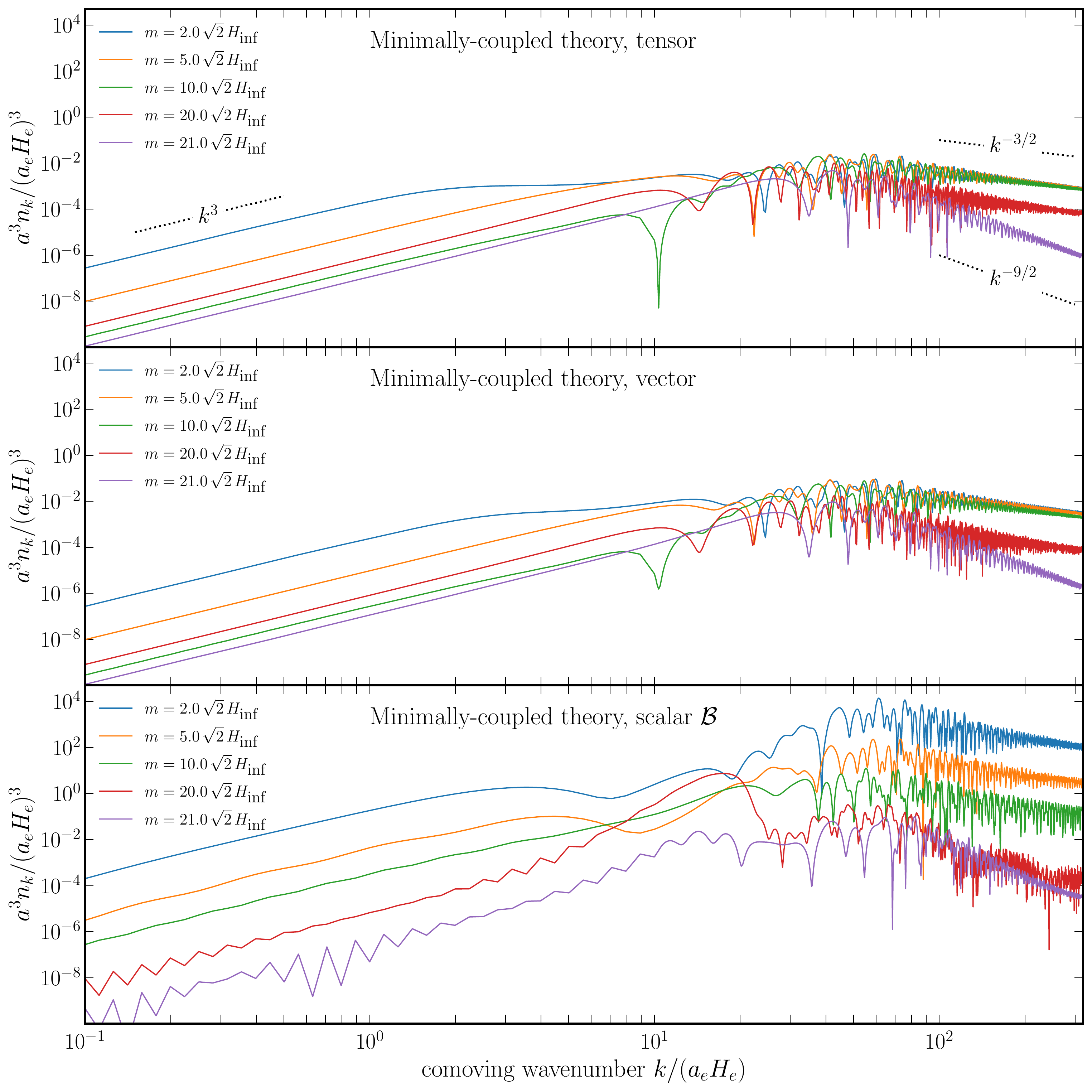}
	\caption{\label{fig:min_spectra}The comoving number density spectrum (per spin degree of freedom) $a^3 n_k$ of gravitationally produced spin-2 particles in a theory of bigravity minimally coupled to matter consisting of scalars driving hilltop inflation.  Dimensionful quantities are normalized using $a_e H_e$, the FRW scale factor and Hubble parameter at the end of inflation; i.e., modes with $k < a_e H_e$ leave the horizon during inflation, and modes with $k > a_e H_e$ remain inside the horizon.  The top, middle, and bottom panels correspond to the tensor sector (helicity $\pm 2$ modes), vector sector (helicity $\pm 1$ modes), and the scalar sector (helicity $0$ mode).  Each panel shows several curves corresponding to different values of the spin-2 field's mass $m$ in units of the inflationary Hubble scale $H_\mathrm{inf}$.  
	}
\end{figure}

For the theory with a minimal coupling to matter, our numerical results are presented in \fref{fig:min_spectra}.  We show the comoving number density spectrum $a^3 n_k$ in units of $(a_e H_e)^3$ such that $a^3 n_k / (a_e H_e)^3 = 1$ corresponds to roughly one particle per Hubble volume at the end of inflation.  The spectrum is expressed as a function of comoving wavenumber $k$ in units of $a_e H_e$, such that $k/(a_e H_e) = 1$ corresponds to modes that are on the Hubble scale at the end of inflation.  The three panels correspond to the degrees of freedom in the tensor sector (top), vector sector (middle), and scalar sector (bottom).  In each panel, the various curves correspond to different choices for $m$, the mass of the spin-2 field, in units of the inflationary Hubble scale $H_\mathrm{inf} \approx 1.1 H_e$, and we take $m / \sqrt{2} H_\mathrm{inf} \geq 2$ to be well clear of the scalar-sector ghost instability.  For the tensor and vector sectors, we show the number density per polarization degree of freedom, and the total number density is larger by a factor of $2$.  

Let us first discuss features that are universal to the tensor and vector sectors.  The tensor and vector sectors have nearly identical spectra  because their equations of motion coincide for non-relativistic modes; see \erefs{eq:model1_tensor_mode_eqn}{eq:model1_vector_mode_eqn}.  All three spectra display similar behavior for asymptotically long-wavelength and short-wavelength modes.  For long-wavelength modes with $k / (a_e H_e) < 1$, the spectra are blue-tilted power laws, $n_k \propto k^n$.  The index of the power law is approximately $n = 3$ for relatively high mass, $m \gtrsim 1.5 H_\mathrm{inf}$, and it decreases as the mass is lowered towards the Higuchi bound at $m = \sqrt{2} H_{\mathrm{inf}} \simeq 1.4 H_\mathrm{inf}$.  The low-$k$ modes leave the horizon long before the end of inflation and re-enter the horizon during early matter domination, so such modes are produced primarily acausally.  The evolution of such modes during inflation may be approximated by Hankel functions~\cite{Chung:2004nh}, and using this approximation in \aref{app:discussion_of_eoms} we show analytically that the low-$k$ behavior of $n_k$ is a power law.  The power-law index is $n = 3 - 2 (9/4 - m^2 / H_{\mathrm{inf}}^2)^{1/2}$ for $m < 3 H_{\mathrm{inf}} / 2$ and the power-law index is $n=3$ for $m > 3 H_{\mathrm{inf}} / 2$, consistent with the behavior we see in \fref{fig:min_spectra}.  It is worth pointing out that the low-$k$ behavior for both the tensor and vector spectra is similar to that for a gravitationally produced, minimally coupled scalar field; after all, the equation of motion for the tensor mode is identical to that of the scalar field.   See \aref{app:discussion_of_eoms} for more details.

For the short-wavelength modes, the spectra exhibit a decreasing power-law envelope and rapid oscillations.  Modes with $k / (a_e H_e) > 1$ never leave the horizon during inflation, and their particle production is most sensitive to the dynamics of the inflaton field at the end of inflation.  After inflation, the inflaton oscillates about the minimum of its potential with an angular frequency $m_\phi$, which imprints oscillatory features onto the spectrum of the gravitationally produced particles with $\Delta k = \order{m_\phi}$.  In this regime, where particle production is governed by coherent inflaton field oscillations on a quadratic potential, gravitational particle production can be described as scattering and annihilation of inflaton particles~\cite{Ema:2018ucl,Chung:2018ayg,Basso:2021whd,Kaneta:2022gug,Basso:2022tpd}.  For $m < m_\phi\simeq 30H_\mathrm{inf} \simeq 20\sqrt{2}H_\mathrm{inf}$, the channel $\phi \phi \to \chi \chi$ dominates, leading to a power-law envelope with $n_k \propto k^{-3/2}$, which is seen in \fref{fig:min_spectra}.  For masses $m \gtrsim m_\phi$, the leading channel is kinematically blocked, and the next open channel, $\phi\phi\phi \to \chi\chi$, dominates for $2m<3m_\phi$, leading to a steeper power-law envelope, $n_k \propto k^{-9/2}$.  The oscillatory features superimposed on the power law are due to interference between different scattering channels from $\phi$ to $\chi$~\cite{Basso:2022tpd}. 

\begin{figure}[t]
	\centering
    \includegraphics[width=0.85\textwidth]{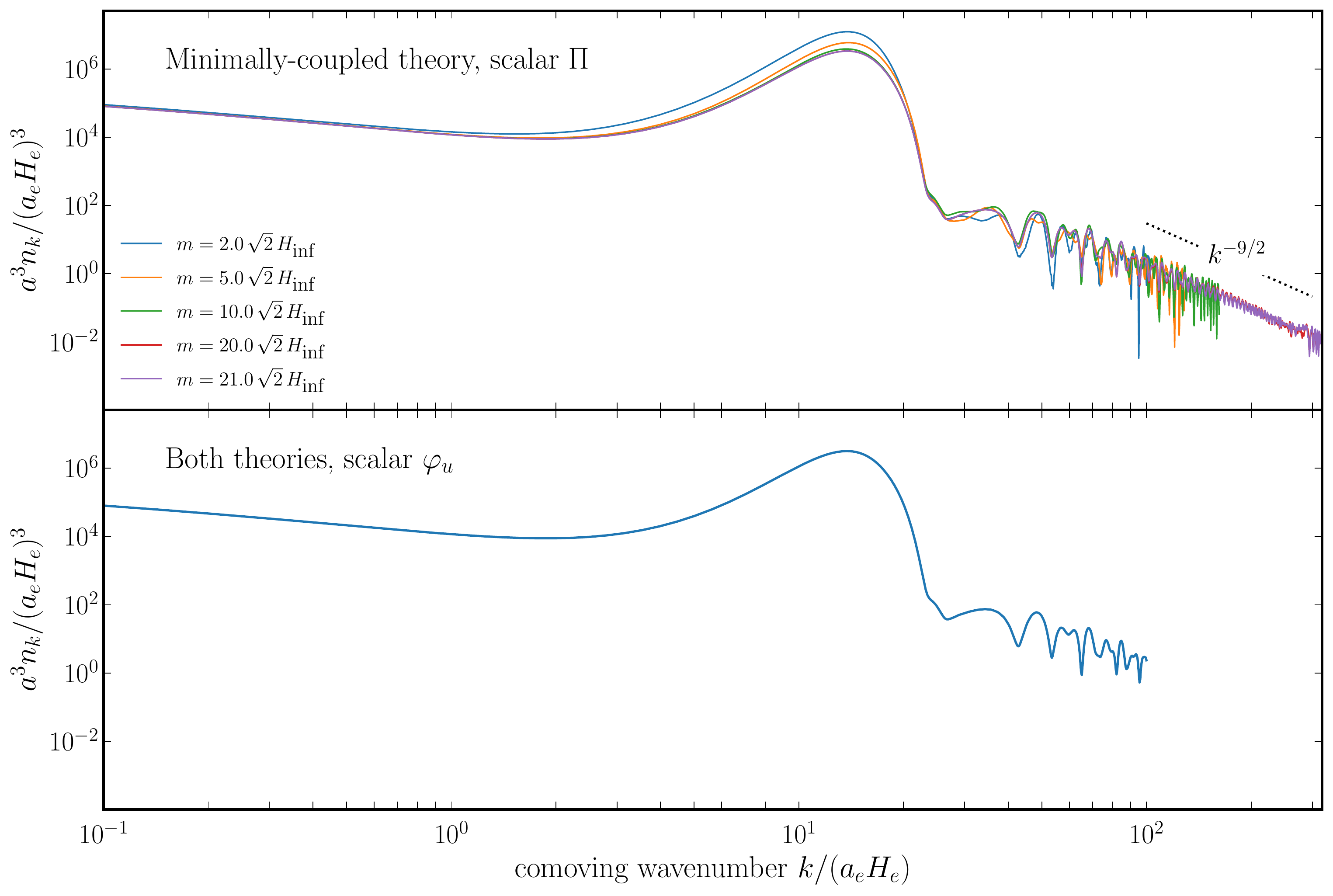} 
	\caption{\label{fig:min_spectra_scalar}The spectrum of perturbation in the inflaton fields.  The notation here is identical to \fref{fig:min_spectra}. 
	}
\end{figure}

Now we focus our attention on the scalar sector.  The scalar sector of the minimally-coupled theory contains two degrees of freedom that experience a time-dependent mixing, and we show the spectra for the field variables that diagonalize the system at late times.  The massive spin-2 scalar degree of freedom $\mathcal{B}$ is shown on \fref{fig:min_spectra} and the inflaton-like degree of freedom $\Pi$ is shown on \fref{fig:min_spectra_scalar}.  Note that we only show spectra for $m / \sqrt{2} H_\mathrm{inf} \geq 2$, since the system of equations has a ghost instability (Higuchi bound) for $m / \sqrt{2} H_\mathrm{inf} \lesssim 1$.  Comparing the three panels of \fref{fig:min_spectra} reveals that the helicity-0 mode of the massive spin-2 field is produced more copiously than the $\pm 2$ or $\pm 1$ polarization modes; similar behavior has been noted previously for spin-1 fields \cite{Graham:2015rva}.  Consequently, the gravitationally-produced massive spin-2 particles are predominantly longitudinally polarized.  Figure \ref{fig:min_spectra_scalar} shows the spectrum of perturbations in the inflaton-like fields, which displays the usual quasi-scale-invariant spectrum toward low $k$ and which is insensitive to the spin-2 mass $m$.  The enhancement around $k/(a_e H_e) \approx 10$ and subsequent harmonic progression of peaks can be understood to arise from parametric resonance associated with the inflaton's non-gravitational self-interaction~\cite{Amin:2014eta}, i.e. $m_\mathrm{eff}^2(\eta) = V^{\prime\prime}(\bar{\phi}(\eta))$, which is absent for the other degrees of freedom.  

\begin{figure}[t]
	\centering
	\includegraphics[width=\textwidth]{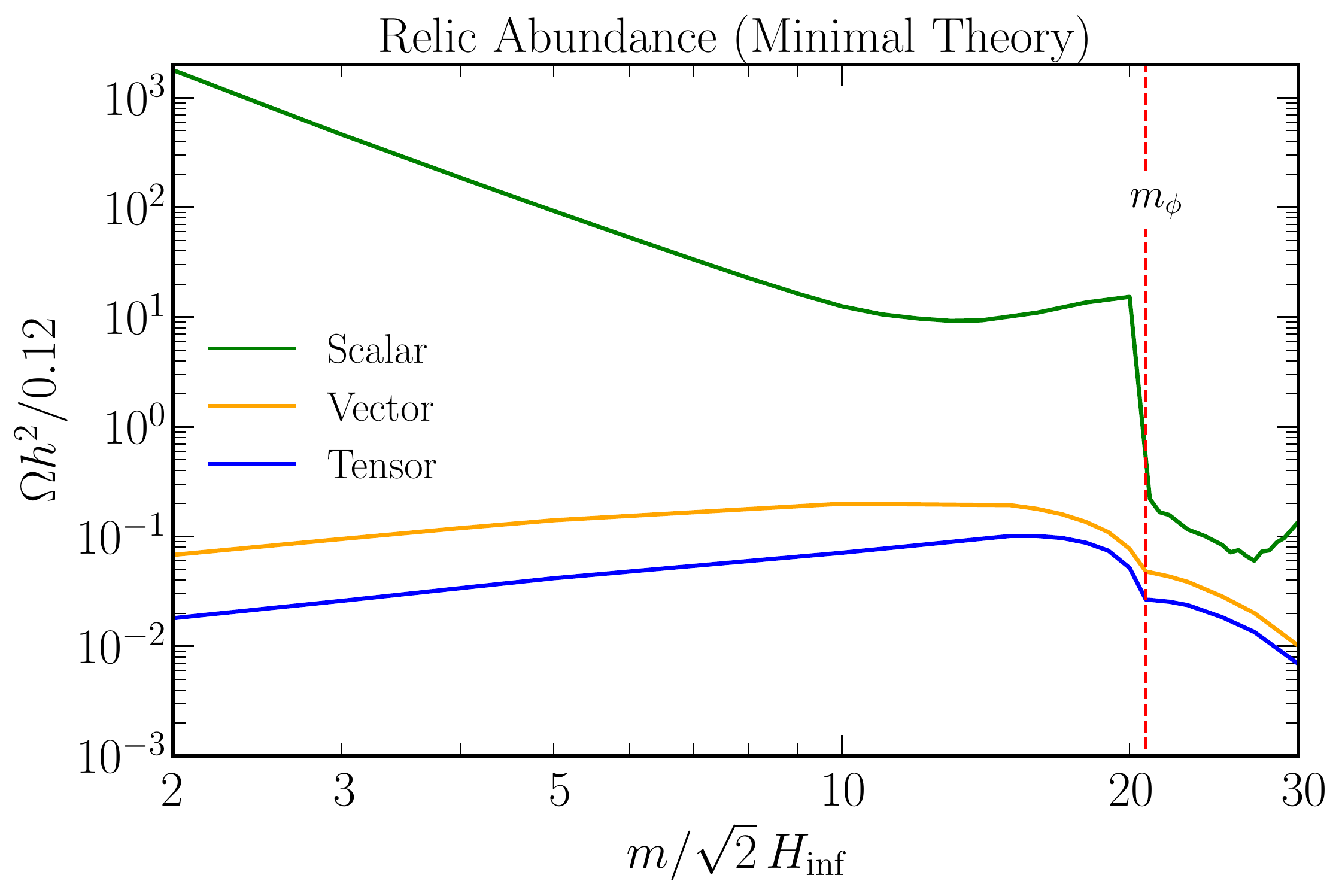}
    \caption{\label{fig:min_relic_by_mass}Relic abundance of gravitationally produced spin-2 particles as a function of the spin-2 mass $m$ for the theory of bigravity with a minimal coupling to matter.  We take $T_{\mathrm{RH}} = 10^{5} \GeV$ and for other values of the reheating temperature one can rescale our numerical results using $\Omega h^2 \propto T_\mathrm{RH}$.  The three curves correspond to the different polarization sectors:  tensor (blue), vector (orange), and scalar (green).  The red dashed line indicates the  inflaton mass $m = m_\phi \approx 30 H_\mathrm{inf}$.  
	}
\end{figure}

Numerical results for the relic abundance $\Omega h^2$ are presented in \fref{fig:min_relic_by_mass} for each of the three sectors as a function of the spin-2 mass $m$.  For this plot we have taken $T_\mathrm{RH} = 10^{5} \GeV$, and the relic abundance for other values of the reheating temperature is obtained by the scaling relation $\Omega h^2 \propto T_\mathrm{RH}$ from \eref{eq:relic_abundance}.  Models with $m > \sqrt{2} H_\mathrm{inf}$ are perfectly healthy, whereas models with $m < \sqrt{2} H_\mathrm{inf}$ have a ghost instability in the scalar sector.  Nevertheless, even for the ghostly models, it is illuminating to investigate gravitational particle production in the tensor and vector sectors, since the analytic scaling behavior as $m \to 0$ is known, and its numerical evaluation provides a check of our methods.  In the tensor sector, the relic abundance goes as $\Omega h^2 \propto m^0$ toward asymptotically small masses, $m \ll H_\mathrm{inf}$, matching known results for a scalar field minimally coupled to gravity; the same behavior occurs in the vector sector, but this cannot be seen from the range of masses shown on the figure.  For intermediate masses with $\sqrt{2} H_\mathrm{inf} < m < m_\phi$, the relic abundance rises linearly with mass, $\Omega h^2 \propto m^1$.  This behavior is understood by recalling from \fref{fig:min_spectra} that the spectra peak at $k \approx 50 a_e H_e$, corresponding to sub-Hubble-scale modes for which gravitational particle production can be described by a scattering $\phi \phi \to \chi \chi$.  For $m \ll m_\phi$ the cross section is insensitive to the mass $m$, implying $a^3 n \propto m^0$ and $\Omega h^2 \propto m^1$~\cite{Ema:2018ucl}, which agrees with the behavior seen in \fref{fig:min_relic_by_mass}.  For large spin-2 masses with $m_\phi \lesssim m$, the $\phi\phi \to \chi\chi$ channel is kinematically blocked, and the relic abundance is abruptly suppressed.  Comparing the three sectors, we see that most particle production occurs in the scalar sector.  

\begin{figure}[t]
	\centering
	\includegraphics[width=\textwidth]{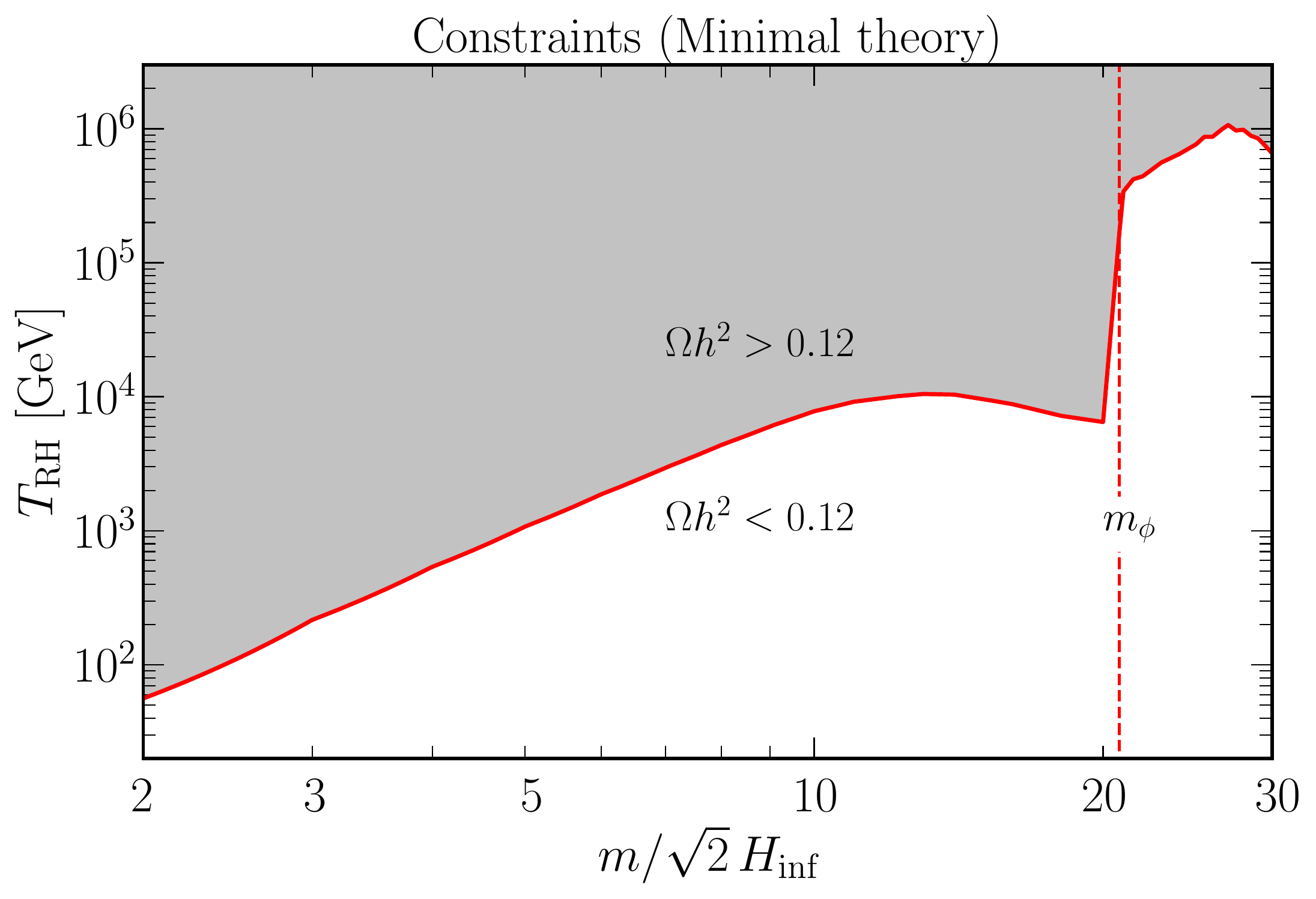} 
    \caption{\label{fig:min_TRH_by_mass}Relic abundance constraints on the parameter space for the theory of bigravity with a minimal coupling to matter.  For parameter points lying on the red curve, the predicted relic abundance of gravitationally produced massive spin-2 particles matches the observed cold dark matter relic abundance $\Omega h^2 = 0.12$. The gray-shaded region above the red curve is excluded due to over-production of massive spin-2 particles.  To avoid a ghost instability, we require the mass $m$ to be above the Higuchi bound $m = \sqrt{2} H_\mathrm{inf}$. 
	}
\end{figure}

Finally we summarize our results on massive spin-2 dark matter in \fref{fig:min_TRH_by_mass}, which shows the two-dimensional parameter space consisting of the spin-2 mass $m$ and the reheating temperature $T_\mathrm{RH}$.  The present-day relic abundance of cold dark matter is $\Omega_\text{\sc dm} h^2 = 0.12 \pm 0.0012$~\cite{Planck:2018vyg}.  Assuming that the massive spin-2 particles are cosmologically long-lived, we sum the three polarization sectors and require $\Omega h^2 \leq \Omega_\text{\sc dm} h^2$ to avoid conflict with the measured dark matter abundance.  Along the red curve on \fref{fig:min_TRH_by_mass}, the massive spin-2 particles can make up all of the dark matter.  The gray shaded region implies an over-production of dark matter, and it is excluded; the unshaded region is viable, and the massive spin-2 particles are a sub-dominant component of the dark matter.  

\subsection{Nonminimal matter coupling}
\label{sub:nonminimal_matter_GPP}

\begin{figure}[t]
	\centering
    \includegraphics[width=\textwidth]{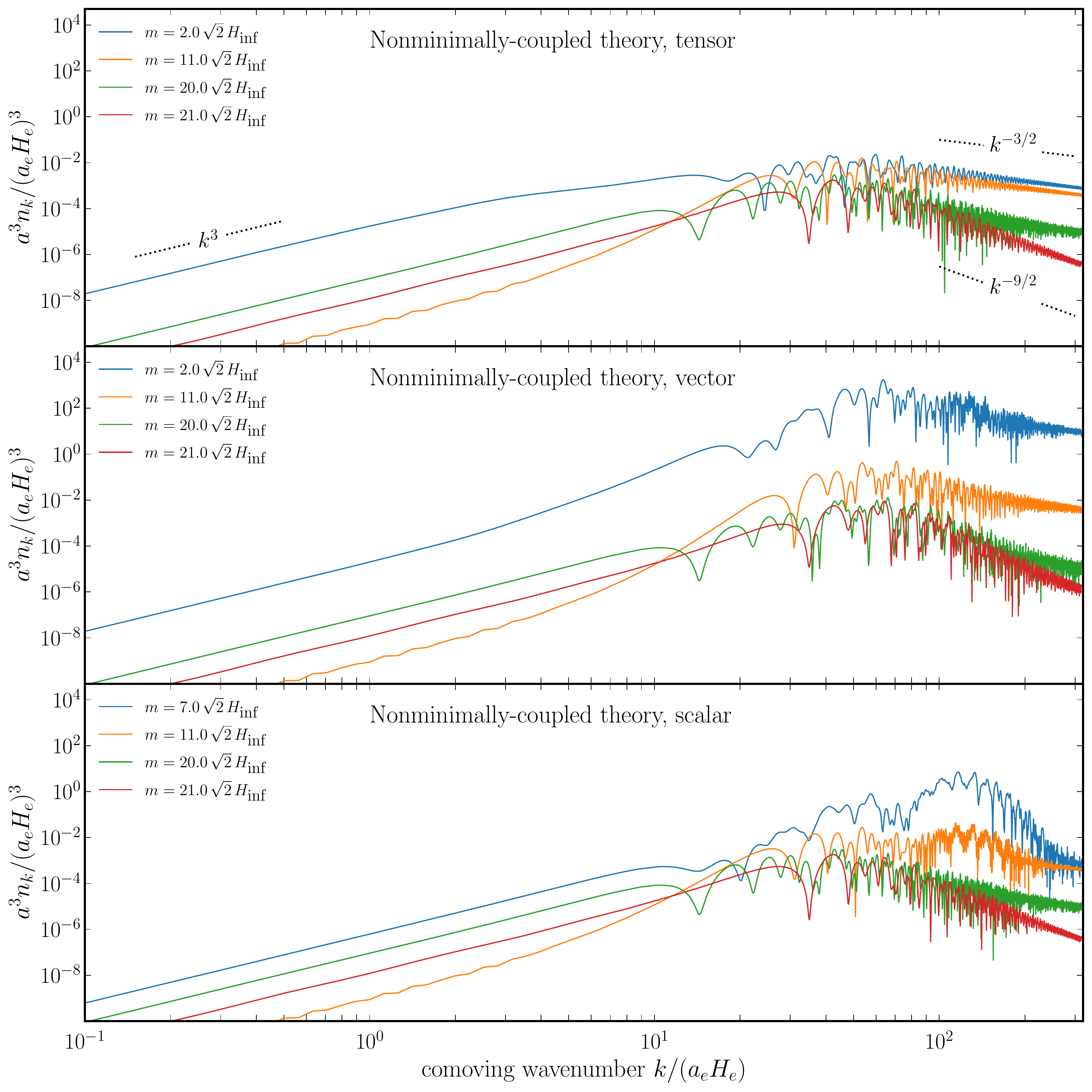}
	\caption{\label{fig:nonmin_spectra}The comoving number density spectrum $a^3 n_k$ of gravitationally produced spin-2 particles in the theory of bigravity with a nonminimal coupling to matter. The notation here is identical to \fref{fig:min_spectra}.  }
\end{figure}

\begin{figure}[t]
	\centering
	\includegraphics[width=\textwidth]{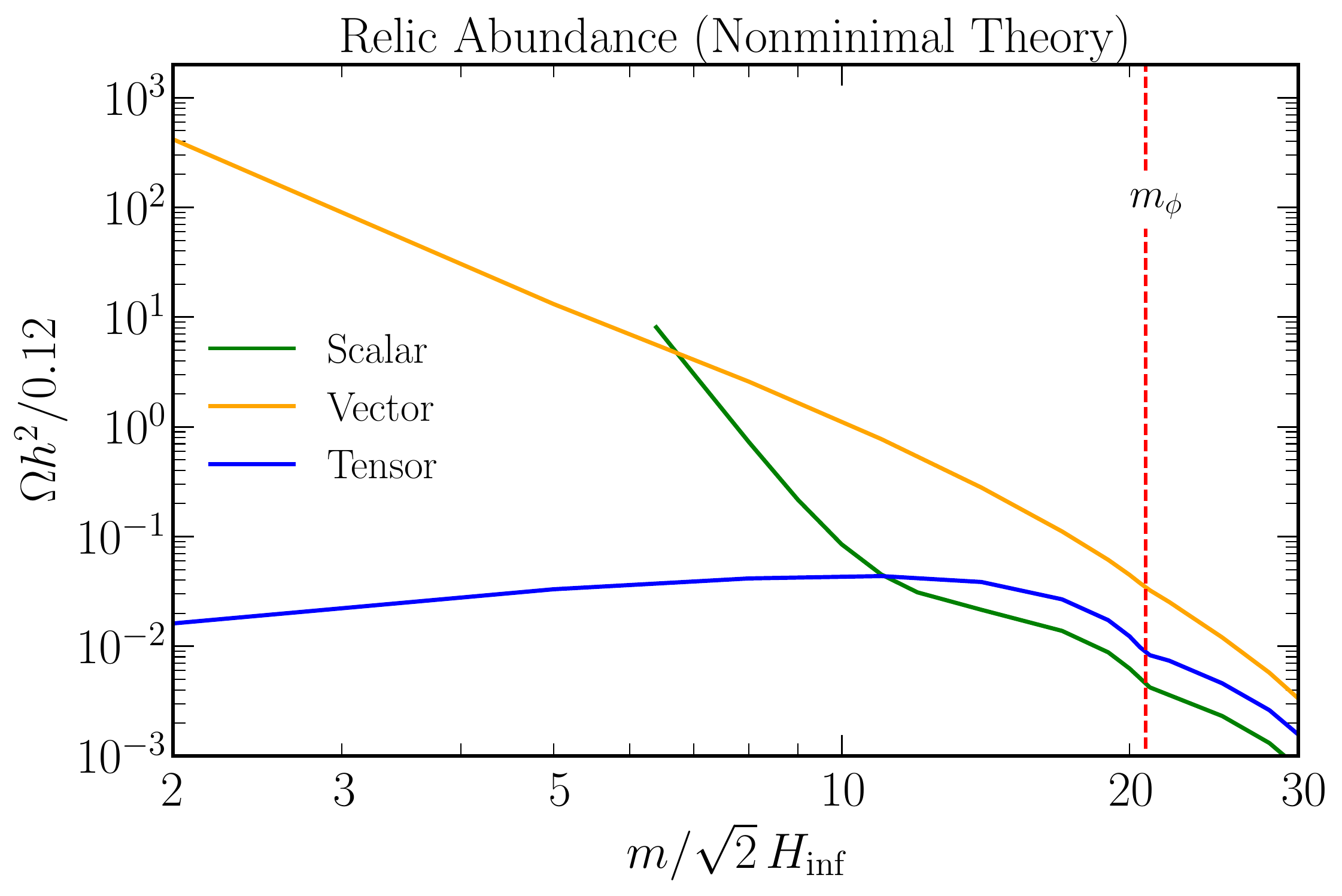} 
	\caption{\label{fig:nonmin_relic_by_mass}Relic abundance of gravitationally produced spin-2 particles versus spin-2 mass $m$ for the nonminimally-coupled theory. The notation here is identical to \fref{fig:min_relic_by_mass}.  
	}
\end{figure}

\begin{figure}[t]
	\centering
	\includegraphics[width=\textwidth]{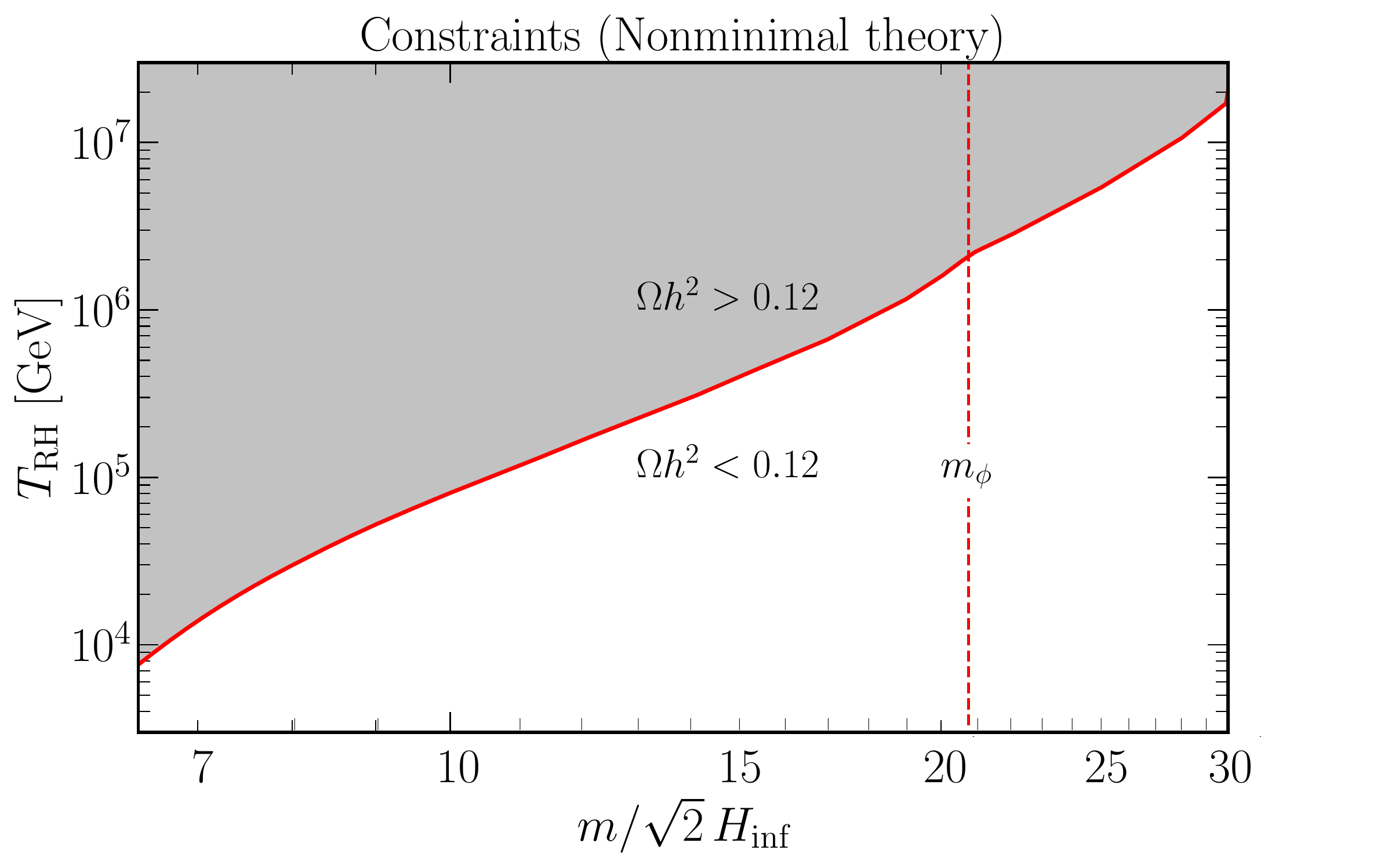} 
    \caption{\label{fig:nonmin_TRH_by_mass}Constraints on the $(m, T_\mathrm{RH})$ parameter space for the nonminimally-coupled theory.  The notation here is identical to \fref{fig:min_TRH_by_mass}.  
	}
\end{figure}

For the theory with a nonminimal coupling to matter, we perform the same analysis that was presented in \sref{sub:nonminimal_matter_GPP} for the minimally-coupled theory.  Our numerical results appear in \fref{fig:nonmin_spectra} that shows the spectra, \fref{fig:nonmin_relic_by_mass} that shows the relic abundance, and \fref{fig:nonmin_TRH_by_mass} that shows the parameter space constraints.  In the remainder of this subsection we discuss each plot in turn.  

The spectra appearing in \fref{fig:nonmin_spectra} for the nonminimally-coupled theory are the analogs of \fref{fig:min_spectra} for the minimally-coupled theory.  Similar to the case of the minimally-coupled theory, the spectra of long-wavelength (low-$k$) modes in the nonminimally-coupled theory are blue-tilted power laws, which go as $a^3 n_k \propto k^3$ for all three sectors (tensor, vector, scalar) and a broad range of spin-2 masses.  In fact, whereas for the minimally-coupled theory the spectra flatten for small masses with $m < 1.5 H_\mathrm{inf}$, this flattening is not seen in the nonminimally-coupled theory where instead the $k^3$ power law persists (not shown on the figure).  We also derive this power law analytically by approximating the modes with Hankel functions during inflation; see \aref{app:discussion_of_eoms} for details of the derivation.  The short-wavelength (high-$k$) modes display the same features that were noted previously in the minimally-coupled theory: the spectrum oscillates under a power-law envelope that transitions from $a^3 n_k \propto k^{-3/2}$ for smaller masses $m < m_\phi$ to the steeper $a^3 n_k \propto k^{-9/2}$ for larger masses $m > m_\phi$ where the annihilation channel $\phi \phi \to \chi \chi$ is kinematically blocked.  

As we discussed in \sref{sub:gradient_instability}, the nonminimally-coupled theory may exhibit a gradient instability in the vector sector, and we have explored this phenomenon with our numerical studies.  Recall that the modes in the vector sector evolve in response to a time-dependent effective squared sound speed $c_s^2(\eta)$, and if there is a period of time during which $c_s^2(\eta) < 0$, the mode equations admit an exponential growth leading to a UV-sensitive spectrum, and $n_k \propto \exp{2 \int k |c_s| \dd{\eta}}$.  In the hilltop model of inflation that we study we find that $m \gtrsim 1.46 H_\mathrm{inf}$ ensures $c_s^2(\eta) > 0$ at all times, and the gradient instability is avoided.  Lowering $m$ toward this threshold leads to an enhancement of particle production in the vector sector, which is seen in \fref{fig:nonmin_spectra} as the $m/\sqrt{2} H_\mathrm{inf} = 2$ curve (blue) in the vector sector panel (middle).  This can be understood in the following way.  For large $k$, the mode equation \eqref{eq:model2_vector_mode_eqn} is approximately $\tilde{\chi}^{\prime\prime} + c_s^2 k^2 \, \tilde{\chi} = 0$, and particle production is enhanced when $(c_s^2)' / c_s^2$ is largest. For models with $m$ close to the threshold $\approx 1.46 H_\mathrm{inf}$, there is a time at which $c_s^2$ drops close to zero from above, and an even smaller $m$ puts $c_s^2$ closer to zero.  In this sense, the large amplitude for the vector sector spectrum at $m / \sqrt{2} H_\mathrm{inf} = 2$ in \fref{fig:nonmin_spectra} foreshadows the onset of the gradient instability.  We have also checked that for $m < 1.46 H_\mathrm{inf}$, exponentially growing mode functions are obtained, although these results do not appear in \fref{fig:nonmin_spectra}.  

For the scalar sector, a ghost instability prevents us from solving the mode equation when the physical momentum $p$ is above the UV cutoff $p_\mathrm{max}$; see \sref{sub:model2_ghost_instability}.  Since $p = k/a > p_\mathrm{max}$ at sufficiently early times for any fixed $k$, all $k$ modes necessarily activate a ghost instability early during inflation.  Nevertheless, we can impose the Bunch-Davies initial condition at a late enough time when the ghost instability is avoided, and study only the $k$ modes for which there is no ghost instability in all subsequent evolution.  
Using \eref{eq:p_max_end_of_inflation}, we find a cutoff $k_\mathrm{max}$ such that the IR modes with $k < k_\mathrm{max}$ are well-behaved, whereas the UV modes with $k \geq k_\mathrm{max}$ run into a singularity during their evolution.
For our model of hilltop inflation, the cutoff is $k_\mathrm{max} / ( a_e H_e ) \approx 0.45 (m/H_\mathrm{inf})^3 \Mpl$.  In \fref{fig:nonmin_relic_by_mass}, we only present spectra for $m \geq 7 \sqrt{2} H_\mathrm{inf}$, corresponding to $k_\mathrm{max} \gtrsim 10^{2.5} a_e H_e$, such that all the modes shown on the figure have $k < k_\mathrm{max}$.   For this range of masses, the spectrum peaks at a wavenumber that is well below the cutoff $k_\mathrm{max}$, and we evaluate the total particle number by integrating $k$ up to the cutoff.  For smaller masses the cutoff drops below the scale at which the spectrum peaks, and the EFT is inapplicable for the study of gravitational particle production.  

By integrating the spectra and using \eref{eq:relic_abundance}, we evaluate the relic abundance $\Omega h^2$.  Unlike the minimally-coupled theory, the massive spin-2 in the nonminimally-coupled theory is stable, and provides a viable dark matter candidate; see \aref{app:decay}.  \Fref{fig:nonmin_relic_by_mass} shows the relic abundance $\Omega h^2$ as a function of the spin-2 mass $m$ for the tensor, vector and scalar sectors at reheating temperature $T_\mathrm{RH} = 10^5 \GeV$.  In the tensor sector, at intermediate masses $3 H_\mathrm{inf} \lesssim m \lesssim 20 H_\mathrm{inf}$, the relic abundance is increasing linearly $\Omega h^2 \propto m^1$, which is the same behavior observed previously in the minimally-coupled theory.  In the vector sector, the relic abundance grows toward smaller $m$, foreshadowing the onset of the gradient instability at the threshold $m \approx 1.46 H_\mathrm{inf}$.  In the scalar sector, we only calculate the relic abundance for $m \geq 7 \sqrt{2} H_\mathrm{inf}$ where the ghost instability is avoided.  In all three sectors, the relic abundance decreases toward large $m$, and there is a break at $m = m_\phi$, where the channel $\phi\phi \to \chi\chi$ is kinematically blocked. Finally, we note the the vector sector dominates the relic abundance for most of the masses shown.

Our constraints on the parameter space of the nonminimally-coupled theory are summarized in \fref{fig:nonmin_TRH_by_mass}.  We sum the relic abundances in the three sectors and compare the predicted $\Omega h^2$ against the measured cold dark matter relic abundance $\Omega_\text{\sc dm} h^2 \approx 0.12$.  Note that we only show results for large values of the spin-2 mass, $m \gtrsim 9 H_\mathrm{inf}$; for smaller masses the scalar sector has a ghost instability at the modes that would contribute predominantly to the total particle number.  Along the red curve, the massive spin-2 particles can make up all of the dark matter, whereas in the gray shaded region, there is an over-abundance, and in the white region the spin-2 particles make up a sub-dominant component of dark matter.  

\section{Summary and conclusion}
\label{sec:summary}

We have studied the phenomenon of gravitational particle production for massive spin-2 particles using the framework of bigravity in the context of hilltop inflation.  We studied two theories of ghost-free bigravity that are distinguished by the coupling of their metrics to matter.  The first theory entails two matter sectors with `mirrored' particle content coupling to each metric, while the second theory consists of a single matter sector coupled to a composite metric.  The first theory can be viewed as a `minimal' coupling of bigravity to matter, as it reproduces expected results for massive gravity, such as the de Sitter limit; the second `nonminimal' theory leads to exotic relations. 

By expanding the actions on a time-dependent FRW background, we isolate the degrees of freedom that transform as scalars, vectors, and tensors under the residual symmetries of spatial translations and rotations.  We derive the equations of motion for these mode functions.  In the next few paragraphs, we offer a brief overview of that procedure.

The starting point in the minimally-coupled model is two metrics $g_{\mu\nu}$ and $f_{\mu\nu}$ and two scalar matter fields $\phi_f$ and $\phi_g$ (\eref{bigL} with $\Lcal_{\star}=0$). The desired final configuration is a massless graviton (2 degrees of freedom), a massive spin-2 field (5 degrees of freedom), and 2 additional scalar degrees of freedom arising from the two scalar matter fields, for a total of 9 degrees of freedom.   

The journey from the starting point to the final configuration involves quite a few twists and turns.  We start by expanding the metrics and scalar fields about backgrounds in a mirrored manner: the backgrounds for $g_{\mu\nu}$ and $f_{\mu\nu}$ are identical, and the backgrounds for $\phi_g$ and $\phi_f$ are proportional, \eref{eq:mirrored}.   We construct new fields $u_{\mu\nu}$ and $v_{\mu\nu}$ describing massless and massive metric perturbations, respectively.  With the construction (\ref{eq:uv_to_hk}) the massive and massive spin-2 sectors decouple.  We perform a similar combination of $\varphi_u$ and $\varphi_v$ in terms of $\phi_g$ and $\phi_f$ in \eref{eq:phigftophiuv}.  The massless state $u_{\mu\nu}$ only propagates two tensor degrees of freedom.  After removing non-propagating fields, the standard SVT decomposition of the massive state $v_{\mu\nu}$ yields two tensor degrees of freedom corresponding to the usual $+,\times$ degrees of freedom associated with the $\pm2$-polarization modes (\eref{eq:model1_tensor_mode_eqn} for the mode equation) and two vector degrees of freedom that can be identified with the $\pm1$-polarization states of $v_{\mu\nu}$ (\eref{eq:model1_vector_mode_eqn} for the mode equation). The surviving scalar degrees of freedom associated with the $0$-polarization state are comprised of one state from the SVT decomposition [$B$ in \eref{eq:model1_LSk}], and one state which is a linear combination of $\varphi_v$ and one state $A$ from the SVT decomposition [the combination of $A$ and $\varphi_v$ is denoted as $\hat{\varphi}_v$ in \eref{eq:model1_phiv_to_phivhat}].  The total scalar Lagrangian consists of (\ref{eq:model1_LSk}) with mixed terms involving $B$ and $\hat{\varphi}_v$, and also (\ref{eq:phi_uphi_u}) describing the scalar field from the massless sector $\Lcal^{(2)}_{\varphi_u \varphi_u}$.  From (\ref{eq:model1_LSk}) we see that $\hat{\varphi}_v$ and $B$ have kinetic mixing and mass mixing.  We define new fields $\Pi$ and $\mathcal{B}$ in terms of $\hat{\varphi}_v$ and $B$ to eliminate kinetic mixing.  The parameter that determines the mixing is defined in \eref{eq:kappa} and presented in graphical form in \fref{fig:kappa}.  The resulting scalar field Lagrangian is given by \eref{eq:model1_LSk_diagonal}.  One final change of field variables to $\chi_\Pi$ and $\chi_\mathcal{B}$ results in canonical kinetic terms with scalar Lagrangian (\ref{eq:L_chi_Pi_and_B}) and mode equations (\ref{eq:EOM_chi_Pi_and_B}). At late times the mode equations for $\chi_\Pi$ and $\chi_\mathcal{B}$ decouple, which allows the identification of $\chi_\Pi$ as the inflaton and $\chi_\mathcal{B}$ as the massive scalar produced by CGGP.

Performing the bookkeeping for the total number of degrees of freedom, we have $2$ tensor degrees of freedom for the massless graviton and $2$ tensor degrees of freedom for the massive spin-2; $2$ vector degrees of freedom from the massive spin-2; and $3$ scalar degrees of freedom arising from the scalar perturbations $\varphi_u$, $\mathcal{B}$, and $\Pi$, for the expected 9 degrees of freedom.  In \fref{fig:min_spectra} we show the spectrum of perturbations for the massive tensor and vector states, along with the scalar state $\mathcal{B}$.  The spectrum of perturbations for $\varphi_u$ and $\Pi$ are presented in \fref{fig:min_spectra_scalar}.  Assuming the massive spin-2 is stable, in \fref{fig:min_relic_by_mass} we show the relative contribution of the tensor, vector, and scalar modes to $\Omega h^2$ for a representative value of the reheat temperature, $T_\mathrm{RH}=10^5\, \mathrm{GeV}$.  Finally, in \fref{fig:min_TRH_by_mass} we show the relic abundance constraints on $T_\mathrm{RH}$ as a function of mass.

For the nonminimally-coupled theory we start with \eref{bigL} with $\Lcal_g = \Lcal_f = 0$, expand the metrics $g_{\mu\nu}$ and $f_{\mu\nu}$ around equal backgrounds as in the minimally-coupled theory, and then form massless states $u_{\mu\nu}$ and massive states $v_{\mu\nu}$.  The composite metric $(g_\star)_{\mu\nu}$ is expressed in terms of the background $\bar{g}_{\mu\nu}$, $u_{\mu\nu}$, and $v_{\mu\nu}$; see \eref{geffex}. The single scalar field $\phi_\star$ is expanded about a background field with perturbation $\varphi_\star$.  Performing a SVT decomposition on the massive state we find the Lagrangian for the canonically-normalized tensor field in \eref{eq:model2_LT_can} with mode equations in \eref{eq:model2_tensor_mode_eqn}.  The Lagrangian for the vector sector resulting from the SVT decomposition is given in \eref{eq:model2_LV}.  After eliminating the nondynamical variable and canonically normalizing the kinetic term leads to the vector  Lagrangian in Fourier space given by \eref{eq:fouriervector} with mode equation \eqref{eq:model2_vector_mode_eqn}.  Since $\varphi_\star$ does not couple to $v_{\mu\nu}$ in the nonminimally-coupled model at the level of the quadratic action there is only one propagating degree of freedom in the scalar sector, $B$.   A new field $\chi$ is defined in terms of $B$ with Lagrangian in Fourier space given by \eref{eq:nonminvectorfourier} with mode equation \eref{eq:nonminvecmode}.  The spectra for the tensor, vector, and scalar modes is presented in \fref{fig:nonmin_spectra}.  Assuming the particle is stable, the relic abundance is presented in \fref{fig:nonmin_relic_by_mass}, again assuming $T_\mathrm{RH}=10^5\, \mathrm{GeV}$.   In \fref{fig:nonmin_TRH_by_mass} we show the relic abundance constraints on $T_\mathrm{RH}$ as a function of mass.  The nonminimally-coupled model propagates the expected 5 degrees of freedom from the massive sector: 2 each in the tensor and vector sector and a single one in the scalar sector.  The model also has 3 degrees of freedom in the massless sector, for a total of 8.

Through the procedures described above, we arrive at the equations of motion for the scalar-, vector-, and tensor-sector mode functions in both theories of bigravity.  For the first theory of bigravity with a minimal coupling to matter, the scalar mode equations admit a ghost instability.  We derive a condition on the spin-2 mass and Hubble parameter for the avoidance of this ghost \eqref{eq:FRW_Higuchi}: $m^2 > 2 H^2 ( 1 - \epsilon )$ where $\epsilon = - H^\prime / (a H^2)$ is the first slow-roll parameter.  This inequality is an FRW-generalization of the Higuchi bound for massive gravity in de Sitter spacetime, and it represents one of the main results of our work.  In the de Sitter limit, our bound reproduces the usual Higuchi bound $m^2 > 2 H^2$; however, after inflation is ended $\epsilon > 1$, and we find that the spin-2 mass bound becomes trivial, $m^2 > 0$.  Many studies of massive gravity and bigravity consider values for the Fierz-Pauli mass $m$ that are comparable to the scale of the dark energy cosmological constant $\Lambda_{cc} \sim (10^{-33} \ \mathrm{eV})^2$.  Our FRW-Higuchi bound implies that the cutoff of such effective theories must fall below the inflationary Hubble scale, to avoid activating the ghost instability during inflation. 

To investigate cosmological gravitational particle production, we employed a hilltop model of inflation that reproduces cosmological observables.  We numerically solved the mode equations on this background along with Bunch-Davies initial conditions.  Special care was taken to treat the scalar sector of the minimally-coupled theory, which includes a mixing of the inflaton perturbations and the scalar perturbations of the massive spin-2 field.  From the late-time solutions of the mode equations, we infered the spectrum and cosmological abundance of gravitationally produced massive spin-2 particles in the two theories of bigravity and for each of three sectors (tensor, vector, scalar).  We developed an analytical understanding of the numerical results, particularly the power law behavior observed in the spectra, and the relations with gradient and ghost instabilities.  

The gravitational production of massive spin-2 particles may have various phenomenological implications for reheating, dark matter, and other cosmological relics.  If these particles are cosmologically long-lived, they provide phenomenologically unique and theoretically compelling candidates for the cold dark matter, which is only known to have gravitational interactions.  In the minimally-coupled theory, the massive spin-2 may decay via the inflaton, and its stability is a model-dependent issue; whereas, for the nonminimally-coupled theory the massive spin-2 is stable at tree level.  Assuming that the massive spin-2 particles are cosmologically long lived, we calculate their relic abundance today and compare with the observed abundance of cold dark matter.  For the minimally-coupled theory, we find that GPP can be responsible for generating all of the dark matter if the spin-2 mass is $\order{1} \lesssim m / H_\mathrm{inf} \lesssim \order{10}$ and the reheating temperature is $\order{10^{2} \GeV} \lesssim T_\mathrm{RH} \lesssim \order{10^6 \GeV}$; see \fref{fig:min_TRH_by_mass}.  For the nonminimally-coupled theory, the avoidance of a ghost instability restricts $10 \lesssim m/H_\mathrm{inf}$; see \fref{fig:nonmin_TRH_by_mass}.  

We have focused on studying the production of spin-2 dark matter during inflation and at the end of inflation, and our work leaves open several avenues for further investigation.  In our minimally-coupled theory of bigravity, the stability and lifetime of the massive spin-2 particles depends on additional model building that was deemed beyond the scope of our work.  It would be interesting to explore what kind of reheating sector (i.e., coupling of radiation to the inflaton) would allow for a cosmologically long-lived massive spin-2.  Along the same line, when the thermal bath is taken into account, another channel opens for gravitational particle production through gravity-mediated thermal freeze in~\cite{Garny:2015sjg,Bernal:2018qlk}.  If these particles are not cosmologically long-lived, they would not provide a candidate for the dark matter, but their out-of-equilibrium decay could be associated with the production of other relics, such as the matter-antimatter asymmetry.  We have focused on two theories of bigravity that admit equal backgrounds for the two metrics, but other non-proportional solutions are available. A calculation of CGPP in such spacetimes may also furnish an explanation for the origin of massive spin-2 particles.  Finally, if the massive spin-2 particle were to leave its imprint on cosmological spectra such as CMB non-gaussianity~\cite{Dimastrogiovanni:2018uqy} (i.e., the ``cosmological collider'' program), this information would provide a powerful new tool for testing these theories.  

\begin{acknowledgments}
We are grateful to Andrew Tolley for illuminating discussions of the FRW-generalized Higuchi bound.  We also thank Matteo Fasiello and Austin Joyce for guidance at the beginning of this project.  The work of E.W.K. was supported in part by the US Department of Energy contract DE-FG02-13ER41958.  A.J.L.~and S.L.~are supported in part by the National Science Foundation under Award No.~PHY-2114024.  R.A.R. is supported by the US Department of Energy grant DE-SC0011941.  Algebraic manipulation of tensor quantities and the SVT decomposition were performed using \texttt{xAct}~\cite{Brizuela:2008ra}, a collection of \texttt{Mathematica} packages available for free on the website \url{http://xact.es/faq.html}.
\end{acknowledgments}

\begin{appendix}

\section{Behavior of long-wavelength modes}
\label{app:discussion_of_eoms}

In \fref{fig:min_spectra} and \fref{fig:nonmin_spectra} of \sref{sec:GPP}, we see that the number density spectra of long-wavelength (low-$k$) modes exhibit power law behavior.  In this appendix, we give a derivation of this power law behavior by studying the evolution of the mode functions $\tilde{\chi}$ during inflation.  

In \sref{sec:SVT}, we presented the equations of motion for all sectors (except for the scalar sector of the minimally-coupled theory) in terms of mode functions $\tilde{\chi}(\eta,k)$ and time-dependent effective frequency $\omega_k^2$.  The frequencies $\omega_k^2$ are functions of background quantities such as $a(\eta)$, $H(\eta)$, $H'(\eta)$, etc.  During inflation, the background spacetime is quasi-de-Sitter, and the effective frequencies have simple limiting forms.  In fact, the frequencies $\omega_k^2$ are approximately constant during inflation, except for a possible jump at the horizon-crossing time $k = a H$.  The frequencies $\omega_k^2$ during inflation are summarized by the formula
\begin{align}
  \omega_k^2 \approx (1 - \delta) k^2 + a^2 (m^2 + \alpha H_{\mathrm{inf}}^2)
  \;.
  \label{eq:omega_k_all}
\end{align}
and \tref{tab:omega_k_coeffs}.  Here, $\delta$ and $\alpha$ are constants of order unity, and $\delta$ may depend on $\mu \equiv m / H_{\mathrm{inf}}$.
\begin{table}[ht]
\centering
\caption{Coefficients for \eref{eq:omega_k_all} during inflation. $\mu \equiv m / H_\mathrm{inf}$.}
\begin{tabular}{| c | c c | c c |}
    \hline
    \multirow{2}{*}{Sector}  
    & \multicolumn{2}{c|}{$k/a \gg (m, H_\mathrm{inf})$}  
    & \multicolumn{2}{c|}{$k/a \ll (m, H_\mathrm{inf})$}  \\
    & \(\alpha\) & \(\delta\) & \(\alpha\) & \(\delta\)    \\
    \hline
    Minimal, tensor     & \(-2\)     & \(0\)  & \(-2\)     & \(0\)                   \\
    Minimal, vector     & \(-6\)     & \(0\) & \(-2\)     & \(\mu^{-2}\)               \\
    Nonminimal, tensor & \(1\)      & \(0\) & \(1\)      & \(0\)                   \\
    Nonminimal, vector & \(-3\)      & \(0\) & \(1\)      & \(1/(3 + \mu^2)\) \\
    Nonminimal, scalar & \(-5\)      & \(0\) & \(1\)      & \(4/(9 + 3 \mu^2)\)   \\
    \hline
  \end{tabular}
  \label{tab:omega_k_coeffs}
\end{table}

At sufficiently early times, $k / a \gg (m, H_\mathrm{inf})$ is satisfied, and \tref{tab:omega_k_coeffs} tells us that $\delta = 0$ and $\omega_k^2 \approx k^2$ for all 5 listed sectors.  This means the mode functions in all 5 sectors should be given the Bunch-Davies initial condition $\tilde{\chi}(\eta) = e^{-i k \eta} / \sqrt{2k}$, as expected.  If $k < a_e H_e$, then there is also a period during inflation such that $k / a \ll (m, H_\mathrm{inf})$, namely the period after the mode left the horizon; the coefficients $\delta$ and $\alpha$ change when this period is entered.  Note that the coefficients in \tref{tab:omega_k_coeffs} for $k / a \gg (m, H_\mathrm{inf})$ are true even for $k > a_e H_e$.

To understand the behavior of long-wavelength modes, we now solve for $\tilde{\chi}(\eta)$ under the assumption that $\delta$ and $\alpha$ are fixed during inflation.
We take \(a(\eta) = -1/(H_{\mathrm{inf}} \eta)\), \(\eta \in (-\infty,0)\), then \eref{eq:omega_k_all} becomes:
\begin{align}
  \omega_k^2 \approx (1 - \delta) k^2 + \frac{1}{\eta^2} (\mu^2 + \alpha)
  \;.
\end{align}
The general solution $\tilde{\chi}(\eta)$ for the above effective frequency is given by Hankel functions:
\begin{align}
  \tilde{\chi}(\eta) &= \sqrt{-\eta}
  \left( 
  C_1 H^{(1)}_\nu(-k \eta \sqrt{1-\delta}) 
  + C_2 H^{(2)}_\nu(-k \eta \sqrt{1-\delta})
  \right) \nonumber \\
  \qq{where} \nu &= \sqrt{\frac14 - \alpha - \mu^2}
  \;.
\end{align}
The solution satisfying the Bunch-Davies initial condition with appropriate normalization is
\begin{align}
 \tilde{\chi}(\eta) &= e^{i \frac{\pi}{2} (\nu + \frac12)} \sqrt{\frac{\pi}{4}} \sqrt{-\eta}\, H^{(1)}_\nu(-k \eta \sqrt{1-\delta})
  \;.
\end{align}
If \(\nu\) is real, then the exponential factor in the front is a merely a phase; if \(\nu\) is imaginary with \(\Im[\nu] > 0\), then the exponential factor contributes to the magnitude of $\tilde{\chi}(\eta)$.

We now discuss the low-\(k\) (\(k \ll a_e H_e\)) behavior of the solutions.  After horizon crossing, we have \(-k\eta = k/(a H_{\mathrm{inf}}) \ll 1\).  If we rename the argument of the Hankel function by \(z \equiv -k \eta \sqrt{1-\delta}\), then \(z \ll 1\) after horizon crossing, and the solution is approximately: 
\begin{align}
\tilde{\chi}(\eta) 
&\approx e^{i \frac{\pi}{2} (\nu - \frac12)} \sqrt{\frac{1}{4\pi}} \sqrt{-\eta} \Big[ e^{-i\pi\nu} \Gamma(-\nu) \Big(\frac{z}{2} \Big)^\nu + \Gamma(\nu) \Big(\frac{z}{2} \Big)^{-\nu} \Big] 
\;.
\end{align}
If \(\nu > 0\), then \(z^{-\nu}\) dominates over \(z^\nu\), and we have:
\begin{align}
\tilde{\chi}(\eta) \approx e^{i \frac{\pi}{2} (\nu - \frac12)} \sqrt{\frac{1}{4\pi}} \sqrt{-\eta} \, \Gamma(\nu) \Big( \frac{-k \eta \sqrt{1-\delta}}{2} \Big)^{-\nu} 
\;.
\end{align}
Note from above that $\tilde{\chi}(\eta)  \sim k^{-\nu}$.
If \(\nu\) is imaginary and \(\Im[\nu] > 0\), then the solution is approximated by:
\begin{align}
\tilde{\chi}(\eta)  &\approx 
e^{\frac{\pi}{2} (- |\nu| - \frac12 i)} \sqrt{\frac{1}{4\pi}} \sqrt{-\eta} \Big[ e^{\pi |\nu|} \Gamma(-i|\nu|) e^{i|\nu| \ln(z/2)} + \Gamma(i|\nu|) e^{-i|\nu| \ln(z/2)} \Big]
\;.
\end{align}
Note that now $\tilde{\chi}$ does not have a power law dependence on $k$, but rather an oscillatory dependence on \(\ln(z/2) \sim \ln(k)\).

Finally, the approximate solutions above inform us about the low-$k$ behavior of the number density spectrum, $n_k$.  Since the long-wavelength modes are frozen outside the horizon and experience negligible particle production after they re-enter the horizon, we expect the Bogoliubov coefficients $| \beta_k |^2$ for these modes at late times to track the corresponding values during inflation.   From \eref{eq:beta_k_def}, we see that $| \beta_k |^2 \sim | \tilde{\chi} |^2$. For \(\nu > 0\), we have $ |\beta_k |^2 \sim k^{-2\nu} $, so the power law for the particle number density is \(n_k \sim k^3 |\beta_k |^2 \sim k^{3-2\nu}\).  For imaginary \(\nu\), we have $ |\beta_k |^2 \sim k^{0} $ and $n_k \sim k^3 |\beta_k|^2 \sim k^{3}$; moreover, due to the interference between the \(e^{\pm i|\nu| \ln(z/2)}\) factors in $\tilde{\chi}$, we expect to see oscillations in \(|\beta_k|^2\) as a function of \(\ln(k)\).  These phenomena are shown in \fref{fig:min_spectra} and \fref{fig:nonmin_spectra} and discussed in \sref{sec:GPP}.

\section{Stueckelberg derivation of FRW Higuchi bound}
\label{app:stueck}

Here we provide an alternative derivation of the FRW-generalized Higuchi bound that appears in \eref{eq:FRW_Higuchi}.  For the theory of bigravity with a minimal coupling to matter, recall  \eref{eq:model1_L2}, i.e. 
\bes{
    \Lcal_\mathrm{massive}^{(2)}
    & = 
    -\tfrac{1}{2} \nabla_\lambda v_{\mu\nu} \nabla^\lambda v^{\mu\nu} 
    + \nabla_\mu v^{\nu\lambda} \nabla_\nu v^\mu_{~ \lambda} 
    - \nabla_\mu v^{\mu\nu} \nabla_\nu v 
    + \tfrac{1}{2} \nabla_\mu v \nabla^\mu v 
    \\ & \qquad 
    + \biggl( \frac{\ddot{a}}{a}+(D-2)\frac{\dot{a}^2}{a^2} \biggr)
    \Bigl( v^{\mu\nu} v_{\mu\nu} -\tfrac{1}{2} v^2 \Bigr) 
    \\ & \qquad 
    - \tfrac{1}{2}m^2 \bigl( v^{\mu\nu} v_{\mu\nu} - v^2 \bigr) 
    \\ & \qquad 
    + \MP^{-1} \Bigl[ \bigl( \nabla_\mu \bar{\phi} \nabla_\nu \varphi_v+\nabla_\nu \bar{\phi} \nabla_\mu \varphi_v \bigr) \bigl( v^{\mu\nu} - \tfrac{1}{2} \bar{g}^{\mu\nu} v \bigr) -V'(\bar{\phi}) \varphi_v v \Bigr] 
    \\ & \qquad 
    - \tfrac{1}{2} \nabla_\mu \varphi_v \nabla^\mu \varphi_v 
    - \tfrac{1}{2} V''(\bar{\phi}) \varphi_v^2
    \;,
}
gives the quadratic action for a massive spin-2 field $v_{\mu\nu}$ on an FRW background sourced by a scalar field with background value $\bar{\phi}$ and perturbation $\varphi_v$: 
For convenience we define
\ba{
    \tilde{R}_{\mu\nu} 
    & \equiv \bar{R}_{\mu\nu} - \frac{1}{\MP^2} \nabla_\mu \bar{\phi} \nabla_\nu \bar{\phi} 
    = \biggl( \frac{\ddot{a}}{a} + (D-2) \frac{\dot{a}^2}{a^2} \biggr) \bar{g}_{\mu\nu} 
    = \Bigl( (D-1) H^2 + \dot{H} \Bigr) \bar{g}_{\mu\nu} \\
    & \equiv \frac{\tilde{R}}{D} \, \bar{g}_{\mu\nu}
    \;,
}
where $D$ is the number of spacetime dimensions.  
We perform the usual Stueckelberg trick, followed by the standard conformal transformation used in massive gravity (see, e.g., \cite{Hinterbichler:2011tt}):
\ba{
    v_{\mu\nu} \rightarrow v_{\mu\nu} + \nabla_\mu A_\nu + \nabla_\nu A_\mu + 2 \nabla_\mu \nabla_\nu \Phi + \frac{2}{D-2} m^2 \bar{g}_{\mu\nu} \Phi 
    \;.
}
We focus on the scalar sector and write every term that contains a $\Phi$:
\bes{
    \Lcal_\mathrm{massive}^{(2)}
    & \supset 
    -2m^2 \biggl[ \frac{D-1}{D-2} m^2 \bar{g}_{\mu\nu} - \bar{R}_{\mu\nu} \biggr] \nabla^\mu \Phi \nabla^\nu \Phi 
    + 2 m^4 \frac{D}{D-2} \biggl[ \frac{D-1}{D-2}m^2-\frac{\tilde{R}}{D} \biggr] \Phi^2 
    \\ & \qquad 
    -4 m^2 \biggl[ \frac{D-1}{D-2} m^2 \bar{g}_{\mu\nu} - \bar{R}_{\mu\nu} \biggr] A^\mu \nabla^\nu \Phi 
    + 2 m^2 \biggl[ \frac{D-1}{D-2} m^2 - \frac{\tilde{R}}{D} \biggr] v \Phi 
    \\ & \qquad 
    - \frac{2m^2}{\MP} \nabla_\mu \bar{\phi} \nabla^\mu \varphi_v \, \Phi - \frac{2m^2}{\MP} \frac{D}{D-2} \Box \bar{\phi} \, \varphi_v \Phi 
    \;.
}
There are also $\varphi_v^2$, $v^2$ and $v \varphi_v$ terms but they are not relevant as the kinetic terms are already diagonal in this language. 

Note that setting $m=0$ causes the $\Phi$ field to drop out of the Lagrangian entirely, as expected from the enhanced gauge symmetry.  
However, even for $m \neq 0$ time derivatives of $\Phi$ are absent from the Lagrangian at a time when 
\ba{
    \frac{D-1}{D-2} \, m^2 \, \bar{g}_{00} - \bar{R}_{00} = 0 
    \;,
}
or equivalently 
\ba{\label{HigAp}
    m^2 
    = \bigl( D-2 \bigr) \bigl( H^2 + \dot{H} \bigr) 
    = \bigl( D-2 \bigr) H^2 \bigl( 1-\epsilon \bigr) 
    \;,
}
where we have written the first slow-roll parameter as $\epsilon = - \dot{H} / H^2 = - H^\prime / a H^2$.  For $D = 4$ spacetime dimensions, \eref{HigAp} is precisely the Higuchi bound from \eref{eq:FRW_Higuchi} that we found using the SVT analysis.  Unlike $m=0$ there is no gauge symmetry at \eref{HigAp} since $\Phi$ and its spatial derivatives don't drop out of the Lagrangian.  Instead, this is just a point where the kinetic term of $\Phi$ passes through zero. 

Alternatively, the gradient terms for $\Phi$ as well as many non-derivative terms containing $\Phi$ vanish at a time when  
\ba{
    \frac{D-1}{D-2} \, m^2-\frac{\tilde{R}}{D} = 0
    \;,
}
or, written equivalently, when
\ba{
\label{HigAp2}
    m^2 = \bigl( D-2 \bigr) H^2 \biggl( 1  - \frac{\epsilon}{D-1} \biggr) 
    \;.
}
When $\epsilon =0$, this coincides with \eqref{HigAp} and the usual Higuchi bound as expected.  We note that this latter expression \eqref{HigAp2} doesn't indicate a bound on a tachyonic instability, since the mixing between $\Phi$ and $\varphi_v$ doesn't also vanish for non-zero $\epsilon$.

\section{Stability of massive spin-2 particles}
\label{app:decay}

In order for the massive spin-2 particle in our theories of bigravity to serve as a dark-matter candidate, it must be stable or at least cosmologically long-lived.  In this appendix, we address the issue of stability.  Terms in the Lagrangian that are linear in the massive spin-2 field $v\indices{_\mu_\nu}$ could potentially mediate its decay (via tree-level Feynman graphs), and our task is to identify whether such terms are present.  

First, we neglect the matter sectors and consider only the bigravity Lagrangian, i.e., the first line of \eref{bigL}, since both the minimally-coupled and nonminimally-coupled theories share these terms.  The action can be written using the massless spin-2 field $u\indices{_\mu_\nu}$ and the massive spin-2 field $v\indices{_\mu_\nu}$ via \erefs{eq:model1_bkg}{eq:uv_to_hk}.  Interactions that would mediate the decay $v \to uu$ can take the form $vuu$ (or $v \to uuu$ via or $vuuu$).  However, we find that the bigravity Lagrangian does not contain any terms that are linear in $v\indices{_\mu_\nu}$ for our choice of parameters: 
\bes{
    & \eval{ \fdv{v\indices{_\mu_\nu}} \left( \frac{M_g^2}{2}\sqrt{-g}R[g] + \frac{M_f^2}{2}\sqrt{-f}R[f] - m^2 M_*^2 \sqrt{-g} V(\mathbb{X}; \beta_n) \right) }_{v=0} 
    \\ & \quad 
    = \eval{ \left(
    \frac{M_g^2}{2} \fdv{g\indices{_\rho_\sigma}}{v\indices{_\mu_\nu}} \fdv{\sqrt{-g}R[g]}{g\indices{_\rho_\sigma}} 
    + \frac{M_f^2}{2} \fdv{f\indices{_\rho_\sigma}}{v\indices{_\mu_\nu}} \fdv{\sqrt{-f}R[f]}{f\indices{_\rho_\sigma}}
    - m^2 M_*^2 \fdv{\sqrt{-g} V(\mathbb{X}; \beta_n)}{v\indices{_\mu_\nu}}
    \right) }_{v=0} 
    \\ & \quad 
    = \eval{ \left(
    \left(\frac{M_g^2}{2}\frac{2M_*}{M_g^2} - \frac{M_f^2}{2}\frac{2M_*}{M_f^2}\right) \delta\indices{_\rho^{(\mu}} \delta\indices{_\sigma^{\nu)}} \fdv{\sqrt{-\mathcal{G}}R[\mathcal{G}]}{\mathcal{G}\indices{_\rho_\sigma}}
    - m^2 M_*^2 \fdv{\sqrt{-g} V(\mathbb{X}; \beta_n)}{v\indices{_\mu_\nu}}
    \right) }_{v=0} 
    \\ & \quad 
    = \eval{
    - m^2 M_*^2 \fdv{\sqrt{-g} V(\mathbb{X}; \beta_n)}{v\indices{_\mu_\nu}}
    }_{v=0} 
    \\ & \quad 
    = 0
    \;.
}
Going from the second to third line, we used $g\indices{_\mu_\nu} = \mathcal{G}\indices{_\mu_\nu} + (2M_*/M_g^2) \, v\indices{_\mu_\nu}$ and $f\indices{_\mu_\nu} = \mathcal{G}\indices{_\mu_\nu} - (2M_*/M_f^2) \, v\indices{_\mu_\nu}$ where $\mathcal{G}\indices{_\mu_\nu} \equiv \bar{g}\indices{_\mu_\nu} + (2/\Mpl) \, u\indices{_\mu_\nu}$.  The expression on the fourth line vanishes upon setting $\Lambda_g / M_g^2 = \Lambda_f / M_f^2$, which is necessary for a proportional background; see \erefs{eq:mirroring_conditions}{eq:a_b_condition}.  
It follows that the bigravity Lagrangian doesn't contain any terms that are linear in $v\indices{_\mu_\nu}$, and it cannot mediate tree-level decays.  See also eq.~(4.12) of \rref{Bonifacio:2017nnt} for complete list of trilinear terms in the bigravity Lagrangian on a Minkowski background.

We now turn our attention to the matter couplings, beginning with the nonminimally-coupled theory.  The matter action appears in the last term of \eref{bigL}.  Varying with respect to the massive metric perturbation $v\indices{_\mu_\nu}$ yields: 
\ba{
    \eval{\fdv{v\indices{_\mu_\nu}} \left(\sqrt{-g_\star} \,\Lcal_\star(g_\star,\phi_\star) \right)}_{v=0}
    = \eval{\fdv{(g_\star)\indices{_\sigma_\rho}}{v\indices{_\mu_\nu}} \fdv{(g_\star)\indices{_\sigma_\rho}} (\sqrt{-g_\star} \,\Lcal_\star(g_\star,\phi_\star)) }_{v=0}
    = 0
    \;,
}
In the last equality we have used \eref{geffex}, which follows from our choice of parameters in \eref{eq:a_b_condition}.
Evidently the matter action of the nonminimally-coupled theory does not contain terms that are linear in the massive spin-2 field $v\indices{_\mu_\nu}$, and it cannot mediate the (tree-level) decay of the massive spin-2 particle.  Note that this argument generalizes to an arbitrary matter sector, containing any number of matter fields, such as the Standard Model particle content.  Our calculation shows that the absence of tree level $v\indices{_\mu_\nu}$ decay channels is an essential feature of the nonminimally-coupled theory; this result was stressed upon in \rref{Schmidt-May:2014xla}.  Also see eq.~(4.59) of \cite{Bonifacio:2017nnt} that provides the $v\phi_\star\phi_\star$ operator coefficient without imposing \eref{eq:a_b_condition}.

\begin{figure}[t]
  \centering
  \includegraphics[width=0.9\textwidth]{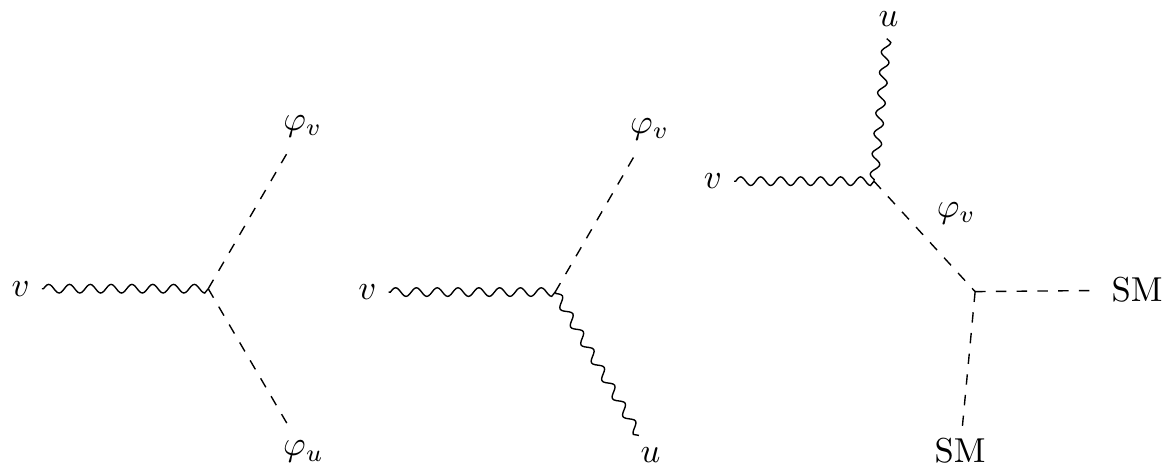}
  \caption{\label{fig:diagram}Direct and indirect decay of \(v\indices{_\mu_\nu}\) in the minimally-coupled theory.}
\end{figure}

Next we consider interactions with matter in the minimally-coupled theory, corresponding to the fourth and fifth terms in \eref{bigL}.
Interactions with the inflaton fields $\phi_g$ and $\phi_f$, include terms such as
\bes{
    \Lcal_{v~\mathrm{decay}} 
    & =\Mpl^{-2}\Big[
    2 v\indices{_\mu_\nu} u\indices{^\mu^\nu} (\nabla_\lambda \varphi_v) (\nabla^\lambda \bar{\phi}) 
    - 4 v\indices{_\mu_\nu} u\indices{^\nu_\lambda} (\nabla^\mu \bar{\phi}) (\nabla^\lambda \varphi_v) 
    \\ & \qquad \qquad  
    + 2 v u\indices{_\mu_\nu} (\nabla^\mu \bar{\phi}) (\nabla^\nu \varphi_v) 
    - 4 v\indices{_\mu_\nu} u\indices{^\nu_\lambda} (\nabla^\lambda \bar{\phi}) (\nabla^\mu \varphi_v) + 2 v\indices{_\mu_\nu} u\indices{^\mu^\nu} \varphi_v V'(\bar{\phi})\Big] 
    \\ & \quad 
    +\Mpl^{-1}\Big[ 2 v\indices{_\mu_\nu} (\nabla^\mu \varphi_u) (\nabla^\nu \varphi_v) 
    - v (\nabla_\mu \varphi_u) (\nabla^\mu \varphi_v) 
    - v \varphi_u \varphi_v V''(\bar{\phi}) \Big] 
    \\ & \quad 
    + ( v \varphi_v \varphi_v \text{ terms} )
  \;,
}
where $\bar{\phi}$ is the inflaton background and where $\varphi_u$ and $\varphi_v$ are the scalar field perturbations. Terms of the form $v u \varphi_v$ and $v \varphi_u \varphi_v$ could mediate the massive spin-2 particle's decay.  For $m > 2 m_\phi$, the decay $v \to \varphi_u \varphi_v$ is kinematically allowed, see \fref{fig:diagram}, and we estimate its rate as 
\ba{\label{eq:Gamma_decay}
    \Gamma 
    \sim \biggl( \frac{m^2}{\Mpl} \biggr)^2 \frac{1}{m}
    \sim \frac{m^3}{\Mpl^2}
    \;.
}
Since the avoidance of a ghost instability during inflation requires $m \gtrsim H_\mathrm{inf}$ (Higuchi bound), the rate is bounded from below as $\Gamma \gtrsim H_\mathrm{inf}^3 / \MP^2$.  Despite the Planck suppression, this large rate would correspond to a decay in the early universe (unless $H_\mathrm{inf}$ were very small, but then CGPP would also be suppressed).  For a smaller mass $m_\phi < m < 2 m_\phi$ the decay to two inflatons is kinematically blocked, but the decay channel $v \to u \varphi_v$ may still occur, although the rate depends on $\nabla\bar{\phi}$ or $V'(\bar{\phi})$, which are small at late times. For an even smaller mass, $m < m_\phi$, decays to on-shell inflatons are kinematically blocked.  Decays via off-shell inflatons into other matter fields may still occur, see \fref{fig:diagram}, although the rate for these channels is subject to additional model dependence.  

Additionally, for the theory of bigravity with a minimal coupling to matter, the massive spin-2 field interacts with the same matter sectors as the massless spin-2 field (graviton).  Consequently, $v\indices{_\mu_\nu}$ may decay directly to Standard Model particles.  
Since the the Higuchi bound requires $m \gtrsim H_\mathrm{inf}$ and since $H_\mathrm{inf}$ is much larger than the Standard Model mass scales, these decay channels cannot be blocked by kinematics.  For the sake of illustration, consider a scalar matter-sector field $\chi$ with matter action 
\ba{\label{eq:dchi_dchi_operator}
    - \sqrt{-g} \, g\indices{^\mu^\nu} \nabla_\mu^{(g)} \chi \nabla_\nu^{(g)} \chi
    = - \sqrt{-\bar{g}} 
    \, \biggl( 1 + \frac12 \frac{2M_*}{M_g^2}\bar{g}\indices{^\mu^\nu} v\indices{_\mu_\nu} + \order{v}^2 \biggr) 
    \, (\nabla\chi)^2 
    \;.
}
Note that the coupling of the $v\chi\chi$ vertex is proportional to $M_\ast / M_g^2 = \alpha / \Mpl$ where $\alpha = M_f / M_g$, and the full decay rate can be estimated as $\Gamma \sim (\alpha / \Mpl)^2 m^3$.  These parametric relations are consistent with earlier work~\cite{Babichev:2016bxi} that studied this scenario in more detail and related the decay of massive spin-2 particles to that of Kaluza-Klein modes~\cite{Han:1998sg}.

\end{appendix}

\bibliographystyle{JHEP}
\bibliography{main}

\providecommand{\href}[2]{#2}\begingroup\raggedright\begin{thebibliography}{10}

\bibitem{Baumann:2022mni}
D.~Baumann, \emph{{Cosmology}}, Cambridge University Press (7, 2022),
  \href{https://doi.org/10.1017/9781108937092}{10.1017/9781108937092}.

\bibitem{Schiappacasse:2016nei}
E.D.~Schiappacasse and L.H.~Ford, \emph{{Graviton Creation by Small Scale
  Factor Oscillations in an Expanding Universe}},
  \href{https://doi.org/10.1103/PhysRevD.94.084030}{\emph{Phys. Rev. D}
  {\bfseries 94} (2016) 084030}
  [\href{https://arxiv.org/abs/1602.08416}{{\ttfamily 1602.08416}}].

\bibitem{Parker:1969au}
L.~Parker, \emph{{Quantized fields and particle creation in expanding
  universes. 1.}}, \href{https://doi.org/10.1103/PhysRev.183.1057}{\emph{Phys.
  Rev.} {\bfseries 183} (1969) 1057}.

\bibitem{Parker:1971pt}
L.~Parker, \emph{{Quantized fields and particle creation in expanding
  universes. 2.}}, \href{https://doi.org/10.1103/PhysRevD.3.346}{\emph{Phys.
  Rev. D} {\bfseries 3} (1971) 346}.

\bibitem{Ford:2021syk}
L.H.~Ford, \emph{{Cosmological particle production: a review}},
  \href{https://doi.org/10.1088/1361-6633/ac1b23}{\emph{Rept. Prog. Phys.}
  {\bfseries 84} (2021) } [\href{https://arxiv.org/abs/2112.02444}{{\ttfamily
  2112.02444}}].

\bibitem{Chung:1998zb}
D.J.H.~Chung, E.W.~Kolb and A.~Riotto, \emph{{Superheavy dark matter}},
  \href{https://doi.org/10.1103/PhysRevD.59.023501}{\emph{Phys. Rev. D}
  {\bfseries 59} (1998) 023501}
  [\href{https://arxiv.org/abs/hep-ph/9802238}{{\ttfamily hep-ph/9802238}}].

\bibitem{Chung:1998ua}
D.J.H.~Chung, E.W.~Kolb and A.~Riotto, \emph{{Nonthermal supermassive dark
  matter}}, \href{https://doi.org/10.1103/PhysRevLett.81.4048}{\emph{Phys. Rev.
  Lett.} {\bfseries 81} (1998) 4048}
  [\href{https://arxiv.org/abs/hep-ph/9805473}{{\ttfamily hep-ph/9805473}}].

\bibitem{Alexander:2020gmv}
S.~Alexander, L.~Jenks and E.~McDonough, \emph{{Higher spin dark matter}},
  \href{https://doi.org/10.1016/j.physletb.2021.136436}{\emph{Phys. Lett. B}
  {\bfseries 819} (2021) 136436}
  [\href{https://arxiv.org/abs/2010.15125}{{\ttfamily 2010.15125}}].

\bibitem{Aoki:2014cla}
K.~Aoki and K.-i.~Maeda, \emph{{Dark matter in ghost-free bigravity theory:
  From a galaxy scale to the universe}},
  \href{https://doi.org/10.1103/PhysRevD.90.124089}{\emph{Phys. Rev. D}
  {\bfseries 90} (2014) 124089}
  [\href{https://arxiv.org/abs/1409.0202}{{\ttfamily 1409.0202}}].

\bibitem{Aoki:2016zgp}
K.~Aoki and S.~Mukohyama, \emph{{Massive gravitons as dark matter and
  gravitational waves}},
  \href{https://doi.org/10.1103/PhysRevD.94.024001}{\emph{Phys. Rev. D}
  {\bfseries 94} (2016) 024001}
  [\href{https://arxiv.org/abs/1604.06704}{{\ttfamily 1604.06704}}].

\bibitem{Babichev:2016bxi}
E.~Babichev, L.~Marzola, M.~Raidal, A.~Schmidt-May, F.~Urban, H.~Veerm\"ae
  et~al., \emph{{Heavy spin-2 Dark Matter}},
  \href{https://doi.org/10.1088/1475-7516/2016/09/016}{\emph{JCAP} {\bfseries
  09} (2016) 016} [\href{https://arxiv.org/abs/1607.03497}{{\ttfamily
  1607.03497}}].

\bibitem{Marzola:2017lbt}
L.~Marzola, M.~Raidal and F.R.~Urban, \emph{{Oscillating Spin-2 Dark Matter}},
  \href{https://doi.org/10.1103/PhysRevD.97.024010}{\emph{Phys. Rev. D}
  {\bfseries 97} (2018) 024010}
  [\href{https://arxiv.org/abs/1708.04253}{{\ttfamily 1708.04253}}].

\bibitem{GonzalezAlbornoz:2017gbh}
N.L.~Gonz\'alez~Albornoz, A.~Schmidt-May and M.~von Strauss, \emph{{Dark matter
  scenarios with multiple spin-2 fields}},
  \href{https://doi.org/10.1088/1475-7516/2018/01/014}{\emph{JCAP} {\bfseries
  01} (2018) 014} [\href{https://arxiv.org/abs/1709.05128}{{\ttfamily
  1709.05128}}].

\bibitem{Armaleo:2019gil}
J.M.~Armaleo, D.~L\'opez~Nacir and F.R.~Urban, \emph{{Binary pulsars as probes
  for spin-2 ultralight dark matter}},
  \href{https://doi.org/10.1088/1475-7516/2020/01/053}{\emph{JCAP} {\bfseries
  01} (2020) 053} [\href{https://arxiv.org/abs/1909.13814}{{\ttfamily
  1909.13814}}].

\bibitem{Armaleo:2020yml}
J.M.~Armaleo, D.~L\'opez~Nacir and F.R.~Urban, \emph{{Pulsar timing array
  constraints on spin-2 ULDM}},
  \href{https://doi.org/10.1088/1475-7516/2020/09/031}{\emph{JCAP} {\bfseries
  09} (2020) 031} [\href{https://arxiv.org/abs/2005.03731}{{\ttfamily
  2005.03731}}].

\bibitem{Manita:2022tkl}
Y.~Manita, K.~Aoki, T.~Fujita and S.~Mukohyama, \emph{{Spin-2 dark matter from
  anisotropic Universe in bigravity}},
  \href{https://arxiv.org/abs/2211.15873}{{\ttfamily 2211.15873}}.

\bibitem{Fierz:1939ix}
M.~Fierz and W.~Pauli, \emph{{On relativistic wave equations for particles of
  arbitrary spin in an electromagnetic field}},
  \href{https://doi.org/10.1098/rspa.1939.0140}{\emph{Proc. Roy. Soc. Lond. A}
  {\bfseries 173} (1939) 211}.

\bibitem{Boulware:1972yco}
D.G.~Boulware and S.~Deser, \emph{{Can gravitation have a finite range?}},
  \href{https://doi.org/10.1103/PhysRevD.6.3368}{\emph{Phys. Rev. D} {\bfseries
  6} (1972) 3368}.

\bibitem{Hassan:2011zd}
S.F.~Hassan and R.A.~Rosen, \emph{{Bimetric Gravity from Ghost-free Massive
  Gravity}}, \href{https://doi.org/10.1007/JHEP02(2012)126}{\emph{JHEP}
  {\bfseries 02} (2012) 126} [\href{https://arxiv.org/abs/1109.3515}{{\ttfamily
  1109.3515}}].

\bibitem{deRham:2014naa}
C.~de~Rham, L.~Heisenberg and R.H.~Ribeiro, \emph{{On couplings to matter in
  massive (bi-)gravity}},
  \href{https://doi.org/10.1088/0264-9381/32/3/035022}{\emph{Class. Quant.
  Grav.} {\bfseries 32} (2015) 035022}
  [\href{https://arxiv.org/abs/1408.1678}{{\ttfamily 1408.1678}}].

\bibitem{DeFelice:2013bxa}
A.~De~Felice, A.E.~G\"umr\"uk\c{c}\"uo\u{g}lu, C.~Lin and S.~Mukohyama,
  \emph{{On the cosmology of massive gravity}},
  \href{https://doi.org/10.1088/0264-9381/30/18/184004}{\emph{Class. Quant.
  Grav.} {\bfseries 30} (2013) 184004}
  [\href{https://arxiv.org/abs/1304.0484}{{\ttfamily 1304.0484}}].

\bibitem{deRham:2014zqa}
C.~de~Rham, \emph{{Massive Gravity}},
  \href{https://doi.org/10.12942/lrr-2014-7}{\emph{Living Rev. Rel.} {\bfseries
  17} (2014) 7} [\href{https://arxiv.org/abs/1401.4173}{{\ttfamily
  1401.4173}}].

\bibitem{Volkov:2011an}
M.S.~Volkov, \emph{{Cosmological solutions with massive gravitons in the
  bigravity theory}},
  \href{https://doi.org/10.1007/JHEP01(2012)035}{\emph{JHEP} {\bfseries 01}
  (2012) 035} [\href{https://arxiv.org/abs/1110.6153}{{\ttfamily 1110.6153}}].

\bibitem{vonStrauss:2011mq}
M.~von Strauss, A.~Schmidt-May, J.~Enander, E.~Mortsell and S.F.~Hassan,
  \emph{{Cosmological Solutions in Bimetric Gravity and their Observational
  Tests}}, \href{https://doi.org/10.1088/1475-7516/2012/03/042}{\emph{JCAP}
  {\bfseries 03} (2012) 042} [\href{https://arxiv.org/abs/1111.1655}{{\ttfamily
  1111.1655}}].

\bibitem{Comelli:2011zm}
D.~Comelli, M.~Crisostomi, F.~Nesti and L.~Pilo, \emph{{FRW Cosmology in Ghost
  Free Massive Gravity}},
  \href{https://doi.org/10.1007/JHEP03(2012)067}{\emph{JHEP} {\bfseries 03}
  (2012) 067} [\href{https://arxiv.org/abs/1111.1983}{{\ttfamily 1111.1983}}].

\bibitem{Volkov:2012cf}
M.S.~Volkov, \emph{{Exact self-accelerating cosmologies in the ghost-free
  bigravity and massive gravity}},
  \href{https://doi.org/10.1103/PhysRevD.86.061502}{\emph{Phys. Rev. D}
  {\bfseries 86} (2012) 061502}
  [\href{https://arxiv.org/abs/1205.5713}{{\ttfamily 1205.5713}}].

\bibitem{Akrami:2012vf}
Y.~Akrami, T.S.~Koivisto and M.~Sandstad, \emph{{Accelerated expansion from
  ghost-free bigravity: a statistical analysis with improved generality}},
  \href{https://doi.org/10.1007/JHEP03(2013)099}{\emph{JHEP} {\bfseries 03}
  (2013) 099} [\href{https://arxiv.org/abs/1209.0457}{{\ttfamily 1209.0457}}].

\bibitem{Koennig:2013fdo}
F.~Koennig, A.~Patil and L.~Amendola, \emph{{Viable cosmological solutions in
  massive bimetric gravity}},
  \href{https://doi.org/10.1088/1475-7516/2014/03/029}{\emph{JCAP} {\bfseries
  03} (2014) 029} [\href{https://arxiv.org/abs/1312.3208}{{\ttfamily
  1312.3208}}].

\bibitem{Comelli:2012db}
D.~Comelli, M.~Crisostomi and L.~Pilo, \emph{{Perturbations in Massive Gravity
  Cosmology}}, \href{https://doi.org/10.1007/JHEP06(2012)085}{\emph{JHEP}
  {\bfseries 06} (2012) 085} [\href{https://arxiv.org/abs/1202.1986}{{\ttfamily
  1202.1986}}].

\bibitem{Koennig:2014ods}
F.~Koennig, Y.~Akrami, L.~Amendola, M.~Motta and A.R.~Solomon, \emph{{Stable
  and unstable cosmological models in bimetric massive gravity}},
  \href{https://doi.org/10.1103/PhysRevD.90.124014}{\emph{Phys. Rev. D}
  {\bfseries 90} (2014) 124014}
  [\href{https://arxiv.org/abs/1407.4331}{{\ttfamily 1407.4331}}].

\bibitem{Lagos:2014lca}
M.~Lagos and P.G.~Ferreira, \emph{{Cosmological perturbations in massive
  bigravity}}, \href{https://doi.org/10.1088/1475-7516/2014/12/026}{\emph{JCAP}
  {\bfseries 12} (2014) 026} [\href{https://arxiv.org/abs/1410.0207}{{\ttfamily
  1410.0207}}].

\bibitem{Akrami:2015qga}
Y.~Akrami, S.F.~Hassan, F.~K\"onnig, A.~Schmidt-May and A.R.~Solomon,
  \emph{{Bimetric gravity is cosmologically viable}},
  \href{https://doi.org/10.1016/j.physletb.2015.06.062}{\emph{Phys. Lett. B}
  {\bfseries 748} (2015) 37}
  [\href{https://arxiv.org/abs/1503.07521}{{\ttfamily 1503.07521}}].

\bibitem{Comelli:2014bqa}
D.~Comelli, M.~Crisostomi and L.~Pilo, \emph{{FRW Cosmological Perturbations in
  Massive Bigravity}},
  \href{https://doi.org/10.1103/PhysRevD.90.084003}{\emph{Phys. Rev. D}
  {\bfseries 90} (2014) 084003}
  [\href{https://arxiv.org/abs/1403.5679}{{\ttfamily 1403.5679}}].

\bibitem{Enander:2014xga}
J.~Enander, A.R.~Solomon, Y.~Akrami and E.~Mortsell, \emph{{Cosmic expansion
  histories in massive bigravity with symmetric matter coupling}},
  \href{https://doi.org/10.1088/1475-7516/2015/01/006}{\emph{JCAP} {\bfseries
  01} (2015) 006} [\href{https://arxiv.org/abs/1409.2860}{{\ttfamily
  1409.2860}}].

\bibitem{EmirGumrukcuoglu:2014uog}
A.~Emir~G\"umr\"uk\c{c}\"uo\u{g}lu, L.~Heisenberg and S.~Mukohyama,
  \emph{{Cosmological perturbations in massive gravity with doubly coupled
  matter}}, \href{https://doi.org/10.1088/1475-7516/2015/02/022}{\emph{JCAP}
  {\bfseries 02} (2015) 022} [\href{https://arxiv.org/abs/1409.7260}{{\ttfamily
  1409.7260}}].

\bibitem{Gumrukcuoglu:2015nua}
A.E.~Gumrukcuoglu, L.~Heisenberg, S.~Mukohyama and N.~Tanahashi,
  \emph{{Cosmology in bimetric theory with an effective composite coupling to
  matter}}, \href{https://doi.org/10.1088/1475-7516/2015/04/008}{\emph{JCAP}
  {\bfseries 04} (2015) 008}
  [\href{https://arxiv.org/abs/1501.02790}{{\ttfamily 1501.02790}}].

\bibitem{Higuchi:1986py}
A.~Higuchi, \emph{{Forbidden Mass Range for Spin-2 Field Theory in De Sitter
  Space-time}}, \href{https://doi.org/10.1016/0550-3213(87)90691-2}{\emph{Nucl.
  Phys. B} {\bfseries 282} (1987) 397}.

\bibitem{Fasiello:2012rw}
M.~Fasiello and A.J.~Tolley, \emph{{Cosmological perturbations in Massive
  Gravity and the Higuchi bound}},
  \href{https://doi.org/10.1088/1475-7516/2012/11/035}{\emph{JCAP} {\bfseries
  11} (2012) 035} [\href{https://arxiv.org/abs/1206.3852}{{\ttfamily
  1206.3852}}].

\bibitem{Fasiello:2013woa}
M.~Fasiello and A.J.~Tolley, \emph{{Cosmological Stability Bound in Massive
  Gravity and Bigravity}},
  \href{https://doi.org/10.1088/1475-7516/2013/12/002}{\emph{JCAP} {\bfseries
  12} (2013) 002} [\href{https://arxiv.org/abs/1308.1647}{{\ttfamily
  1308.1647}}].

\bibitem{deRham:2010kj}
C.~de~Rham, G.~Gabadadze and A.J.~Tolley, \emph{{Resummation of Massive
  Gravity}}, \href{https://doi.org/10.1103/PhysRevLett.106.231101}{\emph{Phys.
  Rev. Lett.} {\bfseries 106} (2011) 231101}
  [\href{https://arxiv.org/abs/1011.1232}{{\ttfamily 1011.1232}}].

\bibitem{MTW:1973}
C.W.~Misner, K.S.~Thorne and J.A.~Wheeler, \emph{{Gravitation}}, {W. H. Freeman
  and Company}, San Francisco (1973).

\bibitem{Hassan:2011hr}
S.F.~Hassan and R.A.~Rosen, \emph{{Resolving the Ghost Problem in non-Linear
  Massive Gravity}},
  \href{https://doi.org/10.1103/PhysRevLett.108.041101}{\emph{Phys. Rev. Lett.}
  {\bfseries 108} (2012) 041101}
  [\href{https://arxiv.org/abs/1106.3344}{{\ttfamily 1106.3344}}].

\bibitem{Schmidt-May:2014xla}
A.~Schmidt-May, \emph{{Mass eigenstates in bimetric theory with matter
  coupling}}, \href{https://doi.org/10.1088/1475-7516/2015/01/039}{\emph{JCAP}
  {\bfseries 01} (2015) 039} [\href{https://arxiv.org/abs/1409.3146}{{\ttfamily
  1409.3146}}].

\bibitem{Mukhanov:2005}
V.~Mukhanov, \emph{{Physical Foundations of Cosmology}}, {Cambridge University
  Press}, {Cambridge, UK} (2005).

\bibitem{Baumann:2009ds}
D.~Baumann, \emph{{Inflation}},  in \emph{{Theoretical Advanced Study Institute
  in Elementary Particle Physics}: {Physics of the Large and the Small}},
  pp.~523--686, 2011, \href{https://doi.org/10.1142/9789814327183_0010}{DOI}
  [\href{https://arxiv.org/abs/0907.5424}{{\ttfamily 0907.5424}}].

\bibitem{Hasegawa:2017hgd}
F.~Hasegawa, K.~Mukaida, K.~Nakayama, T.~Terada and Y.~Yamada, \emph{{Gravitino
  Problem in Minimal Supergravity Inflation}},
  \href{https://doi.org/10.1016/j.physletb.2017.02.030}{\emph{Phys. Lett. B}
  {\bfseries 767} (2017) 392}
  [\href{https://arxiv.org/abs/1701.03106}{{\ttfamily 1701.03106}}].

\bibitem{Kolb:2021xfn}
E.W.~Kolb, A.J.~Long and E.~McDonough, \emph{{Catastrophic production of slow
  gravitinos}}, \href{https://doi.org/10.1103/PhysRevD.104.075015}{\emph{Phys.
  Rev. D} {\bfseries 104} (2021) 075015}
  [\href{https://arxiv.org/abs/2102.10113}{{\ttfamily 2102.10113}}].

\bibitem{Deser:1983mm}
S.~Deser and R.I.~Nepomechie, \emph{{Gauge Invariance Versus Masslessness in De
  Sitter Space}},
  \href{https://doi.org/10.1016/0003-4916(84)90156-8}{\emph{Annals Phys.}
  {\bfseries 154} (1984) 396}.

\bibitem{Deser:1983tm}
S.~Deser and R.I.~Nepomechie, \emph{{Anomalous Propagation of Gauge Fields in
  Conformally Flat Spaces}},
  \href{https://doi.org/10.1016/0370-2693(83)90317-9}{\emph{Phys. Lett. B}
  {\bfseries 132} (1983) 321}.

\bibitem{Comelli:2015pua}
D.~Comelli, M.~Crisostomi, K.~Koyama, L.~Pilo and G.~Tasinato, \emph{{Cosmology
  of bigravity with doubly coupled matter}},
  \href{https://doi.org/10.1088/1475-7516/2015/04/026}{\emph{JCAP} {\bfseries
  04} (2015) 026} [\href{https://arxiv.org/abs/1501.00864}{{\ttfamily
  1501.00864}}].

\bibitem{Cook:2015vqa}
J.L.~Cook, E.~Dimastrogiovanni, D.A.~Easson and L.M.~Krauss, \emph{{Reheating
  predictions in single field inflation}},
  \href{https://doi.org/10.1088/1475-7516/2015/04/047}{\emph{JCAP} {\bfseries
  04} (2015) 047} [\href{https://arxiv.org/abs/1502.04673}{{\ttfamily
  1502.04673}}].

\bibitem{deSalas:2015glj}
P.F.~de~Salas, M.~Lattanzi, G.~Mangano, G.~Miele, S.~Pastor and O.~Pisanti,
  \emph{{Bounds on very low reheating scenarios after Planck}},
  \href{https://doi.org/10.1103/PhysRevD.92.123534}{\emph{Phys. Rev. D}
  {\bfseries 92} (2015) 123534}
  [\href{https://arxiv.org/abs/1511.00672}{{\ttfamily 1511.00672}}].

\bibitem{Planck:2018vyg}
{\scshape Planck} collaboration, \emph{{Planck 2018 results. VI. Cosmological
  parameters}},
  \href{https://doi.org/10.1051/0004-6361/201833910}{\emph{Astron. Astrophys.}
  {\bfseries 641} (2020) A6}
  [\href{https://arxiv.org/abs/1807.06209}{{\ttfamily 1807.06209}}].

\bibitem{Kumekawa:1994gx}
K.~Kumekawa, T.~Moroi and T.~Yanagida, \emph{{Flat potential for inflaton with
  a discrete R invariance in supergravity}},
  \href{https://doi.org/10.1143/PTP.92.437}{\emph{Prog. Theor. Phys.}
  {\bfseries 92} (1994) 437}
  [\href{https://arxiv.org/abs/hep-ph/9405337}{{\ttfamily hep-ph/9405337}}].

\bibitem{Izawa:1996dv}
K.I.~Izawa and T.~Yanagida, \emph{{Natural new inflation in broken
  supergravity}},
  \href{https://doi.org/10.1016/S0370-2693(96)01638-3}{\emph{Phys. Lett. B}
  {\bfseries 393} (1997) 331}
  [\href{https://arxiv.org/abs/hep-ph/9608359}{{\ttfamily hep-ph/9608359}}].

\bibitem{hairer_norsett_wanner_2009}
E.~Hairer, S.P.~Nørsett and G.~Wanner, \emph{Solving ordinary differential
  equations I: Nonstiff problems}, Springer (2009).

\bibitem{Kolb:2020fwh}
E.W.~Kolb and A.J.~Long, \emph{{Completely dark photons from gravitational
  particle production during the inflationary era}},
  \href{https://doi.org/10.1007/JHEP03(2021)283}{\emph{JHEP} {\bfseries 03}
  (2021) 283} [\href{https://arxiv.org/abs/2009.03828}{{\ttfamily
  2009.03828}}].

\bibitem{Ling:2021zlj}
S.~Ling and A.J.~Long, \emph{{Superheavy scalar dark matter from gravitational
  particle production in $\alpha$-attractor models of inflation}},
  \href{https://doi.org/10.1103/PhysRevD.103.103532}{\emph{Phys. Rev. D}
  {\bfseries 103} (2021) 103532}
  [\href{https://arxiv.org/abs/2101.11621}{{\ttfamily 2101.11621}}].

\bibitem{Chung:2004nh}
D.J.H.~Chung, E.W.~Kolb, A.~Riotto and L.~Senatore, \emph{{Isocurvature
  constraints on gravitationally produced superheavy dark matter}},
  \href{https://doi.org/10.1103/PhysRevD.72.023511}{\emph{Phys. Rev. D}
  {\bfseries 72} (2005) 023511}
  [\href{https://arxiv.org/abs/astro-ph/0411468}{{\ttfamily
  astro-ph/0411468}}].

\bibitem{Ema:2018ucl}
Y.~Ema, K.~Nakayama and Y.~Tang, \emph{{Production of Purely Gravitational Dark
  Matter}}, \href{https://doi.org/10.1007/JHEP09(2018)135}{\emph{JHEP}
  {\bfseries 09} (2018) 135}
  [\href{https://arxiv.org/abs/1804.07471}{{\ttfamily 1804.07471}}].

\bibitem{Chung:2018ayg}
D.J.H.~Chung, E.W.~Kolb and A.J.~Long, \emph{{Gravitational production of
  super-Hubble-mass particles: an analytic approach}},
  \href{https://doi.org/10.1007/JHEP01(2019)189}{\emph{JHEP} {\bfseries 01}
  (2019) 189} [\href{https://arxiv.org/abs/1812.00211}{{\ttfamily
  1812.00211}}].

\bibitem{Basso:2021whd}
E.E.~Basso and D.J.H.~Chung, \emph{{Computation of gravitational particle
  production using adiabatic invariants}},
  \href{https://doi.org/10.1007/JHEP11(2021)146}{\emph{JHEP} {\bfseries 11}
  (2021) 146} [\href{https://arxiv.org/abs/2108.01653}{{\ttfamily
  2108.01653}}].

\bibitem{Kaneta:2022gug}
K.~Kaneta, S.M.~Lee and K.-y.~Oda, \emph{{Boltzmann or Bogoliubov? Approaches
  compared in gravitational particle production}},
  \href{https://doi.org/10.1088/1475-7516/2022/09/018}{\emph{JCAP} {\bfseries
  09} (2022) 018} [\href{https://arxiv.org/abs/2206.10929}{{\ttfamily
  2206.10929}}].

\bibitem{Basso:2022tpd}
E.~Basso, D.J.H.~Chung, E.W.~Kolb and A.J.~Long, \emph{{Quantum interference in
  gravitational particle production}},
  \href{https://arxiv.org/abs/2209.01713}{{\ttfamily 2209.01713}}.

\bibitem{Graham:2015rva}
P.W.~Graham, J.~Mardon and S.~Rajendran, \emph{{Vector Dark Matter from
  Inflationary Fluctuations}},
  \href{https://doi.org/10.1103/PhysRevD.93.103520}{\emph{Phys. Rev. D}
  {\bfseries 93} (2016) 103520}
  [\href{https://arxiv.org/abs/1504.02102}{{\ttfamily 1504.02102}}].

\bibitem{Amin:2014eta}
M.A.~Amin, M.P.~Hertzberg, D.I.~Kaiser and J.~Karouby, \emph{{Nonperturbative
  Dynamics Of Reheating After Inflation: A Review}},
  \href{https://doi.org/10.1142/S0218271815300037}{\emph{Int. J. Mod. Phys. D}
  {\bfseries 24} (2014) 1530003}
  [\href{https://arxiv.org/abs/1410.3808}{{\ttfamily 1410.3808}}].

\bibitem{Garny:2015sjg}
M.~Garny, M.~Sandora and M.S.~Sloth, \emph{{Planckian Interacting Massive
  Particles as Dark Matter}},
  \href{https://doi.org/10.1103/PhysRevLett.116.101302}{\emph{Phys. Rev. Lett.}
  {\bfseries 116} (2016) 101302}
  [\href{https://arxiv.org/abs/1511.03278}{{\ttfamily 1511.03278}}].

\bibitem{Bernal:2018qlk}
N.~Bernal, M.~Dutra, Y.~Mambrini, K.~Olive, M.~Peloso and M.~Pierre,
  \emph{{Spin-2 Portal Dark Matter}},
  \href{https://doi.org/10.1103/PhysRevD.97.115020}{\emph{Phys. Rev. D}
  {\bfseries 97} (2018) 115020}
  [\href{https://arxiv.org/abs/1803.01866}{{\ttfamily 1803.01866}}].

\bibitem{Dimastrogiovanni:2018uqy}
E.~Dimastrogiovanni, M.~Fasiello and G.~Tasinato, \emph{{Probing the
  inflationary particle content: extra spin-2 field}},
  \href{https://doi.org/10.1088/1475-7516/2018/08/016}{\emph{JCAP} {\bfseries
  08} (2018) 016} [\href{https://arxiv.org/abs/1806.00850}{{\ttfamily
  1806.00850}}].

\bibitem{Brizuela:2008ra}
D.~Brizuela, J.M.~Martin-Garcia and G.A.~Mena~Marugan, \emph{{xPert: Computer
  algebra for metric perturbation theory}},
  \href{https://doi.org/10.1007/s10714-009-0773-2}{\emph{Gen. Rel. Grav.}
  {\bfseries 41} (2009) 2415}
  [\href{https://arxiv.org/abs/0807.0824}{{\ttfamily 0807.0824}}].

\bibitem{Hinterbichler:2011tt}
K.~Hinterbichler, \emph{{Theoretical Aspects of Massive Gravity}},
  \href{https://doi.org/10.1103/RevModPhys.84.671}{\emph{Rev. Mod. Phys.}
  {\bfseries 84} (2012) 671} [\href{https://arxiv.org/abs/1105.3735}{{\ttfamily
  1105.3735}}].

\bibitem{Bonifacio:2017nnt}
J.~Bonifacio, K.~Hinterbichler, A.~Joyce and R.A.~Rosen, \emph{{Massive and
  Massless Spin-2 Scattering and Asymptotic Superluminality}},
  \href{https://doi.org/10.1007/JHEP06(2018)075}{\emph{JHEP} {\bfseries 06}
  (2018) 075} [\href{https://arxiv.org/abs/1712.10020}{{\ttfamily
  1712.10020}}].

\bibitem{Han:1998sg}
T.~Han, J.D.~Lykken and R.-J.~Zhang, \emph{{On Kaluza-Klein states from large
  extra dimensions}},
  \href{https://doi.org/10.1103/PhysRevD.59.105006}{\emph{Phys. Rev. D}
  {\bfseries 59} (1999) 105006}
  [\href{https://arxiv.org/abs/hep-ph/9811350}{{\ttfamily hep-ph/9811350}}].

\end{thebibliography}\endgroup

\end{document}